\documentclass[%
 reprint,
nofootinbib,
 amsmath,amssymb,
 aps,
 showkeys,
 showpacs
]{revtex4-1}

\usepackage[utf8]{inputenc}

\usepackage{graphicx}%
\usepackage{dcolumn}
\usepackage{bm}

\usepackage{amsfonts}
\usepackage{textcomp}
\usepackage[british]{babel}
\usepackage{hyphenat}
\usepackage{color}
\def\ds {\ensuremath{\displaystyle}}
\def\lsk {\ensuremath{\mathcal{L}_{\rm Skyrme}}}

\def\ug       {\ensuremath{\!&=&\!}}
\def\C        {\ensuremath{\mathbb{C}}}
\def\R        {\ensuremath{\mathbb{R}}}
\def\N        {\ensuremath{\mathbb{N}}}

\def\ie       {i.e.} 
\def\re       {\ensuremath{{\rm Re}}}
\def\im       {\ensuremath{{\rm Im}}}
\def\bfit     {\ensuremath{b_{\rm fit}}}
\def\hh       {\hspace{10 mm}}
\def\hhh      {\hspace{10 mm}}
\def\g        {\ensuremath{\gamma}}
\def\e        {\ensuremath{\epsilon}}
\def\ee       {\ensuremath{e^+e^-}}
\def\pp       {\ensuremath{p\overline{p}}}
\def\qth      {\ensuremath{q^2_{\rm theo}}}
\def\qphy     {\ensuremath{q^2_{\rm phys}}}
\def\bfr      {\begin{flushright}}
\def\efr      {\end{flushright}}
\def\bm       {\begin{minipage}}
\def\em       {\end{minipage}}
\def\P		  {Pad\'e}
\def\lt       {\left(}
\def\rt       {\right)}
\def\lq       {\left[}
\def\rq       {\right]}
\renewcommand{\r}[1]{{\color{black} #1}}

\renewcommand{\=}[1]{& #1 &}
\newcommand{\ov}[1]{\overline{#1}}
\newcommand{\be}{\begin{eqnarray}}
\newcommand{\en}{\end{eqnarray}}
\newcommand{\nen}{\nonumber\end{eqnarray}}
\newcommand{\no}{\nonumber}
\newcommand{\n}[1]{\left\lVert#1\right\rVert}


\begin{document}

\preprint{APS/123-QED}

\title{Analytic continuation of nucleon electromagnetic form factors in the time-like region}% 
\author{Pedro Alberto}
\affiliation{%
 CFisUC, Physics Department of the University of Coimbra, Coimbra, Portugal}%
\author{Alessandro Drago}
\affiliation{%
 Dipartimento di Fisica e Scienze della Terra, Universit\`a di Ferrara and INFN Sezione di Ferrara, Ferrara, Italy}%
\author{Alessio Mangoni}
\affiliation{%
INFN Sezione di Perugia, Perugia, Italy}%
\author{Simone Moretti}
\affiliation{%
%Departamento de Fisica Aplicada, Universidad de Salamanca, Salamanca, Spain
Fachbereich Physik, Universit\"at Konstanz, 78457 Konstanz, Germany
}%
\author{Simone Pacetti}
 \email{simone.pacetti@pg.infn.it}
\affiliation{%
Dipartimento di Fisica e Geologia, Universit\`a degli Studi di Perugia and INFN Sezione di Perugia, Perugia, Italy}%
\begin{abstract}
The possibility to compute nucleon electromagnetic form factors in the time-like region by analytic continuation of their space-like expressions, obtained in the framework of a generic model of nucleons, has been explored. We have developed a procedure to solve analytically Fourier transforms of the nucleon electromagnetic current and hence to obtain form factors defined in all kinematic regions and fulfilling the first-principles requirements of analyticity and unitarity. The results obtained in the particular case of the Skyrme model are discussed and compared to data, both in space-like and time-like region.
\end{abstract}

% General theory of currents, Electromagnetic form factors, Skyrmions
\maketitle
%
%\tableofcontents
%
%
%  Article body
%
%

\section{Introduction}
\label{intro}
\begin{figure}[h!]
\begin{center}
\includegraphics[width=75mm]{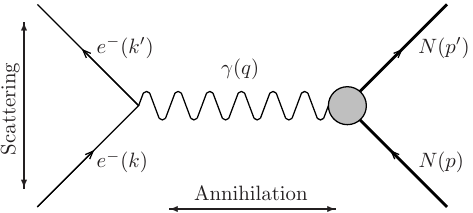}
\caption{Feynman diagram in Born approximation for $e^+e^-\leftrightarrow N\ov{N}$ and $eN\rightarrow eN$. The solid disc at the nucleon vertex symbolizes FFs.}
\label{art-fig:feyn1}
\end{center}
\end{figure}

Nucleon electromagnetic form factors~\cite{sakurai} (FFs) are Lorentz scalar functions of the squared four-momentum transfer of the photon, $q^2$, that parametrize those degrees of freedom of the nucleon electromagnetic current, which are not constrained by Lorentz and gauge invariance. 
\\
They represent a unique source of information about the internal structure of nucleons. In particular, in the non-relativistic limit (low $q^2$), FFs can be interpreted as the Fourier transforms of the electric charge and magnetic momentum spatial distributions of the nucleon.
\\
From the point of view of quantum field theory, being related to the electromagnetic current and hence only to the Born amplitude (one-photon exchange), FFs embody the resummation of all high order processes with two nucleons and one photon as external particles.
\\
Such high-order processes represent the connection with quantum chromodynamics (QCD) of FFs, that indeed could be described in terms of hadronic loops, involving virtual mesons and baryons. Due to the large number of hadron ``species'' to be accounted for and also to the unknown couplings among them, a direct calculations of FFs in the framework of QCD, especially in the low-$q^2$ regime, is a very hard task.
 \\
Nevertheless, interesting results have been obtained by lattice calculations~\cite{lattice} and effective model approaches, such as: chiral perturbation theory~\cite{chipt}, chiral soliton models~\cite{pedro}, large-$N_c$ approximation~\cite{largeNc} and holographic QCD~\cite{holographic}. 
However, in the majority of these cases, the obtained FF descriptions are restricted to the only space-like (SL) region.
\\
In general, two kinds of FFs could be identified: 
\begin{itemize}
	\item SL FFs (SLFFs), related to the elastic scattering process $e^-N\to e^-N$ ($N$ and $e^-$ stand for nucleon and electron respectively), which occurs with $q^2<0$ (see, for instance, Refs.~\cite{Perdrisat} and~\cite{egle-simone});
	\item time-like (TL) FFs (TLFFs), related to the annihilation processes $\ee\leftrightarrow N\ov{N}$, where $q^2>(2M_N)^2$, $M_N$ is the nucleon mass (for a review see  Ref.~\cite{Salme} and references therein). 
\end{itemize}
%\\
The scattering and annihilation processes are related by crossing symmetry, which, considering only the Born approximation, see Fig.~\ref{art-fig:feyn1}, implies that SLFFs and TLFFs represent values, for negative and positive $q^2$ respectively, of a unique function of $q^2$, simply named FF.
\\
As a consequence, in order to understand the meaning of FFs, especially in the TL region, where the interpretation in terms of Fourier transforms of spatial distributions fails, we must adopt descriptions or parametrizations defined in the whole kinematic region.
Moreover, new FF data, coming from different experiments\footnote{BESIII~\cite{besiii} at BEPCII in Beijing, China; SND~\cite{snd} and CMD3~\cite{cmd3} at VEPP-2M in Novosibirsk, Russia; PANDA at FAIR in Darmstadt, Germany~\cite{panda}.}, should help in shedding light especially in the more puzzling TL region.
\vspace{-0mm}\\
Various techniques and procedures have been proposed to develop such a SL-TL unified description of nucleon FFs. Many of them make use of dispersion relations, see for example Refs.~\cite{Holzwarth1,Holzwarth2,Hammer}, others suggest new models (for instance, in Ref.~\cite{Egle}, a semi-phenomenological microscopic model is proposed), and some others use analytic continuation methods to extend, to all values of $q^2$, parameterizations usually defined only in the SL or TL region~\cite{Pacetti}. 
\\
We will expound here a procedure, originally formulated in Ref.~\cite{Bardini}, that allows to make the analytic continuation to the whole $q^2$ complex plane of a  parametrization of FFs, initially conceived for a particular reference frame. 
More in detail, nucleon FFs are computed in the Breit frame as Fourier transforms of the time and space components of the electromagnetic current. This particular representation is defined only in the SL region, \ie, the Fourier integrals, which depend on $q^2$, converge only for SL four-momenta.
However, if such a representation can be analytically solved, that is, the Fourier transforms are obtained as analytic functions of $q^2$, 
instead of discrete numeric values at each four-momentum transfer,
then the FF parameterizations should be valid in all non-singular points of the $q^2$ complex plane.
\\
In particular, the nucleon electromagnetic current has been computed in the framework of the Skyrme model \cite{Skyrme,Braaten}, by solving numerically a set of non linear differential equations.
The most relevant aspect of the procedure outlined here, consists in 
assigning to these numerical solutions opportune analytic expressions, so that their Fourier transforms embody the properties required for FFs by analyticity and unitarity.
\\
The structure of the article is the following: in the second section we briefly introduce FFs in SL and TL regions and describe their analytic properties. In the third section we review the Skyrme model and calculate SLFFs. In the fourth section we illustrate the method of analytic continuation and the obtained results. In closing, we discuss the main issues of these results, also in comparison with all available FF data. 
\subsection{Space-like form factors}
\label{subsec:sl-ffs}
In the scattering channel, Fig.~\ref{art-fig:feyn1} vertical direction, the Feynman amplitude of the nucleon vertex, $N\to \gamma^*N$, is parametrized as~\cite{Foldy}
\be 
\langle N'(p')| J^{\mu}(0)|N(p)\rangle &=& \bar{u}(p')\left(
\tilde F^N_{1}(q^2)\gamma^{\mu}
\right.\no\\
&&\left.+\,i\frac{\sigma^{\mu\nu}q_{\nu}}{2M_N}\tilde F_2^N(q^2)\right)
u(p) \label{foldy} \,,
\label{eq:dirac-pauli}
\en
where the four-momenta follow the labelling of Fig.~\ref{art-fig:feyn1} and $\tilde F^N_1(q^2)$ and $\tilde F^N_2(q^2)$ are the so-called Dirac and Pauli FFs (the ``tilde'' indicates their SL definition). They are Lorentz scalar functions and, as a consequence of the hermiticity of the current operator $J^\mu$ and the time reversal symmetry, are real for $q^2\le 0$. At $q^2=0$ the Dirac FF is normalized to the nucleon charge $Q_N$, in units of the positron charge, while the Pauli FF is normalized to the anomalous magnetic moment $\kappa_N$, in units of the Bohr magneton $\mu_B$,
\be
\tilde F^N_1(0)=Q_N\,,\hh \tilde F^N_2(0)=\kappa_N\,.
\label{norm:f1f2}
\en
In the special frame, called Breit frame, where there is no energy exchange, hence: $p=(E,-\vec{q}/2)$, $p'=(E,\vec{q}/2)$ and
$q=(0,\vec{q})$, the time and space components of the current expectation value, Eq.~(\ref{foldy}), reduce to
\be
\begin{array}{rcl}
\langle N'(p')| J^{0}(0)|N(p)\rangle \!\!\!\ug\! \ds
\tilde F^N_1(q^2)+\frac{q^2}{4M_N^2}\tilde F^N_2(q^2)\,,
\\
&&\\
\langle N'(p')| \vec{J}(0)|N(p)\rangle \!\!\!\ug\!
\bar{u}(p')\vec \gamma\, u(p)\Big(\tilde F^N_1(q^2)\!+\!\tilde F^N_2(q^2)\Big)
.\\
\end{array}\hspace{-3mm}
\label{eq:current-ffs}
\en
These combinations of the Dirac and Pauli FFs, representing the Fourier transforms of charge and magnetization spatial distributions of the nucleon, define the electric and magnetic Sachs FFs~\cite{Sachs}
\be 
\tilde G_E^N(q^2)&=&\tilde F_{1}^N(q^2)+ \frac{q^2}{4M_N^2}\tilde F_{2}^N(q^2)\,,\no\\
&& \label{eq:sachs}\\
\tilde G_M^N(q^2)&=&\tilde F_1^N(q^2)+\tilde F_2^N(q^2)\,,
\no
\en
that, following Eq.~(\ref{norm:f1f2}), are normalized at $q^2=0$ as
\be 
\tilde G^N_E(0)=Q_N\,,
\hh 
\tilde G^N_M(0)=Q_N+\kappa_N\equiv \mu_N\,,
\nen
where $\mu_N$ is the total magnetic moment of the nucleon.
Isoscalar (isospin $I=0$) and isovector (isospin $I=1$) components are obtained by the following combinations of proton and neutron FFs
\be 
\begin{array}{rcl c rcl}
\ds \tilde F^{S}_{1,2}\ug \ds
\frac{\tilde F^{p}_{1,2}+\tilde F^{n}_{1,2}}{2}\,,&\hspace{3mm} &
\tilde F^{V}_{1,2}\ug \ds
\frac{\tilde F^{p}_{1,2}-\tilde F^{n}_{1,2}}{2}\,,\\
&& && &&\\
\ds \tilde G^{S}_{E,M}\ug \ds
\frac{\tilde G^{p}_{E,M}+\tilde G^{n}_{E,M}}{2}\,,& &
\tilde G^{V}_{E,M}\ug \ds
\frac{\tilde G^{p}_{E,M}-\tilde G^{n}_{E,M}}{2}\,. \\
\end{array}
\label{eq:SV-pn}
\en
\subsection{Time-like form factors}
\label{subsec:TLFFs}
In case of annihilation, Fig.~\ref{art-fig:feyn1} horizontal direction,
following the notation of Eq.~(\ref{eq:dirac-pauli}),
the amplitude for the nucleon-antinucleon production, $\gamma^*\to N\ov{N}$, is
\be 
\langle N(p)\bar{N}(p')|J^{\mu}(0)|0\rangle &=&
\bar{u}(p)\left(\bar{F}^N_1(q^2)\gamma^{\mu}\right.\no\\
&&\left.+i\frac{\sigma^{\mu\nu}q_{\nu}}{2M_N}\bar{F}_2^N(q^2)\right)v(p')\,,
\nonumber
\en
where $\bar{F}^N_1(q^2)$ and $\bar{F}^N_2(q^2)$ are the Dirac and Pauli FFs in the TL region, as indicated by the over-bar. Even in this case, the hermiticity of $J^\mu$ and the time reversal symmetry would imply real TLFFs. However this would be true only if the TL photon had not enough virtual mass, $q^2$, to produce physical particles as intermediate states. Otherwise, when the values of $q^2$ exceed the mass squared of the lightest allowed intermediate state, the amplitude, and hence the TLFFs, become complex. The rising of a finite imaginary part is a consequence of unitarity and can be formally demonstrated by considering the optical theorem. 
\\
 Since the lightest hadronic physical (on-shell particles) state, allowed by quantum number conservation, is the two-pion one, the imaginary part of the amplitude is different from zero starting from the so-called theoretical threshold $\qth= (2M_\pi)^2$, where $M_\pi$ is the pion mass. 
 In light of this non-vanishing imaginary part, the hermiticity of the current operator and the time reversal symmetry enforce, for the FFs, instead of reality, the Schwarz reflection principle and hence a discontinuity across the half line $(\qth,\infty)$. Such a portion of the TL region is then excluded from the analyticity domain or, in other words, it represents a branch cut.
\\
From the experimental point of view, the extraction of TLFF data involves additional difficulties with respect to SLFFs. First of all,
TLFFs are complex so, to have a complete determination, moduli and phases, or imaginary and real parts, should be measured. However, even by using polarization observables~\cite{polarization-tl}, only relative phases between electric and magnetic Sachs FFs are accessible.
\\
Moreover, since TL data are extracted from the cross section of the annihilation processes $\ee\leftrightarrow N\ov{N}$, TLFFs can be measured only for $q^2$ values above the so-called physical threshold $\qphy= (2M_N)^2$. It follow that the TL interval $[0,(2M_N)^2]$, where TLFFs are still well defined and receive also the most important contributions from hadronic intermediate states, is not experimentally accessible and for that reason it is called ``unphysical region''.
\\
As already stated~\cite{Drell}, taking advantage from crossing relations, SLFFs and TLFFs are interpreted as limit values, over the negative and positive real axis, respectively, of unique functions, $F_{1,2}^N(q^2)$, defined in the whole $q^2$ complex plane with the discontinuity cut $\big(\qth,\infty\big)$, due to unitarity (optical theorem). In more detail
\be 
\left\{
\begin{array}{rcl l l} 
 \!\!\tilde F^N_{i}(q^2)\ug\ds\lim_{\e\to 0}F^N_{i} (q^2\!\pm\! i\e)   \hspace{4mm}&q^2<0& \mbox{(SL)}\\
&&&\\
 \!\!\bar F^N_{i}(q^2)\ug\ds\lim_{\e\to 0}F^N_{i} (q^2\!\pm\! i\e)   &0\le q^2\le \qth &\mbox{(TL)}\\
&&&\\
 \!\!\bar F^N_{i}(q^2)\ug\ds\lim_{\e\to 0}F^N_{i} (q^2\!+\!i\e)   &q^2>\qth
 & \mbox{(TL)}\\\end{array}\right.\,, 
\nen
with $i=1,2$ and the limits in the SL region and in the portion of TL region up to the theoretical threshold can be taken indifferently from above or below the real axis, because there is no discontinuity there. On the other hand the limit values of FFs around the cut, in the TL region, depend on which edges of the cut is considered. In particular, as a consequence of the Schwarz reflection principle, the FF values in the upper and lower edges are complex conjugates, \ie, as $\e\to0^+$,
\be
F_{1,2}(q^2+i\e)=F_{1,2}^*(q^2-i\e)\,,\hh q^2\ge \qth\,.
\nen
We omitted tilde and over-bar because such a relation does hold in both SL and TL regions.  
\subsection{Analytic properties of form factors}
\label{subsec:analyticity}
Analyticity and unitarity, as well as perturbative QCD (pQCD), determine important model-independent features of FFs (some of which have already been touched upon in the previous section). Any reliable model of FFs must be able to reproduce such fundamental features, that concern the analytic structure of FFs as functions of the complex four-momentum square and their asymptotic behavior, \ie, the power law that rules their vanishing as 
$|q^2|\to\infty$. 
\\
Below we list, without proof, the main properties of FFs that will be addressed and discussed in the next sections.
\begin{itemize}
\item Form factors are function of $q^2$, analytic in the whole complex plane except for the branch cut $(\qth,\infty)$. Physical FFs are defined as the values of such functions for real  $q^2$. Moreover, from the experimental point of view, FFs are measurable only for
$q^2<0$ (SL region), and $q^2\ge\qphy$ (a subset of TL region).
The TL interval $(0,\qphy)$, being experimentally forbidden for FFs, is called unphysical region. 
\item At high momentum transfer we can invoke the pQCD or the quark counting rule~\cite{Brodsky} to infer the FF asymptotic behavior. In particular, in the scattering channel in order to maintain the nucleon entirety, the four-momentum transferred by the virtual photon must be shared among the three valence quarks via gluon-exchanges. The minimal number of gluons to be exchanged is two and hence the FFs must contain terms with, at least, two gluon propagators that entail the power law behavior
\be 
G^N_{E,M}(q^2)\sim \left(\frac{1}{q^2}\right)^2\,,\hh
q^2\to-\infty\,,
\nen
where the limit is in the SL region. However, such a power law can be extended also to the TL region by considering the Phragm\'{e}n-Lindel\"{o}f theorem~\cite{Titmarch}, that applies to FFs because of their analyticity and boundedness.
\item 
A very powerful consequence of analytic properties of FFs is the possibility of using a particular analytic continuation tool based on the Cauchy theorem~\cite{arfken}, \ie, the dispersion relations for the imaginary part 
\be  
F(q^2)=\frac{1}{\pi}\int_{\qth}^{\infty} \frac{{\rm Im} \left[F(q'^2)\right]}{q'^2-q^2-i\epsilon}\,dq'^2 \,,
\label{eq:DR}
\en 
valid for $q^2\not\in(\qth,\infty)$ and where the symbol $F$ stands for a generic FF. The threshold value to be used as lower limit of the dispersion relation integral depends on the isospin of the considered FF. In case of isovector components, intermediate states with only even numbers of pions are allowed, hence, as already seen, the threshold is $\qth=(2M_\pi)^2$, while for isoscalar components $q'^2_{\rm theo}=(3M_\pi)^2$. Obviously using the lower threshold $\qth$ is always correct, since the imaginary parts of the isoscalar FFs are null for $q^2\le q'^2_{\rm theo}$.   
\item 
Another interesting issue, that emerges by considering the definition of $G_E^N$ and $G_M^N$, and assuming analyticity for the Dirac and Pauli FFs,
is the identity $G_E^N(4M_N^2)=G_M^N(4M_N^2)$. On the other hand,
the electric and magnetic FFs could be different at the physical threshold only if $F_1^N$ and $F_2^N$ were singular there\footnote{Interesting discussions about the threshold value of EMFFs are developed in Ref.~\cite{noi-meissner}.}~\cite{Bardini}. Such an identity implies that, at the threshold $\qphy$, the nucleon vertex is described by a unique FF, \ie, there is only one degree of freedom and the cross section, loosing its dependence on the scattering angle, becomes isotropic. In other words, even though angular momentum conservation allows S and D waves for the $N\overline{N}$ system produced by one virtual photon (Born approximation), at the production threshold the D-wave contribution must vanish, so that only the isotropic S-wave survives. 
\\
In principle, the identity $G_E^N(4M_N^2)=G_M^N(4M_N^2)$ can be verified experimentally by measuring, for instance, the ratio $G_E^N/G_M^N$ at the physical threshold. However, in a symmetric \ee\ collider, the possibility to reach or even get very close to the threshold is prevented by physical limitations. Indeed, in case of the annihilation process $\ee\to\pp$, the proton and the antiproton are produced almost at rest in the laboratory frame and hence they have no enough momentum to reach the detector. 
\\
In the last twenty years, the so-called ``initial state radiation technique'', developed at the flavor factories, allowed to avoid this limitation, so that values of the ratio $G_E^N/G_M^N$~\cite{BaBar} have been measured very close to the physical threshold. These data, together with older measurements performed in the crossed channel $\pp\to\ee$~\cite{PS170}, agree with threshold-isotropy requirement $G_E^p(4M_N^2)=G_M^p(4M_N^2)$, but do not exclude possible, small D-wave contributions.
\end{itemize}
\section{The nucleon model}
\label{sec:nucleon-model}
We use the Skyrme model~\cite{Skyrme} as test bed for our analytic continuation procedure. In the framework of such a model, which is described in detail in the next section, the charge and magnetization spatial distributions of nucleons are obtained with no further phenomenological or experimental constraint.\\ It represents a typical example of those models that, by providing an exclusively SL description of FFs, are particularly suitable to be treated with our ``analyticization'' procedure.
\subsection{The Skyrme model}
\label{sec:skyrme}
The Skyrme model was introduced by Tony Skyrme in 1960 as a model for strong interactions~\cite{Skyrme}. The basic and innovative idea was that fermions could emerge as particular, stationary and quantized solutions of a non-linear field theory with only boson fields. Stationary solutions of this kind are usually called {\itshape solitons}, the quantized ones, associated to the Skyrme Lagrangian, are instead called {\itshape skyrmions}.\\
The interest in this model increased when 't Hooft and Witten proposed the $1/N_c$ expansion of QCD~\cite{Witten1, Hofft} and Witten showed that the Skyrme model led to a Lagrangian which was equivalent to that of the $1/N_c$ expansion.\\
The first application of this model is due to Adkins, Nappi and Witten~\cite{Adkins1,Adkins2}, who computed some static quantities for nucleons by obtaining a quite acceptable ($\sim30\%$) agreement with the measured values. Such an agreement strengthened the conviction to being on the right track to achieve an effective approximation of QCD at low energy.
\\
In order to build up a representation of nucleon SLFFs, we will follow the work of Braaten, Tse and Willcox~\cite{Braaten}, that, in 1986, for the first time, used the Skyrme model to compute nucleon FFs.
\subsection{Skyrme Lagrangian}
\label{subsec:skyrme-L}
The Skyrme Lagrangian, which is based on the Lagrangian of the so-called {\it non-linear $\sigma$-model}~\cite{sigma-model}, has an $SU(2)_L\times SU(2)_R$ chiral symmetry which is spontaneously broken to $SU(2)$. Assuming that the isoscalar and isovector fields $\sigma$ and $\vec{\pi}$, as a consequence of the symmetry breaking, are linked by the relation
\be
\sigma^2+\vec{\pi}^2=F_\pi^2\,,
\nen
where $F_\pi=108$ MeV~\cite{Adkins2} is the weak pion decay constant, see Tab.~\ref{tab1}, the Skyrme Lagrangian can be written in terms of the only $SU(2)$ field
\be
U(\vec{r})=\frac{1}{F_\pi}\,
\Big(\sigma(\vec{r})+ i\, \vec{\tau} \cdot \vec{\pi}(x)\Big)\equiv
\exp\left(i\,\vec{\tau} \cdot \vec{F}(\vec{r})\right)\,,
\label{eq:U}
\en
where: $\vec\tau$ is the vector of Pauli matrices and the function $\vec{F}(\vec{r})$ is the ``axis-angle'' representation of the chiral field $U(\vec{r})$.
\begin{table}[h!]
\begin{center}
%\bm{80 mm}
\normalsize
\renewcommand{\arraystretch}{1.3}
\begin{tabular}{l|r|r}
Quantity (units) &  \multicolumn{1}{c|}{This work} & \multicolumn{1}{c}{Experimental}\\
\hline
\hline
$F_{\pi}$ (MeV) & (fixed)\hfill\ 108 & $\sim 186$ \\
\hline
$M_{\pi}$ (MeV) & (fixed)\hfill\ 138 & $\sim 138$  \\
\hline
$g$  & (fixed)\hfill\ 4.84 & -  \\
\hline
$M$ (MeV) & 937 & $\sim 938$ \\
\hline
$\langle r^2_p \rangle^{1/2}_E$ (fm) & 0.88 & $0.8775(51)$  \\
\hline
$\langle r^2_n \rangle_E $ (fm$^2$) & $-0.31$ & $-0.1161(22)$\\
\hline
$\langle r^2_p \rangle^{1/2}_M$ (fm) & 0.79 &$ 0.777(16)$ \\\hline
$\langle r^2_n \rangle^{1/2}_M$ (fm) & 0.82 &$ 0.862(9)$\\\hline
$\mu_p$ ($\mu_B$) & 1.97 & $2.792847356(23)$  \\\hline
$\mu_n$ ($\mu_B$) & $-1.24$ & $-1.9130427(5)$ \\\hline
$\mu_p/\mu_n$ & $-1.59$ & $-1.459898075(5)$ \\
\end{tabular}
\caption{Parameters and static quantities obtained in the framework of the Skyrme model compared with their experimental values~\cite{pdg}. Our results, being obtained in the same conditions, reproduce quite well those of Ref.~\cite{Adkins2}. The $\sim$1\textperthousand\ difference in the nucleon mass is probably due to a slightly different normalization range for the chiral angle $F(r)$. We used $0\le r\le 8$ fm, while, in Ref.~\cite{Adkins2}, there is no indication on that. The symbol $\mu_B$ stands for the Bohr magneton.}
\label{tab1}
%\em
\end{center}
\end{table}\\
The complete Skyrme Lagrangian density, that will be used in the following, reads
\be 
\lsk &=&\frac{F_{\pi}^2}{16}\,{\rm Tr}\big(D_{\mu}U D^{\mu}U^{\dagger}\big) 
\no\\
&&+\frac{1}{32 g^2}\, {\rm Tr}\left(\big[D_{\mu}UU^{\dagger},D_{\nu}UU^{\dagger}\big]^2\right)\no\\
&&+ \mathcal{L}_{\rm WZ} + \frac{ F_{\pi}^2\,M_{\pi}^2}{8} \, {\rm Tr}\left(U+U^\dagger-2\right)\,, \label{eq:Ld-sk}
\en
where $D_{\mu}$ is the covariant derivative, which includes the electromagnetic interaction. Besides the usual kinetic term, the second contribution, which is quadratic in the field derivative and represents a repulsive short-range potential with a coupling $g$, has been introduced {\itshape ad hoc} by Skyrme in order to have stationary solutions. The contribution $\mathcal{L}_{\rm WZ}$, called Wess-Zumino term~\cite{Zumino}, which accounts for the QCD anomalies, is written as a non-gauge invariant coupling between the photon and a conserved topological current $B_\mu$~\cite{Witten3}, \ie,
\be 
&&\mathcal{L}_{\rm WZ}=-\frac{e}{2} A_{\mu}B^{\mu} \,, 
\no\\
 && \mbox{with: }B^{\mu}=\frac{1}{24\pi^2}\epsilon^{\mu\nu\lambda\sigma} {\rm Tr}(U^\dagger\partial_{\nu}U\partial_{\lambda}U^\dagger\partial_{\sigma}U)\,.
 \no\label{B}
\en 
The topological charge associated to $B_\mu$ corresponds to the baryon number $B$, hence the baryons are identified as those solutions with $B=1$.
Finally, the last contribution of Eq.~(\ref{eq:Ld-sk}) is a mass term, which explicitly breaks the chiral symmetry and it is treated perturbatively. 
\\
We consider the particular class of solutions obtained by specializing 
the axis-angle function, $\vec{F}(\vec{r})$ of Eq.~(\ref{eq:U}), according to the so-called {\it hedgehog} ansatz:
$\vec{F}(\vec{r})=F(r)\, \vec{r}/|\vec{r}|\equiv F(r)\,\hat r$. In this way, the space of the surviving $SU(2)$ symmetry, which is the isospin, takes the radial configuration of the $r$-space. In other words, in a given position $\vec{r}$, the isospin vector has the same direction and orientation of the position vector $\vec{r}$. The intensity of the axis-angle function $\vec{F}(\vec{r})$, indicated with the symbol $F(r)$, is called chiral angle. Stable solutions, the skyrmions, are stationary minima of the energy, obtained by solving the Euler-Lagrange equation, which is a non-linear differential equation for the chiral angle $F(r)$, {see, e.g., the differential equation obtained from  Eq.~(8) of Ref.~\cite{Adkins2}}.
\\
The further step, that is the skyrmion quantization, consists in quantizing collective modes, translations and rotations, in the isospin space. This can be done by using, in the Lagrangian density \lsk, instead of the field $U(\vec{r})$ of Eq.~(\ref{eq:U}), its time-dependent version
\be
U(\vec{r},t)=A(t)\,U\big(\vec{r}-\vec{X}(t)\big)\,A^\dagger(t)\,,
\nen
where $A(t)$ is a uniform $SU(2)$ matrix and $\vec{X}(t)$ is the skyrmion center-of-mass position vector. As a consequence of rotational and translational invariance, the resulting Lagrangian depends only on the derivatives of $A(t)$ and $\vec{X}(t)$, and it reads
\be
L_{\rm Skyrme}& =& \int  d^3\vec{r}\,\lsk
\no\\
& = & -M+\frac{M\,\dot{\vec{X}}}{2}+\Lambda\,{\rm Tr}\big(\dot{A}^\dagger \dot{A}\big)\,,
\label{eq:L-sk}
\en 
where $M$ and $\Lambda$ are mass and moment of inertia of the skyrmion.
\\
Owing the hedgehog ansatz, the rotational operator
is related, not only to isospin, but also to spin. The skyrmion can be interpreted as a nucleon by requiring the rotational operator to have a semi-integer eigenvalue, so that, spin and isospin are both quantized to 1/2~\cite{Finkel}. The Hamiltonian for the quantized skyrmion, in terms of its three-momentum and spin operators $\vec{P}$ and $\vec{S}$, is 
\be 
H_{\rm Skyrme}=M+\frac{1}{2M}\vec{P}^2+\frac{1}{2\Lambda}\vec{S}^2\,.
\label{eq:H-sk}
\en
Since, $\vec{P}$, $S_3$ and $I_3$ (third components of the spin and isospin) are mutually commuting operators (they also commute with the Hamiltonian), the system described by the Hamiltonian of Eq.~(\ref{eq:H-sk}) is manifestly non-relativistic and
 has the eigenstate $|\vec{p},s_3,t_3\rangle$, where $\vec{p}$, $s_3$ and $t_3$ are the corresponding eigenvalues. Moreover, as already noticed, there is also the conserved topological charge $B$. Thus the nucleon is identified as the eigenstate of $H_{\rm Skyrme}$ with: $s_3=t_3=1/2$ and $B=1$.
\\
The skyrmion mass $M$, appearing in Eqs.~(\ref{eq:L-sk}) and~(\ref{eq:H-sk}), represents the minimum of the energy, obtained by solving the Euler-Lagrange equation of the Skyrme Lagrangian, Eq.~(\ref{eq:Ld-sk}), to find the solitonic solution, \ie, the chiral angle $F(r)$. Parameters and static quantities obtained in this work are reported in Tab.~\ref{tab1}.
\subsection{Electromagnetic form factors in Skyrme model}
\label{subsec:em-current}
Nucleons FFs were firstly obtained in the framework of Skyrme model, by Braaten, Tse and Willcox~\cite{Braaten}. The starting point consisted in deducing the most general expression for the electromagnetic current, at a given order in some expansion parameter,
that fulfilled all symmetries and constraints of the model. Due to the
equivalence, discussed in Sec.~\ref{sec:skyrme}, of the Skyrme and an $SU(N_c)$-QCD Lagrangian, in the large-$N_c$ limit, the natural expansion parameter turns out to be $1/N_c$. For instance, the 	Skyrme Hamiltonian of Eq.~(\ref{eq:H-sk}), being both $M$ and $\Lambda$ of order $N_c$, contains terms up to the first order in the $1/N_c$ expansion.
\\
The expression of the electromagnetic current, at the leading $1/N_c$ order, is written in terms of position, momentum, spin and isospin operators, and also two model-dependent four-vector functions of $\vec{X}^2$ (eight scalar functions), to be furthermore specified by imposing symmetries (hedgehog ansatz included) and physical constraints (e.g.: $B=1$). Only two, $b(\vec{X}^2)$ and $t(\vec{X}^2)$, out of the eight scalar functions, survive the characterization procedure, hence, using relations and definitions of Eqs.~(\ref{eq:current-ffs}) and~(\ref{eq:sachs}) in the Breit frame, nucleon SLFFs can be obtained as the Fourier transforms {(e.g., apart from constant normalization factors, see Eqs~(2.5-2.8) of Ref.~\cite{Holzwarth1})}
\begin{subequations}
\be 
G_E^S(Q^2) \ug \frac{1}{2}\int d^3 \mathbf{r} j_0(Qr)b(r)\,, \label{ff1}\\
G_E^V(Q^2) \ug \frac{1}{3 \Lambda} \int d^3 \mathbf{r} j_0(Qr) r^2 t(r)\,, 
\label{ff2}\\
G_M^S(Q^2) \ug \frac{M}{ 2\Lambda Q} \int d^3 \mathbf{r} j_1(Qr) r b(r)\,,
\label{ff3}\\
G_M^V(Q^2) \ug \frac{2M}{3 Q} \int d^3 \mathbf{r} j_1(Qr) r t(r)\,,
\label{ff4}
\en\label{eq:FTs}%
\end{subequations}
where $j_i(x)$ is the $i$-th spherical Bessel function and $Q^2=-q^2$, with $\sqrt{Q^2}\equiv Q>0$ in the SL region. From now on we will consider a unique FF definition in SL and TL regions and hence we will use the same symbols, omitting tilde and over-bar.
\\
In the Skyrme model the functions $b(r)$ and $t(r)$, 
{also called baryon and moment-of-inertia densities, respectively,}  are defined only in terms of chiral angle $F(r)$. Their expressions, obtained by comparing the electromagnetic current, as extracted from the Lagrangian of Eq.~(\ref{eq:Ld-sk}), with the general ``educated'' parametrization discussed so far, are
\be 
\begin{array}{rcl}
b(r) &=& \ds-\frac{F'(r)}{2\pi^2}\,\frac{\sin^2\left[F(r)\right]}{r^2}\,,\\
&&\\
t(r)&=&\ds\frac{F_{\pi}^2}{4}\frac{\sin^2\left[F(r)\right]}{r^2}\\
&&\\
&&
+\ds\frac{1}{g^2}\frac{\sin^2\left[F(r)\right]}{r^2}\left(\left[F'(r)\right]^2+\frac{\sin^2\left[F(r)\right]}{r^2}\right)\,.\\
\end{array}
\label{eq:b-t}
\en
{Apart from constant normalization factors, $b(r)$ and $t(r)$ coincide with the functions $B_0(r)$ and $B_1(r)$ fo Eqs.~(2.9) and~(2.10) of Ref.~\cite{Holzwarth1}.}
The chiral angle $F(r)$ is obtained as solution of the Euler-Lagrange equation, a non-linear differential equation of second order,
which follows from the functional minimization of the skyrmion mass-energy computed with the Lagrangian density \lsk\ of Eq.~(\ref{eq:Ld-sk}) and the representation of the chiral field $U(\vec{r})$ given in Eq.~(\ref{eq:U}). 
\\
Since the function $F(r)$ and hence $b(r)$ and $t(r)$, are known only numerically, the Fourier transforms of Eq.~(\ref{eq:FTs}) allow to compute nucleon FFs only in their convergence domain, which corresponds to the SL region, \ie, $Q^2>0$ and $Q>0$. Indeed, negative values of $Q^2$ would imply pure imaginary $\sqrt{Q^2}$, so that the Bessel spherical functions become exponentially divergent as $r\to\infty$.
\\
It follows that there is no possibility to perform analytic continuations of the SLFFs in the TL region, if $b(r)$ and $t(r)$ are known only numerically. In Sec.~\ref{sec:an-cont} we will develop a procedure to overcome this limitation.
\\
The order in the $1/N_c$ expansion of the FF expressions given in  Eq.~(\ref{eq:FTs}) can be easily inferred by the presence of factors $M$ and $\Lambda$ at numerator or denominator. In particular, the isoscalar electric and magnetic FFs, Eqs.~(\ref{ff1}) and~(\ref{ff3}) are of order zero, the isovector electric FF, Eq.~(\ref{ff2}) is of order one, while
the isovector magnetic FF, Eq.~(\ref{ff4}), is of order minus one, \ie\ $\mathcal{O}(N_c)$. Such a heterogeneity should prevent the possibility to combine these expressions, following Eq.~(\ref{eq:SV-pn}), to obtain proton and neutron FFs, as actually has been done in Sec.~\ref{subsec:results}, in order to compare our results with the available data. The fair agreement, which has been obtained, could be an indication that the non-leading contributions to the nucleon FFs, in the $1/N_c$ expansion, are sub-dominant as if there were additional suppression factors.
\subsection{Relativistic corrections}
\label{subsec:rel-corr}
\begin{figure}[h!]
\begin{center}
\includegraphics[width=75mm]{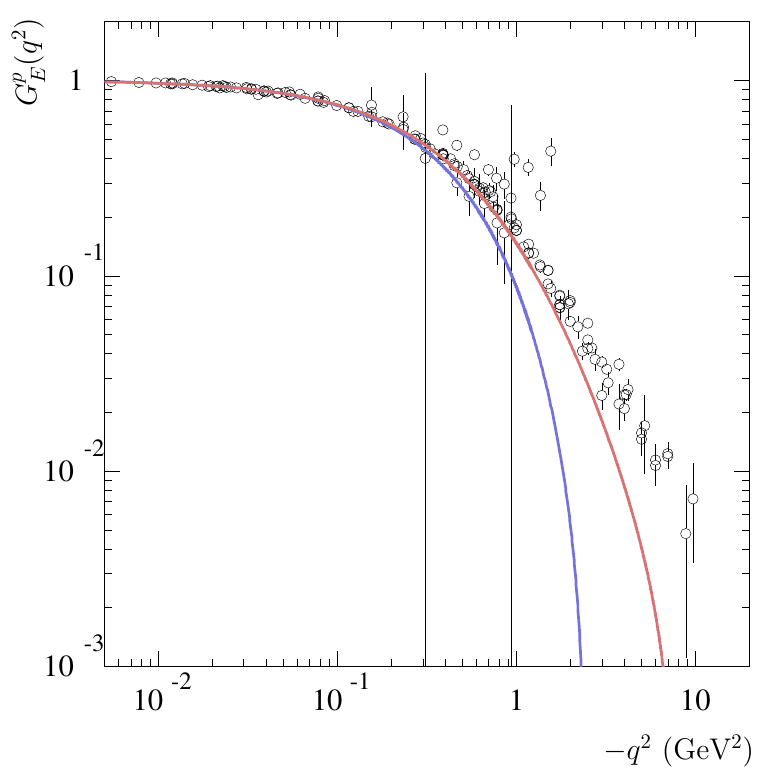}
\\
\includegraphics[width=75mm]{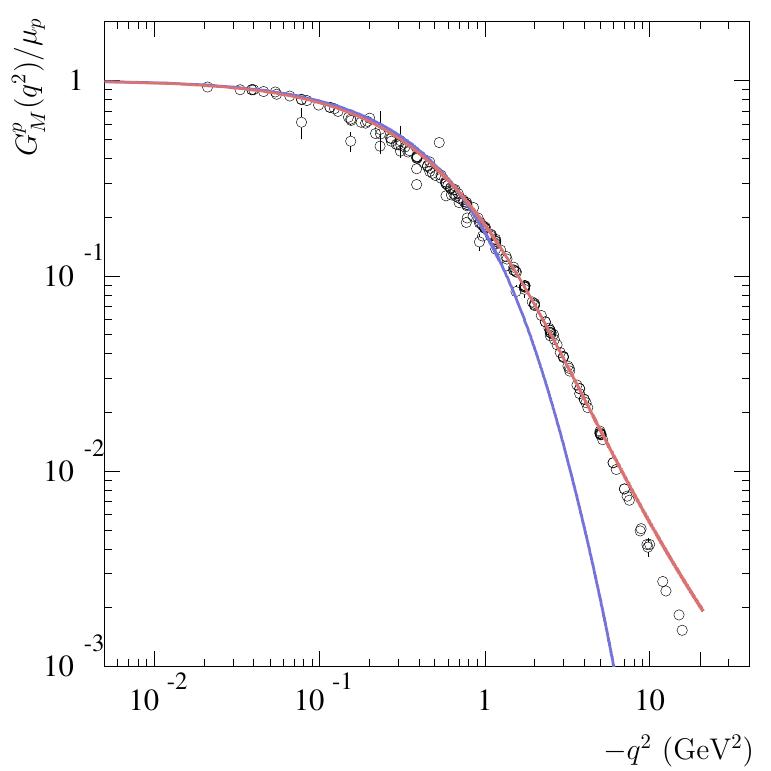}
\vspace{0mm}
\caption{The electric FF (upper panel) and magnetic (lower panel) proton FF normalized to the proton magnetic moment in the SL region. Red and violet curves represent the results with and without relativistic corrections, respectively, while the empty circles are the world data sets~\cite{Perdrisat}. The systematic error due to the technique of multipoint \P\ approximation is negligible.}
\label{art-fig:SL-p}
\end{center}
\end{figure}

To extend the results obtained for the SLFFs to high $Q^2$, relativistic corrections have to be included. However, the procedure to obtain relativistic skyrmions is still debated and different methods are present in literature. To include relativistic corrections in our FF parameterizations we will follow Ref.~\cite{Ji}. In particular, in the SL region, relativistic FF expressions are obtained from the non-relativistic ones as
\be
\begin{array}{rcl}
 G_E^{N,\rm rel}(Q^2) &=& \ds G_E^N\!\left (\frac{Q^2}{1+\frac{Q^2}{4M_N^2}}\right)\,,\\
 &&\\
 G_M^{N,\rm rel}(Q^2) &=& \ds \frac{1}{1+\frac{Q^2}{4M_N^2}} G_M^{N}\!\left (\frac{Q^2}{1+\frac{Q^2}{4M_N^2}}\right) \,,\\
\end{array}
\label{slcor}
\en
while, in the TL region~\cite{Bardini}, using $q^2=-Q^2$,
\be 
%\begin{array}{rcl}
G_E^{N,\rm rel}(q^2) &\!=\!& \left\{\begin{array}{ll}\ds
\!\!\!G_E^{N}(q^2) & q^2\le 4M_N^2\\
&\\
\!\!\!G_E^{N}\!\ds\left[4M_N^2\left(\frac{4M_N^2}{q^2}\!-\!2\right)\!\right]
&q^2> 4M_N^2 \\
\end{array}
\right.
%\,,%%%%%\no
\nen
\be\label{tlcor}\en
%%%%%
\be
 G_M^{N,\rm rel}(q^2)&\!=\!&
 \left\{\begin{array}{ll}\ds
\!\!\!G_M^{N}(q^2) & q^2\le 4M_N^2\\
&\\
 \!\!\! \ds\frac{4M_N^2}{q^2} G_M^{N}\!\left[\!4M_N^2\left(\!\!\frac{4M_N^2}{q^2}\!-\!2\!\right)\!\!\right]
&q^2> 4M_N^2\\
  \end{array}\right.  \hspace{-6mm}
%%\hspace{2mm} Q^2\to-\infty\,,
%%\end{array}
\nen
where the relativistic forms are labelled by the superscript ``rel''.
\\
It is important to stress the fact that, unless one considers fine tuned, non-relativistic FF expressions, with zeros of particular orders at particular finite values of $Q^2$, such corrections are incompatible with pQCD predictions concerning the asymptotic behavior. 
\\
Moreover, since the asymptotic SL and TL limits for a given FF are
different, for instance $G_E^{N,\rm rel}(Q^2)$ tends to the non relativistic value $G_E^N(4M_N^2)$ as $Q^2\to\infty$ and to $G_E^N(8M_N^2)$ as $Q^2\to-\infty$, these corrections do not even verify the Phragm\'en-Lindel\"of theorem.
\\
The effect of relativistic corrections is shown in Fig.~\ref{art-fig:SL-p},
in case of SL electric and magnetic proton FFs. Their inclusion improves the agreement with data at higher $Q^2$, even though, at very high momenta, as expected, the agreement worsens.
\section{Analytic continuation and results}
\label{sec:an-cont}
Once the profiles $b(r)$ and $t(r)$ are known, FFs are computed as their Fourier transforms. However, since only numerical expressions of $b(r)$ and $t(r)$ can be obtained, the applicability of the representations given in Eqs.~(\ref{ff1}-\ref{ff4}) is limited to their domain of convergence, \ie,  the SL region. Indeed, TL transferred momenta would imply divergent real exponentials in the Fourier integrals. The only way to overcome such a limitation and hence to obtain FF values also in the TL region, consists in performing an analytic continuation of the original representation. The simplest procedure is represented by a direct, analytical computation of the integral, which would return, for the FFs, well defined expressions, depending on the variable $Q^2$, that are manifestly analytic.
In this case the direct approach is prevented by the lack of analytic forms for the profiles $b(r)$ and $t(r)$. Nevertheless, the fact that they are well known in a wide range of $r$, allows us to define simple analytic functions that approximate them with an accuracy that, in principle, could be indefinitely improved by increasing the number of free parameters to be settled. Moreover, the structure of such fit functions is inferred by the knowledge of the profiles in the two limits: $r\to 0$ and $r\to\infty$.
To expound the analytic continuation procedure that we have developed, 
we consider in detail the case of the electric isoscalar FF $G_E^S(Q^2)$. 
Its integral representation, given in Eq.~(\ref{ff1}), can be also written as
\be
G_E^S(Q^2)
=\frac{\pi}{iQ} \int_0^{\infty}  \left (e^{iQr}-e^{-iQr} \right ) rb(r)dr\,,\label{fff}
\en
it contains only the profile $b(r)$.
\\
The behaviors in the limits $r\to 0$ and $r\to\infty$ are well known because the corresponding differential equations can be analytically solved and we get
\be\begin{array}{lcr}
b(r) \ds\mathop{\sim}_{r\to 0} h_0+ h_2 r^2 \,,\hspace{5mm}
b(r) \ds\mathop{\sim}_{r\to \infty} h_\infty\, e^{-3 M_{\pi}r} \frac{1}{r^5}\,, \end{array} 
\label{eq:asy-b}
\en
where $h_0$, $h_2$ and $h_\infty$ are free parameters. 
{The asymptotic falloff of $b(r)$, as $r\to\infty$, follows from its definition of Eq.~\eqref{eq:b-t}, as a consequence of the asymptotic behavior of the chiral phase $F(r)$, of its sinus, as well as its first derivative, i.e.,
\be
F(r)\mathop{\propto}_{r\to\infty}\frac{e^{-M_\pi r}}{r} \Rightarrow
\left\{\begin{array}{l}\displaystyle
	\sin\left[F(r)\right]\mathop{\propto}_{r\to\infty}\frac{e^{-M_\pi r}}{r}\\
	\\
\displaystyle	F'(r)\mathop{\propto}_{r\to\infty}-M_\pi\frac{e^{-M_\pi r}}{r}\\
\end{array}
\right.,
\label{eq:asy-F}
\en
as can be found, for instance, in Ref.~\cite{Adkins2}.}
The exponential, which guarantees the fast vanishing of the profile, plays a crucial role in characterizing the analytic structure of the FF, so that the fit function has been defined as
\be 
b_{\rm fit}(r)=\frac{P_n(r)}{P_m(r)}\,e^{-3M_{\pi}r}
=\frac{\sum_{i=0}^n a_i r^i}{\sum_{j=0}^m b_j r^j}\,e^{-3M_{\pi}r} \,,\label{fitt}
\en
where $P_n(r)$ and $P_m(r)$ are polynomials of degrees $n$ and $m$ respectively ($m,n\in\N$), with real coefficients $\{a_k\}_{k=0}^n$ and $\{b_j\}_{j=0}^m$, with: $a_n\not=0$, $b_m\not=0$ and $b_0\not=0$. 
\\
The conditions given in Eq.~(\ref{eq:asy-b}) imply: $(a_1-a_0 b_1/b_0-3M_\pi a_0)=0$ and $m-n=5$, respectively. The rational part of $b_{\rm fit}(r)$ is a meromorphic function with a finite number $M\le m$ of distinct poles in the $r$ complex plane, hence it can be written as the Mittag-Leffler expansion~\cite{ahlfors}
\be
\frac{P_n(r)}{P_m(r)}=\sum_{j=1}^M\left[\sum_{k=-\mu_j}^{-1} C_k^{(j)}\,(r-z_j)^{k}
\right]\,,
\no\en
where $\{z_j\}_{j=1}^M \subset\C$ and $\{\mu_j\}_{j=1}^M\subset \N$ represent the sets of poles and of the corresponding multiplicities, while $C_k^{(j)}$ is the $k$-th coefficient of the Laurent series about the $j$-th pole $z_j$. If the polynomial $P_m(z)$ has only order-one zeros, which, moreover, do not coincide with those of $P_n(z)$, then the function $b_{\rm fit}(r)$ has only simple poles, \ie, $\mu_j=1$ for all $j=1,2,\ldots,M$, and $M=m$. In this case the Mittag-Leffler expansion reduces to
\be 
\frac{P_n(r)}{P_m(r)}=\sum_{j=1}^{m}\frac{R_{z_j}}{r-z_j}\,, 
\label{eq:M-L}
\en
where the coefficient $R_{z_j}$ is the residue of the $j$-th pole
\be
R_{z_j}=C_{-1}^{(j)}={\rm Res }\left [\frac{P_n(r)}{P_m(r)},z_j\right ] 
=\frac{\ds\sum_{i=0}^n a_i z_j^i}{\ds\sum_{k=1}^m k\, b_k z_j^{k-1}}\,,\label{eq:residue0}
\en
with $j=1,2,\ldots,m$.
We have an additional condition on the parameters $z_j$: they can not be positive real numbers, because $b(r)$ has no poles for $r>0$. \\%*******\\
The fit function $b_{\rm fit}(r)$, per se, has apparently
 no physical content, because its parameters are not directly connected to physical properties of the system under consideration.
Nevertheless, as will be discussed in the following, the poles 
$z_j$ play a crucial role, indeed they define the analytic structure of FFs in the $q^2$ complex plane.
 \\
The orders $m$ and $n$ of the polynomials and hence the number of free parameters that define $b_{\rm fit}(r)$, have been chosen by following a criterion which combines the higher accuracy in the description with the smaller redundancy of parameters (a given parameter is defined redundant when its inclusion does not improve the accuracy of the fit). The only constraint on the orders $m$ and $n$ of the polynomials is about their difference, that must be: $m-n=5$. \\
A satisfactory fit has been obtained with $n=6$ and $m=11$. The resulting electric isoscalar FF is
\be 
G_E^S(Q^2)
&=&
\sum_{j=1}^8\frac{\pi\tilde{R}_{z_j}}{iQ}\left [\int_0^{\infty} \frac{e^{(iQ-3M_{\pi})r}}{r-z_j}dr\right.\no\\
&&\left.
-\int_0^{\infty} \frac{e^{(-iQ-3M_{\pi})r}}{r-z_j}dr \right],
\label{ff4-2}
\en
where, having only poles of order one,
\be 
\tilde{R}_{z_j}={\rm Res }\left [\frac{rP_n(r)}{P_m(r)},r=z_j\right ]=z_jR_{z_j}\,.
\label{eq:residue}
\en
All integrals appearing in Eq.~(\ref{ff4-2}) belong to the same class
\be
H(\alpha \beta)\equiv \int_0^{\infty} \frac{e^{-\alpha r}}{r+\beta}dr\,,
\hspace{2mm}
\mbox{with: }\left\{\begin{array}{l}
\re(\alpha)>0\\
\\
\beta\not\in(-\infty,0]\\
\end{array}\right.\hspace{-2mm}.
\label{eq:G}
\en
The conditions on the parameters $\alpha$ and $\beta$ ensure the convergence of the integral that, as can be easily seen by making the substitution $w=\alpha\, r$, depends only on the product $\alpha\beta$. In particular, the integrals of Eq.~(\ref{ff4-2}) can be obtained with: $\alpha=3M_\pi \pm iQ$ and $\beta=-z_j$. These assignments, having no poles on the positive real axis and being $Q>3M_\pi$, automatically fulfill the convergence conditions.
\\
In the $\alpha\beta$ domain defined in Eq.~(\ref{eq:G}), the function $H(\alpha \beta)$ has also the following representation (see App.~\ref{app:b})
\be 
H(\alpha \beta)&=&e^{\alpha\beta}E_1(\alpha\beta)
\no\\
&=& e^{\alpha\beta}\Bigg[\!-\!\gamma \!-\!\ln(\alpha\beta)
\!+\!\!\sum_{k=1}^{\infty}\frac{(-1)^{k+1}(\alpha\beta)^k}{kk!} \Bigg],\label{eq:G2}\hspace{4mm}
\en
where $E_1(z)$ is the exponential integral function or ''ExpIntegral'' and $\gamma$ is the Euler-Mascheroni constant~\cite{stegun}.
Finally, using Eqs.~(\ref{eq:G}) and~(\ref{eq:G2}), the representation of Eq.~(\ref{fff}) can be integrated to obtain
\be 
G_E^S(Q)&\!=\!&\frac{\pi}{iQ}\sum_{j=1}^{8} \tilde{R}_{z_j} \Big\{ e^{(iQ-3M_{\pi})z_j}E_1\left[(iQ-3M_{\pi})z_j\right] \Big.
\no\\
&&
\Big.-e^{(-iQ-3M_{\pi})z_j}E_1\left[(-iQ-3M_{\pi})z_j\right]\Big\}\,. 
\label{gg}
\en
Since the function $E_1(z)$ is analytic in the whole $z$ complex plane with a cut along the negative real axis\footnote{This is a typical logarithmic branch cut as can be seen in the representation of $E_1(z)$ given in Eq.~(\ref{eq:G2}).}, it is now possible to extend the parametrization for $G_E^S$ to the TL region, by making the substitution $Q\rightarrow iq$ ($q>0$), so that
\be 
G_E^S(iq )
&\!=\!&
-\frac{\pi}{q}\sum_{j=1}^{8} \tilde{R}_{z_j} \Big\{
e^{(-q-3M_{\pi})z_j}E_1\left[(-q-3M_{\pi})z_j\right]\Big.
\no\\
&&\Big.
 -e^{(q-3M_{\pi})z_j}E_1\left[(q-3M_{\pi})z_j\right]\Big\}\,. 
\label{gtl}
\en
\subsection{The branch cut in the $q^2$ complex plane}
\label{subsec:cut}
The properties of the representation obtained for $G_E^S$ and, in particular, the presence of branch cuts, as well as their location in the $q^2$ complex plane, depend on the analytic structure of the ExpIntegral functions. Following the derivation given in App.~\ref{app:c}, we obtain for $G_E^S$, in the SL region, the expression
\be
G_E^S(Q)&\!=\!&
\frac{2\pi}{Q}\sum_{j=1}^{l}\tilde{R}_{r_j}\im\left\{
 H\left[(iQ-3M_{\pi})r_j\right] \right\}
 \label{eq:ges-easy}\\
 &&+
\frac{2\pi}{Q}\sum_{j=1}^{h}\im\Big\{\tilde{R}_{c_j}
 H\left[(iQ-3M_{\pi})c_j\right]\Big.
 \no\\
\Big. 
 && -\tilde{R}_{c_j}
 H\left[(iQ-3M_{\pi})^*c_j\right] \Big\}+
 \frac{4\pi^2}{Q}\sum_{j=1}^{h}\theta(x_j)
 \no\\
&& \times\re\left[
\theta\!\left(\!Q-3M_\pi\frac{y_j}{x_j}\!\right)\theta(y_j)
\tilde R_{c_j}
e^{(iQ-3M_\pi)c_j}\right. \no\\
&&+
\left.\theta\!\left(\!Q+3M_\pi\frac{y_j}{x_j}\!\right)\theta(-y_j)\tilde R_{c_j}e^{(-iQ-3M_\pi)c_j}
\right]\,,
\no\en
where $r_j$ and $c_j$ are real and complex poles, $x_j$ and $y_j$, in the arguments of the Heaviside theta functions, represent real and imaginary parts of $c_j$, while $\tilde{R}_{r_j}$ and $\tilde{R}_{c_j}$ are the residues. For a detailed description see App.~\ref{app:c}. The parametrization of Eq.~(\ref{eq:ges-easy}) is explicitly real in the SL region, i.e., for $Q>0$. In the TL region the $G_E^S$ expression becomes
\be 
G_E^S(iq)&\!=\!&-\frac{2\pi}{q}\sum_{j=1}^{h} 
\re\left\{ \tilde{R}_{c_j} H\left[(-q-3M_{\pi})c_j\right]
\right.\label{eq:ges-easy-tl}\\
&&\left.-\tilde{R}_{c_j} H\left[(q-3M_{\pi})c_j\right]\right\}
\no\\ 
&&
 -\frac{\pi}{q}\sum_{j=1}^{l} \tilde{R}_{r_j} \Big\{H\left[(-q-3M_{\pi})r_j\right]
 \Big.\no\\
 \Big. &&
 -H\left[(q-3M_{\pi})r_j\right]\Big\}
\no\\ 
&&
+\frac{2i\pi^2}{q}\sum_{j=1}^h \theta(q-3M_\pi)\theta(-x_j)
\no\\
&&
\!\!\!\!\!\times\!\left[\tilde R_{c_j}e^{(q-3M_\pi)c_j}\theta(y_j)\!+\!
\tilde R_{c_j}^*e^{(q-3M_\pi)c_j^*}\theta(-y_j)\right]\,.
\no
\en
While the first term is real, the second and the third could have non vanishing imaginary parts. In particular, the second term contains $H$ functions, that embed a logarithmic structure and hence, for negative arguments, have non-zero imaginary parts. Having only negative real poles, $r_j<0$, and $q>0$, the argument $(-q-3M_{\pi})r_j$ is always positive, whereas $(q-3M_{\pi})r_j$ can be negative when $q>3M_\pi$, following exactly the theoretical requirement (see Sec.~\ref{subsec:analyticity}). A further imaginary contribution, given by the last term, is due to non real poles with negative real part.
Finally, the TL imaginary part of $G_E^S$, which is non vanishing only if there are poles with negative real parts, is given by
\be 
\im[G_E^S(iq)]&\!=\!& \frac{2\pi^2}{q}\theta(q-3M_\pi)
\no\\
&&\times\left\{
\sum_{j=1}^h \theta(-x_j) 
\re\left[\tilde R_{c_j}e^{(q-3M_\pi)c_j}\right]
\right.\no\\ 
&&\left.
-\frac{1}{2}\sum_{j=1}^{l} \tilde{R}_{r_j} e^{(q-3M_{\pi})r_j}
\right\}\,.
\label{eq:im-ges}
\en
In summary.
\begin{itemize}
\item
The profile $b(r)$ is obtained as numerical solution of a differential equation.
\item
Such a solution is fitted with $\bfit(r)$, a product of a rational function and an exponential, that fulfills the requirements for $r\to 0$ and $r\to\infty$ and moreover it has only simple poles not belonging to the positive real axis.
\item
The rational part of $\bfit(r)$, being a meromorphic function, can be written as a Mittag-Leffler sum and hence its Fourier transform, which is the FF $G_E^S(Q)$, is a sum of Fourier transforms of simple poles, $z_j$, multiplied by an exponential, \ie, ExpIntegral functions with arguments: $(\pm iQ-3M_\pi)z_j$.
\item
%
%	Round2-1
%
The analyticity domain of $G_E^S(Q)$, especially when there is at least one pole with negative real part, is exactly that expected for FFs on the basis of first principles, i.e., the $q^2$ complex plane with the branch cut $\left((3M_\pi)^2,\infty\right)$. Nevertheless, only logarithmic and not square-root branch cuts can be generated. See Sec.~\ref{subsec:cuts} for a detailed treatment.
\end{itemize}
The isovector FFs are obtained with the same procedure described in detail for $G_E^S$, \ie, by fitting the profile function $t(r)$ with a ratio of polynomials and an exponential deduced from the solution of the asymptotic differential equations. However, in this case, to account for the two and four-pion coupling, two exponential contributions are considered
\be
t(r)=\frac{P_{n'}(r)}{P_{m'}(r)}e^{-2M_{\pi}r}+\frac{P_{n''}(r)}{P_{m''}(r)}e^{-4M_{\pi}r} \,,
\label{eq:ta-tb}
\en
with $m'-n'=4$ and $m''-n''=6$, as a consequence of the asymptotic behavior of the chiral angle $F(r)$, see Eqs.~\eqref{eq:b-t} {and~\eqref{eq:asy-F}}. Following the line of reasoning used to study $G_E^S$, this expression leads to two different branch cuts, that generate from the two thresholds: $q^2_{\rm theo}=(2M_\pi)^2$ and ${q''}^2_{\rm theo}=(4M_\pi)^2$. 
\\
\r{Rigorous tests of analyticity for the parameterizations in connection with the radial profiles  of the Skyrme model are presented in App.~\ref{app:d}.}
\subsection{Asymptotic behavior}
\label{subsec:asy-beha}
As already discussed, the asymptotic behavior of the FFs obtained with this procedure is completely determined by the relativistic corrections of Eqs.~(\ref{slcor}) and~(\ref{tlcor}). Nevertheless, it is interesting to study the high-$Q$ behavior of the non-relativistic (uncorrected) FF expressions. Such a behavior can be derived from the asymptotic expansion of the ExpIntegral function~\cite{bleistein}
\be
E_1(z)=\frac{e^{-z}}{z}\sum_{k=0}^{n-1}(-1)^k \frac{k!}{z^k}+\mathcal{O}\left [(n-1)!|z|^{-n}\right ]\,,
\label{eq:asym}
\en
with $z \to\infty$.
The rigorous treatment is given in App.~\ref{app:e}, where the SL and TL asymptotic behaviors are obtained for general profile functions, but not taking into account the branch cut corrections discussed in Sec.~\ref{subsec:cut}. However, as we will see in detail in the case of $G_E^S$, such corrections do not spoil the power law behavior driven by the expansion of Eq.~(\ref{eq:asym}).
Using Eq.~(\ref{eq:g0-sl}), the SL isoscalar FF in the high-$Q$ regime can be written as the series of increasing powers of $Q^{-1}$
\be
G_E^S(Q) \mathop{\sim}_{Q\to \infty}
\sum_{k=0}^\infty g_{SL}^{(k)}(Q)\,,
 \label{eq:sl-series}\en
with: $g_{SL}^{(0)}(Q)\sim Q^{-2}$ and
$g_{SL}^{(k)}(Q)\sim (Q^{-2})^{2\,{\rm Int}[(k+1)/2]}$, for $k\ge 1$.
In particular, when $Q\to\infty$, the first four terms behave as
\be
\begin{array}{rcl}
g_{SL}^{(0)}(Q)
\!&\sim &\!
\ds-\frac{2\pi}{Q^2}
\sum_{j=1}^{m}\re\left(\frac{\tilde{R}_{j}}{z_j}\right)
\,; \\
g_{SL}^{(1)}(Q)
\!&\sim &\!
\ds
\frac{12\pi \,M_\pi}{Q^4}
\sum_{j=1}^{m}\re\left(\frac{\tilde{R}_{j}}{z_j^{2}}\right)
\,;\\
g_{SL}^{(2)}(Q)
\!&\sim &\!
\ds
\frac{4\pi }{Q^4}
\sum_{j=1}^{m}\re\left(\frac{\tilde{R}_{j}}{z_j^{3}}\right)
\,;\\
g_{SL}^{(3)}(Q)
\!&\sim &\!
\ds
\frac{144\pi\,M_\pi }{Q^6}
\sum_{j=1}^{m}\re\left(\frac{\tilde{R}_{j}}{z_j^{4}}\right)\,.\\
\end{array}
\label{eq:ges-k}
\en
Each of them depends on the corresponding derivative (the $k$-th derivative for the $k$-th term) of the rational function $r\,\bfit(r)e^{3M_\pi r}$ evaluated in the origin, \ie,
\be
\left.\frac{d^k}{dr^k}\left(
r\,\bfit(r)e^{3M_\pi r}\right)\right|_{r=0}
&=&
\left.
\frac{d^k}{dr^k} \frac{rP_n(r)}{P_m(r)}\right|_{r=0}
\no\\
&=&
\frac{d^k}{dr^k}\left. \sum_{j=1}^m\frac{\tilde R_j}{r-z_j}\right|_{r=0}
\no\\
&=&-k!\sum_{j=1}^m\frac{\tilde R_j}{z_j^{k+1}}\,,
\nen
with $k=0,1,\ldots$.
Since the function $r\,\bfit(r)e^{3M_\pi r}$ vanishes in the origin, having no poles there, the first contribution ($k=0$) in Eq.~(\ref{eq:ges-k}) is also vanishing and hence 
$g_{SL}^{(1)}(Q)$ and $g_{SL}^{(2)}(Q)$, that are of the same order in $Q$, \ie\ $Q^{-4}$, become the leading terms.
The TL expression for $G_E^S(iq)$, given in Eq.~(\ref{eq:ges-easy-tl}), apart from the factor $q^{-1}$, has two kinds of contributions: the first depends on the functions $H(z)$, while the second, which accounts for the branch cut corrections, has an exponential behavior. More in detail, there are two exponentials that, being complex conjugates, have the same modulus and hence the same asymptotic behavior. Their moduli scale like $\sim e^{-q\,x_j}$ when $q\to\infty$, where $x_j$ is the real part of the $j$-th pole. However, such contributions are weighted by three Heaviside theta functions, one of which ensures the strict positivity of $x_j$, hence all the exponentials are vanishing as $q\to\infty$ and the asymptotic behavior is dominated by the only terms which contain the $H(z)$ functions. In light of that, the asymptotic behavior of $G_E^S(iq)$ can be described in terms of the series
\be 
G_E^S(iq)
 \ds\mathop{\sim}_{q\to\infty}
\sum_{k=0}^\infty g_{TL}^{(k)}(q)\,,
 \no
\en
similar to that of Eq.~(\ref{eq:sl-series}), where the functions $g_{TL}^{(k)}(q)$ are defined by the TL expansion of Eq.~(\ref{eq:g0-tl}), and are related to the corresponding SL terms as $g_{TL}^{(k)}(q)=g_{SL}^{(k)}(iq)$. In other words, the TL asymptotic behavior follows the same power law as in the SL region. The leading contributions, in both regions, are determined by the behavior of the profile function in the origin $r=0$.
\\
However, while for the electric isoscalar FF the obtained behavior agrees with the perturbative QCD prediction, \ie, $G_E^S(Q)\ds\mathop{\sim}_{Q\to\infty} Q^{-4}$ and $G_E^S(iq)\ds\mathop{\sim}_{q\to\infty} q^{-4}$, for all the other three FFs, see Eqs.~(\ref{eq:asy-sca}) and~(\ref{eq:asy-vec}), we achieved the faster vanishing behaviors 
\be
&& G_E^V(Q),\,G_M^S(Q),\,G_M^V(Q)
\ds\mathop{\sim}_{Q\to\infty} 
Q^{-6}\,,\no\\
&&\label{eq:asy-6} \\
&& G_E^V(iq),\,G_M^S(iq),\,G_M^V(iq)
\ds\mathop{\sim}_{q\to\infty} 
q^{-6}\,.
\nen
The only possibility to recover the expected power laws should be that to consider a profile function having in the origin a zero of a lower order. For instance, in case of 
$G_M^V$, the profile function is $f(r)=r^3 t(r)$, see Eq.~(\ref{ff4}), and, as $r\to 0$, $f(r)\propto r^{l}$,  with $l=3$, because the density $t(r)$ is finite and non vanishing in the origin. This power, $l=3$, determines (see Eq.~(\ref{eq:asy-01}) and~(\ref{eq:asy-02})) the asymptotic behavior as given in Eq.~(\ref{eq:asy-6}). On the other hand, the perturbative QCD expectation, \ie, the power laws $Q^{-4}$ and $q^{-4}$, in SL region and TL region respectively, would be obtained only with $l=2$, which means that $t(r)$ should have a simple pole in the origin.
\subsection{Results}
\label{subsec:results}
To have a direct comparison with data, results are primarily given for the electric and magnetic Sachs FFs of proton and neutron, $G_{E,M}^p$ and $G_{E,M}^n$, even though the primary outcomes of this procedure, see Eqs.~(\ref{eq:FTs}), are their isospin components $G_{E,M}^S$ and $G_{E,M}^V$. These two sets of FFs are related by the linear combinations given in Eq.~(\ref{eq:SV-pn}). Moreover, SLFFs and TLFFs will be given as functions of $Q^2$ and $q^2$ respectively, with the simple convention $G_{E,M}^{S,V}(\pm q^2)\equiv G_{E,M}^{S,V}(\mp Q^2)$.
\\
All FFs have been obtained by means of the procedure outlined in Secs.~\ref{sec:an-cont},~\ref{subsec:rel-corr} and in App.~\ref{app:a}, such a procedure does embody a certain degree of uncertainty mainly due to the multipoint \P\ approximation technique. The consequent systematic error has been accounted for by using two different sets of interpolation points. It is interesting to notice that this systematic error is perceptible only for TL results. In fact, the two curves that are obtained for each FF, corresponding to the two sets of interpolation points, are superimposed and hence indistinguishable in the SL region,  while they form a band with a finite width in the TL region.
\\
Moreover, all the SLFFs, but for $G_E^n$ which is vanishing at $Q^2=0$, are normalized to the so-called dipole FF
\be
G_D(Q^2)=\left(1+\frac{Q^2}{M_D^2}\right)^{-2}\,,
\label{eq:dipole}
\en
with $M_D^2=0.71$ GeV$^2$. Such a FF, with only one free parameter, the dipole mass $M_D$, describes quite well the SL data on $G_E^p(Q^2)$, $G_M^p(Q^2)/\mu_p$ and $G_M^n(Q^2)/\mu_n$, as can be seen in Fig.~\ref{art-fig:GEp-GMp} and in the lower panel of Fig.~\ref{art-fig:GEn-GMn}, where indeed the data (empty circles) spread out around the unity.
\subsubsection{Space-like region}
\label{subsubsec:res-sl}
\begin{figure}[h]
\begin{center}
\includegraphics[width=75mm]{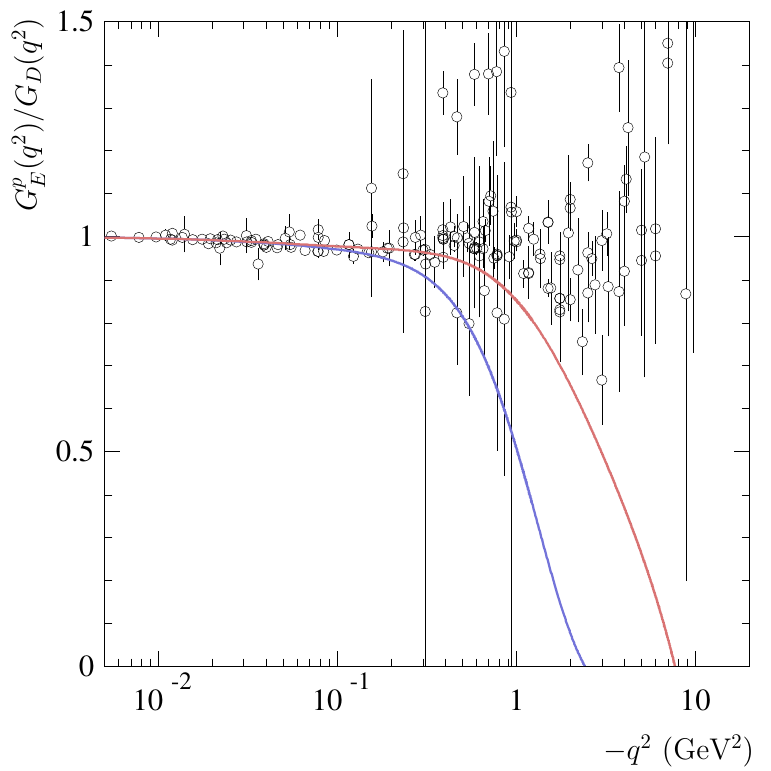}
\\
\includegraphics[width=75mm]{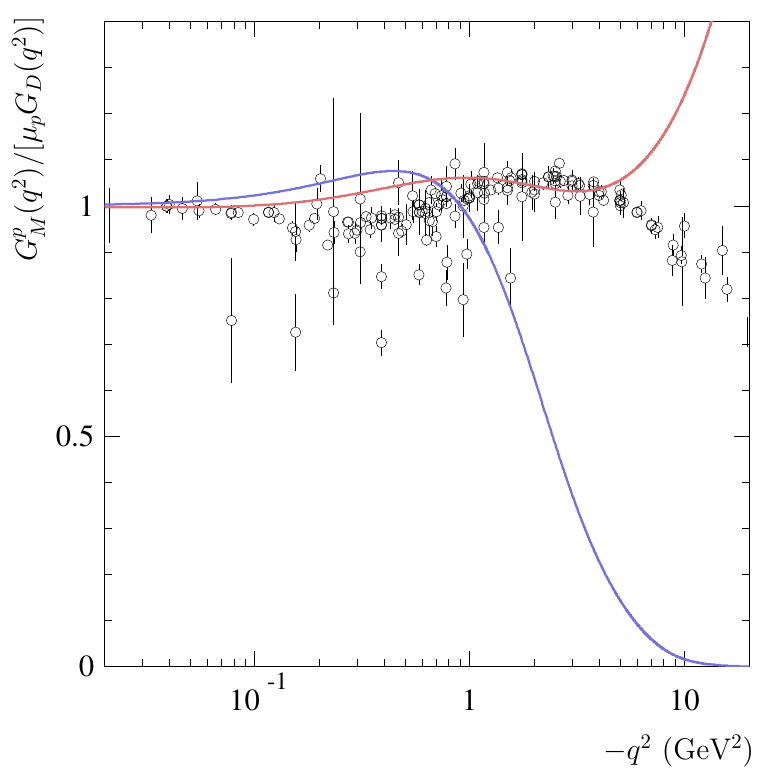}
\vspace{0mm}
\caption{The electric (upper panel) and magnetic (lower panel) proton FFs in the SL region, normalized to the dipole FF and $G_M^p$ also to the magnetic moment, are compared with the world data sets, empty circles, from Ref.~\cite{Perdrisat}. The red and violet curves represent the predictions for the FFs obtained including and not including relativistic corrections as given in Eq.~(\ref{slcor}). The data sets are the same of Fig.~\ref{art-fig:SL-p}, however, due to the different scales, logarithmic and linear, and to the dipole normalization, the errors appears larger. The systematic error due to the technique of multipoint \P\ approximation is negligible.}\label{art-fig:GEp-GMp}
\end{center}
\end{figure}

\begin{figure}[h]
\begin{center}
\includegraphics[width=75mm]{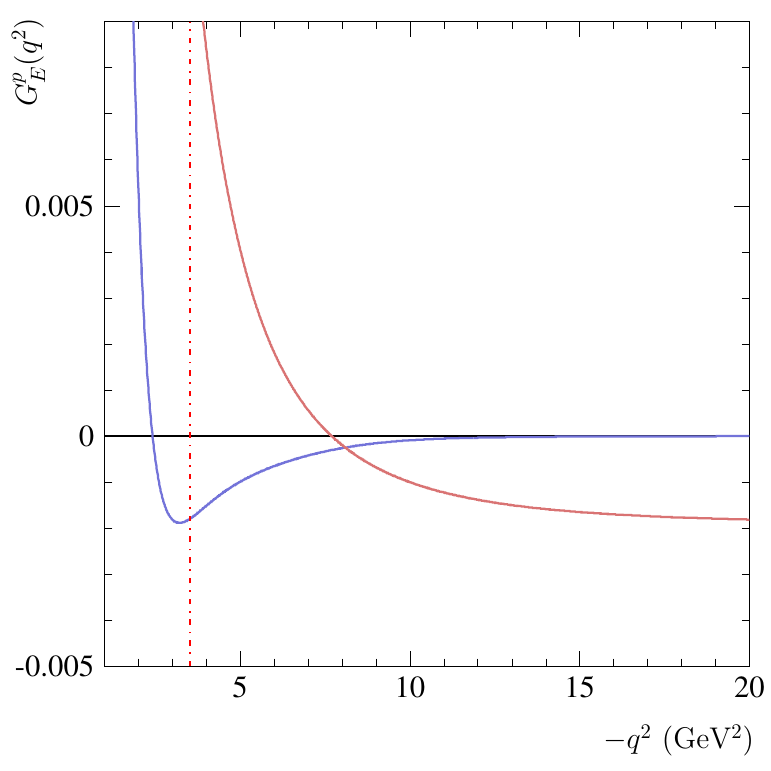}
\\
\includegraphics[width=75mm]{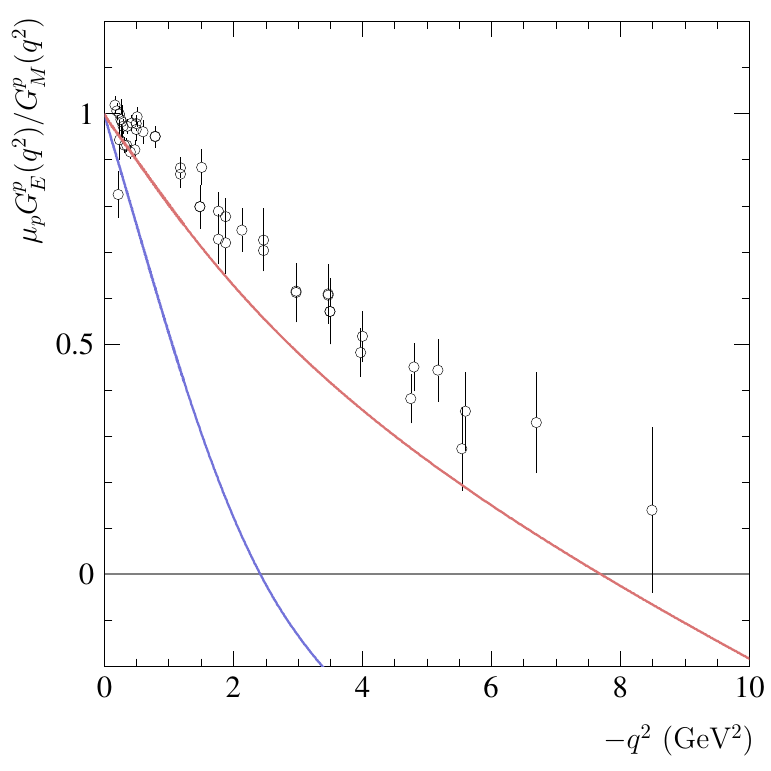}
\vspace{0mm}
\caption{The electric (upper panel) proton FF and the ratio electric to magnetic (lower panel) proton FF in the SL region. The ratio is compared with the data, empty circles, obtained by means of polarization observables~\cite{Perdrisat}. The red and violet curves represent the predictions that include and do not include relativistic corrections, respectively. The red dash-dotted line in the upper panel indicates the value $Q^2=(2M_N)^2$.}
\label{art-fig:GEp-Rp}
\end{center}
\end{figure}

\begin{figure}[h]
\begin{center}
\includegraphics[width=75mm]{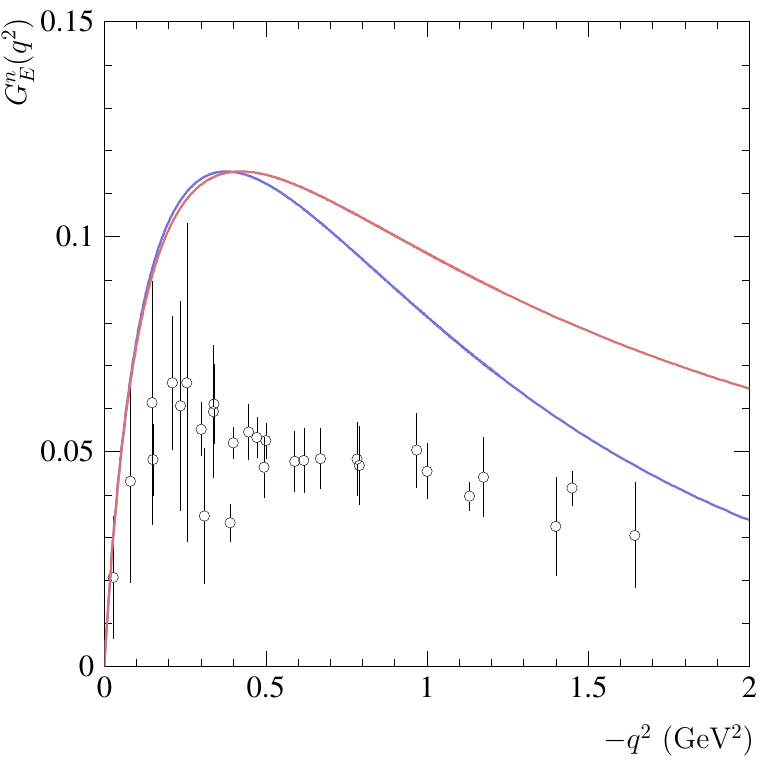}
\\
\includegraphics[width=75mm]{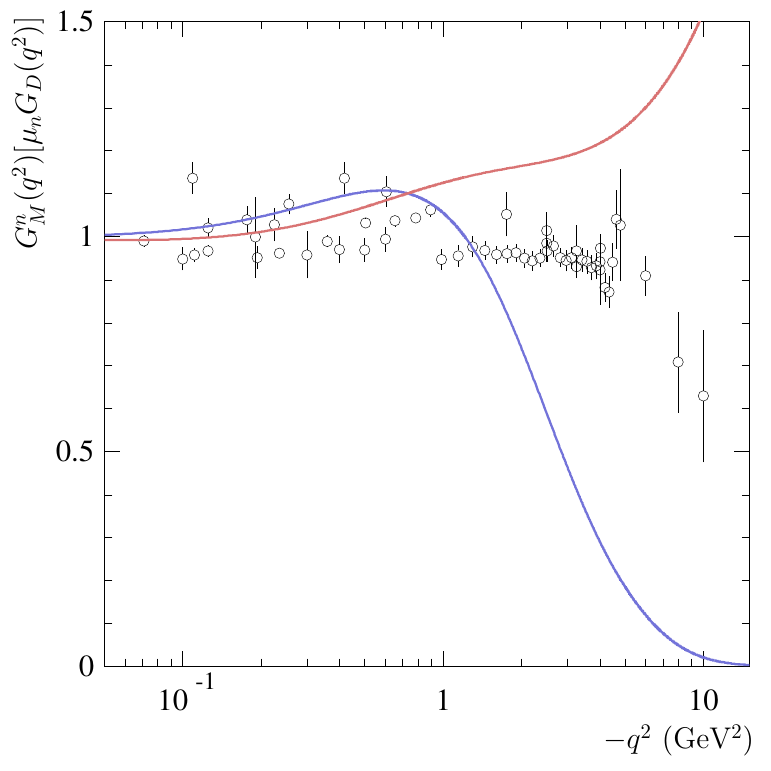}
\vspace{0mm}
\caption{The electric (upper panel) and magnetic (lower panel) neutron FFs in the SL region compared with the world data set, empty circles, from Ref.~\cite{Perdrisat}. The red and violet curves represent the predictions for the FFs obtained including and not including relativistic corrections as given in Eq.~(\ref{slcor}).}
\label{art-fig:GEn-GMn}
\end{center}
\end{figure}

Figures~\ref{art-fig:GEp-GMp}-\ref{art-fig:GEn-GMn} show predictions (red and violet curves) and data (empty circles) for the SL electric and magnetic FFs of proton and neutron. 
In particular, red and violet curves represent the predictions including and not including the relativistic correction described in Eq.~(\ref{slcor}).
In the case of the proton, Fig.~\ref{art-fig:GEp-GMp}, the relativistic-uncorrected predictions, also thanks to the constrained unitary normalization at $Q^2=0$, describe quite well data up to $Q^2\simeq~0.4$~GeV$^2$. Above this limit the predictions 
start to decrease faster than the dipole.
Such a behavior is expected in case of the magnetic FF, in fact, as shown in Eq.~(\ref{eq:asy-6}), the power law that rules its high-$Q^2$ vanishing is $Q^{-6}$. On the other hand, the electric FF, due to the contribution of $G_E^S$, see Eq.~(\ref{eq:asy-sca}), should tend to zero as $Q^{-4}$, i.e., at the same rate as the dipole.
%
%  GEp and GMp
%
\vspace{-0mm}
%
%  GEp-study and GEp/GMp
%
\\
The obtained faster vanishing behavior is due to the presence of a zero for $G_E^p(Q^2)$, at $Q_0^2\simeq 2.3$ GeV$^2$,  see the violet curve in the upper panel of Fig.~\ref{art-fig:GEp-Rp}, so that $G_E^p(Q^2)\to 0^-$ (from below), as $\mathcal{O}\left(Q^{-4}\right)$, when $Q^2\to\infty$.
The agreement with data is improved by including the relativistic corrections, red curve in Fig.~\ref{art-fig:GEp-GMp}. In particular in case of $G_E^p$, left panel of Fig.~\ref{art-fig:GEp-GMp}, the prediction follows the trend of the data, i.e., the dipole behavior, up to $Q^2\simeq 1$~GeV$^2$ and then it drops down. This is a consequence of the $Q^2$-dilation nature of the relativistic correction, that moves the zero for $G_E^p(Q^2)$ from $Q_0^2$ to 
$Q_{0,\rm rel}^2=4M_N^2/(4M_N^2/Q_0^2-1)\simeq 7.7$ GeV$^2$, see the red curve in the lower panel of Fig.~\ref{art-fig:GEp-Rp}, and hence the quick descent is shifted at higher $Q^2$.
\\
As already discussed in Sec.~\ref{subsec:rel-corr}, the asymptotic behavior of the electric FF is drastically modified by the relativistic corrections, in fact, $G_E^{p,\rm rel}(Q^2)$ tends to the finite value $G_E^p(4M_N^2)$, i.e.,
\be
G_E^{p,\rm rel}(Q^2)\mathop{\longrightarrow}_{Q^2\to\infty}G_E^p(4M_N^2)\simeq 0.012\,.
\no\en
The fact that such a value is very close to zero, see the vertical line in the upper panel of Fig.~\ref{art-fig:GEp-Rp}, and that the uncorrected electric FF scales as the dipole makes the corrected FF closer to the data.
\\
Also the prediction for the magnetic proton FF, lower panel of Fig.~\ref{art-fig:GEp-GMp}, improves its agreement with data up to $Q^2\simeq 4$ GeV$^2$, when the relativistic corrections are considered. 
In this case, at high $Q^2$, the prediction gets larger than data, see also the lower panel of Fig.~\ref{art-fig:SL-p}, and its steep rising, from $Q^2\simeq 5$ GeV$^2$, is a consequence of the normalization to the dipole. Moreover, asymptotically $G_M^{p,\rm rel}$ goes like $Q^{-2}$, so that the ratio to the dipole grows like $Q^2$. Contrary to the case of $G_E^p$,
no zeros are found for $G_M^p$. 
\\
The lower panel of Fig.~\ref{art-fig:GEp-Rp} shows the ratio between electric and magnetic proton FFs normalized to the proton magnetic moment, the red and violet curves are the predictions with and without relativistic corrections, while the empty circles represent the data extracted from polarization transfer observables in $e$-$p$ scattering~\cite{Perdrisat}. 
Such experimental values show an unexpected linear decreasing trend, whose extrapolation would give a zero at $Q^2\simeq 10$~GeV$^2$, which is close to the obtained value $Q_{0,\rm rel}^2\simeq 7.7$ GeV$^2$.
%
%  GEn and GMn
%
\\
Electric and magnetic FFs of neutron are shown in Fig.~\ref{art-fig:GEn-GMn} in comparison with the data. The two predictions, also in this case, refer to the relativistically corrected (red curve) and uncorrected (violet  curve) results.
Apart from the low-$Q^2$ region, where the normalization forces the predictions to follow the experimental points, the agreement with data appears worse with respect to what has been found for the proton. The inclusion of relativistic corrections does not improve the accordance with data, in particular, the agreement is even worsened in case of $G_E^n$, left panel of Fig.~\ref{art-fig:GEn-GMn}. Finally, no zeros are found for $G_E^n$ and $G_M^n$. 
\\
Concerning the asymptotic behavior of neutron SLFFs, the same conclusions driven for the proton can be considered. In particular, as $Q^2\to\infty$, the uncorrected predictions for $G_E^n$ and $G_M^n$ scale as $Q^{-4}$ and $Q^{-6}$, respectively, while the corrected behaviors are 
\be
\left\{\begin{array}{rcl}
G_E^{n,\rm rel}(Q^2)&\ds\mathop{\longrightarrow}_{Q^2\to\infty}&
G_E^n(4M_N^2)\simeq 0.009\\
G_M^{n,\rm rel}(Q^2)&\ds\mathop{\longrightarrow}_{Q^2\to\infty}&
\ds\frac{4M_N^2}{Q^2}G_M^n(4M_N^2)\\
\end{array}\right.\,.
\no\en
\subsubsection{Time-like region}
\label{subsubsec:res-tl}
\begin{figure}[h]
\footnotesize 
\bm{\columnwidth}
\includegraphics[height=80mm]{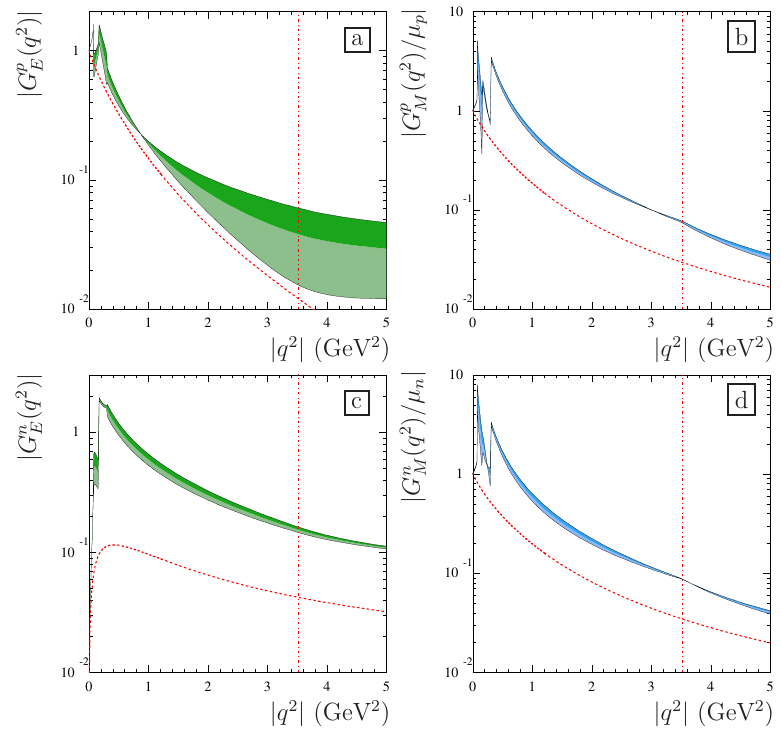}
\em
\vspace{-0mm}
\caption{Panels a and b: moduli of TL (green and blue bands) and SL (dashed red curve) electric and normalized magnetic FFs of the proton. Panels c and d: moduli of TL (green and blue bands) and SL (dashed red curve) electric and normalized magnetic FFs of the neutron. The symbol $|q^2|$ stands for positive (TL) and negative  (SL) $q^2$. The red dash-dotted line indicates the physical threshold $\qphy=(2M_N)^2$. Only relativistically corrected values have been considered.}
\label{art-fig:SL.vs.TL}
\end{figure}

\begin{figure}[h]
\begin{center}
\includegraphics[height=80.mm]{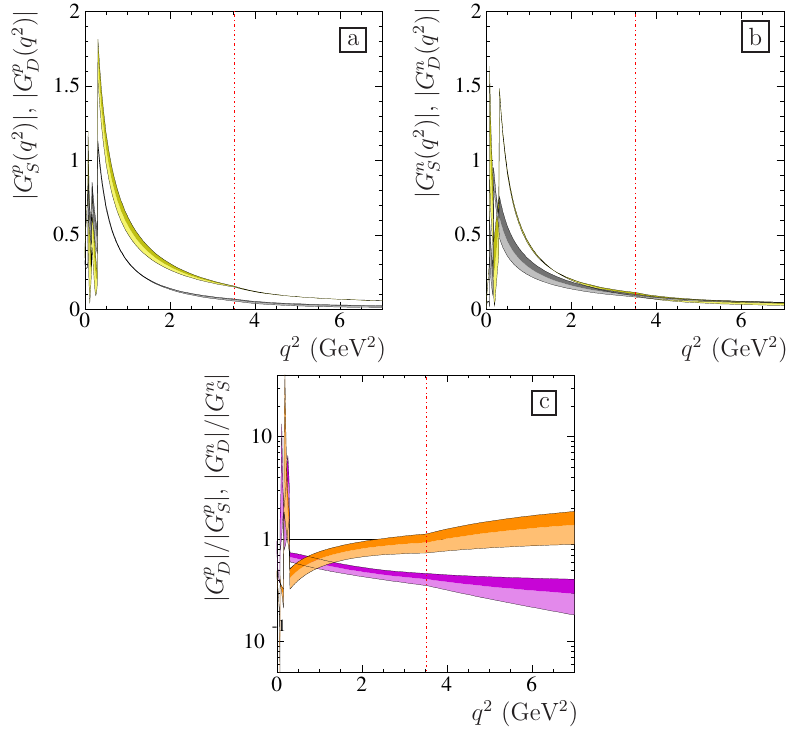}
\vspace{0mm}
\caption{Panel a: moduli of the S-wave (yellow band) and D-wave (grey band) proton FFs. Panel b: moduli of the S-wave (yellow band) and D-wave (grey band) neutron FFs. Panel c: moduli of the ratios of D-wave and S-wave FFs of proton (violet band) and neutron (orange band). The red dash-dotted line indicates the physical threshold $\qphy=(2M_N)^2$.} 
\label{art-fig:GSGDTL-mod}
\end{center}
\end{figure}

\begin{figure}[h!]
\begin{center}
\includegraphics[width=75mm]{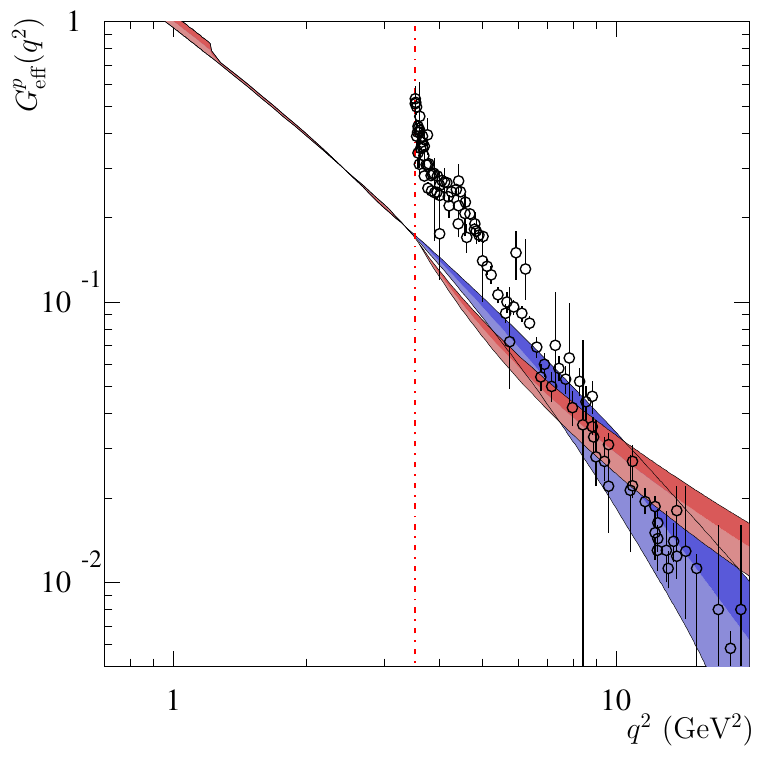}
\\
\includegraphics[width=75mm]{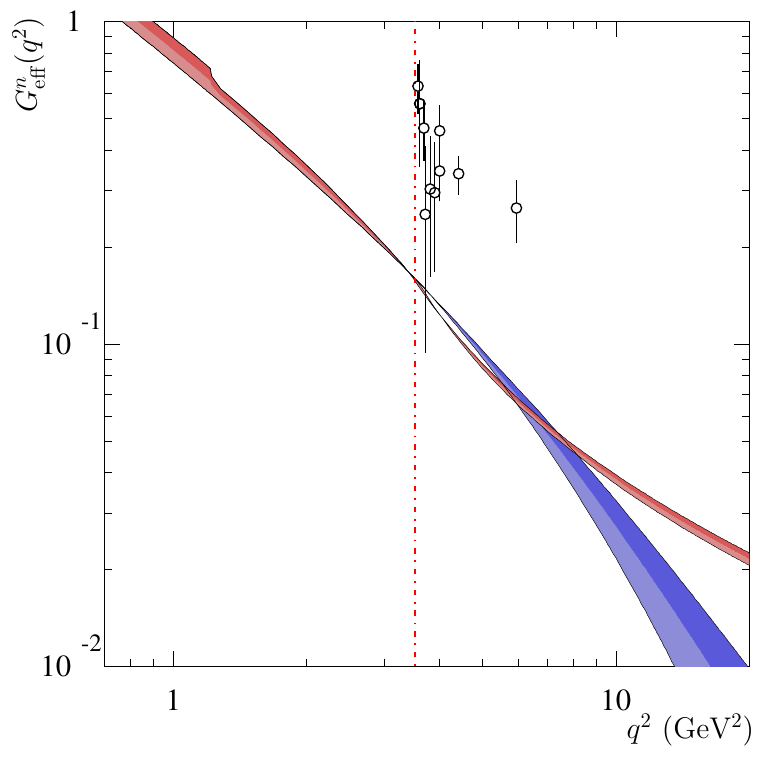}
\vspace{0mm}
\caption{Upper panel: effective FF of the proton. Lower panel: effective neutron FF. Red and violet bands represent the relativistically corrected and the uncorrected values, respectively. The red dash-dotted line indicates the physical threshold $\qphy=(2M_N)^2$ and the empty points are the world data sets form Ref.~\cite{egle-simone} and references therein.
}
\label{art-fig:Geff}
\end{center}
\end{figure}

\begin{figure}[t]
\begin{center}
\includegraphics[width=75mm]{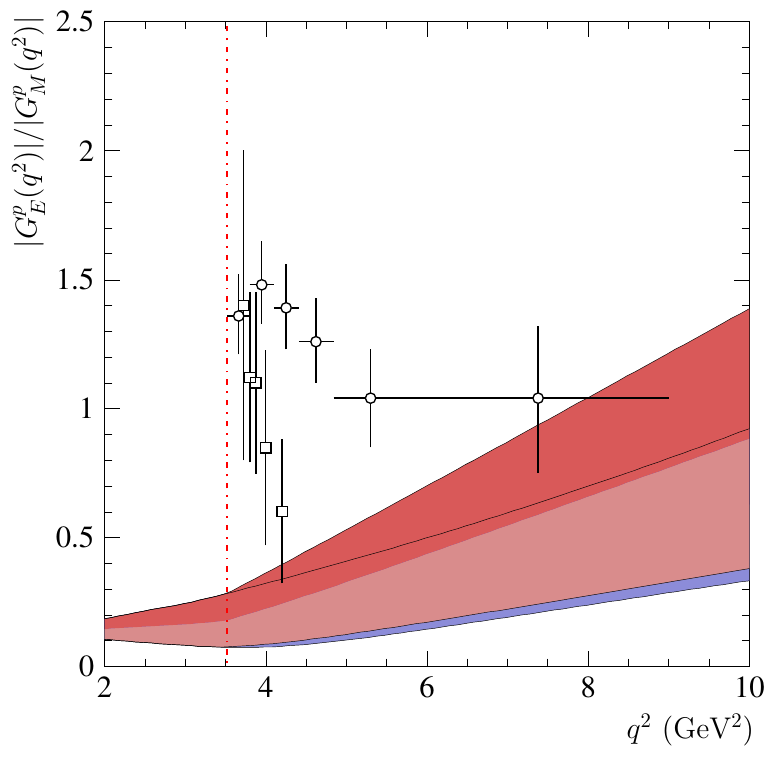}
\vspace{-3mm}
\caption{Modulus of the ratio electric to magnetic proton FF, 
red and violet bands represent predictions with and without relativistic corrections. Two incompatible sets of data are shown: the circles are from the BaBar Collaboration~\cite{babar-ratio} and the squares from the Lear Collaboration~\cite{lear-ratio}. The red dash-dotted line indicates the physical threshold $\qphy=(2M_N)^2$.}
\label{art-fig:RpTL}
\end{center}
\end{figure}

Results and data in the TL region will be described as functions of the positive, squared four-momentum transfer $q^2=-Q^2>0$. As extensively discussed in Sec.~\ref{subsec:TLFFs}, starting from the theoretical threshold $q^2_{\rm theo}=(2M_\pi)^2$, FFs develop non-vanishing imaginary parts due to the coupling of the virtual photon, which now has enough virtual mass, with hadronic intermediate states. It follows that, in this kinematical region, the nucleon structure is described by four real functions, i.e., real and imaginary parts of the electric and magnetic FFs. 
\\
%
%  GEp and GMp TL
%
\begin{figure}[h!]
\bm{\columnwidth}
\includegraphics[height=80mm]{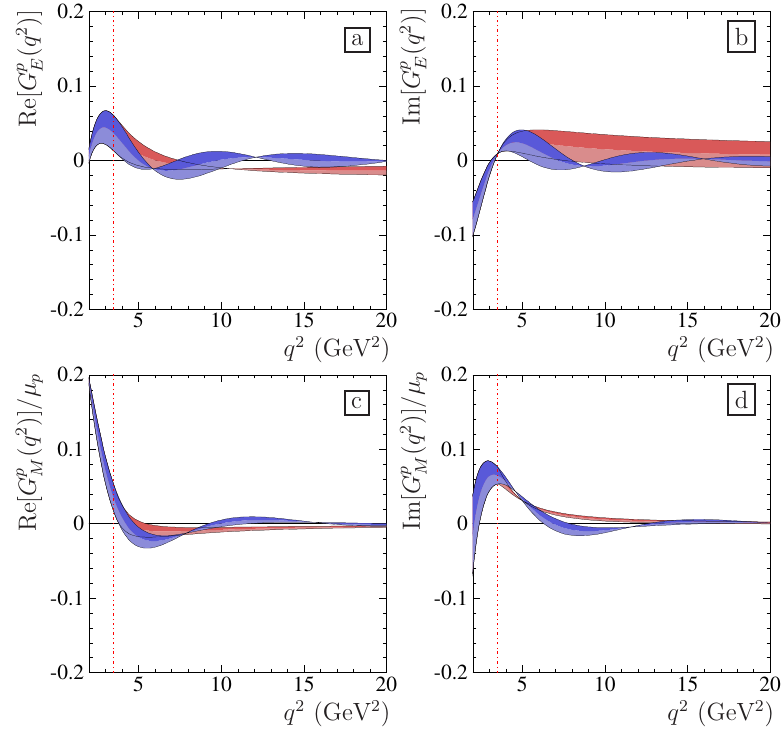}
\em
\vspace{-0mm}
\caption{Panels a and b: real and imaginary part of the proton electric FF. Panel c and d: real and imaginary part of the proton magnetic FF normalized to the magnetic moment $\mu_p$. The red dash-dotted line indicates the physical threshold $\qphy=(2M_N)^2$. The bands, red and violet include and not include relativistic corrections respectively, are given by the combination of the curves obtained by considering two different discretization procedures (see text).}
\label{art-fig:GMpTL}
\vspace{0mm}
\end{figure}
%
%  GEn and GMn TL
%
\begin{figure}[h!]
\footnotesize 
\bm{\columnwidth}
\includegraphics[height=80mm]{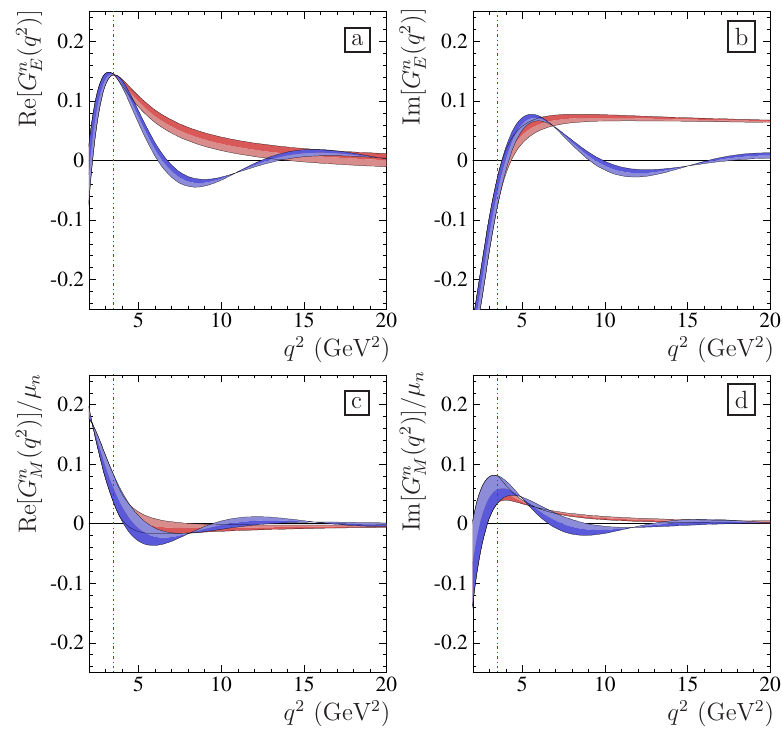}
\em
\vspace{-0mm}
\caption{Panels a and b: real and imaginary part of the neutron electric FF. Panel c and d: real and imaginary part of the neutron magnetic FF normalized to the magnetic moment $\mu_n$. The red dash-dotted line indicates the physical threshold $\qphy=(2M_N)^2$. Color scheme of Fig.~\ref{art-fig:GMpTL}.}
\label{art-fig:GMnTL}
\end{figure}\vspace{0mm}\\
Figures~\ref{art-fig:GMpTL} and~\ref{art-fig:GMnTL} show real and imaginary parts of TLFFs for proton and neutron respectively, including (red band) and not including (violet band) the relativistic corrections, as given in Eq.~(\ref{tlcor}).
Such corrections become effective only above the physical threshold $q^2=(2M_N)^2$. For all these quantities there are no available data and moreover, even in case of an ideal experiment able to exploit also polarization observables in annihilation processes, only relative phases between $G_E^N$ and $G_M^N$ would be accessible, besides their moduli.
In Fig.~\ref{art-fig:SL.vs.TL} the relativistically corrected, TL (solid band) and SL (dash-dot red curve) moduli of the four nucleon FFs are represented as functions of $|q^2|$. 
Apart from the very first portion of the unphysical region, $0\le q^2\le 1$ GeV$^2$, where the opening of the logarithmic branch cuts
manifests itself in bumpy behaviors, TLFFs are smooth decreasing functions of $q^2$. Moreover, as it is shown in Fig.~\ref{art-fig:SL.vs.TL}, TLFFs are systematically larger than their SL counterparts at $|Q^2|=|q^2|$. Such a discrepancy contrasts with the Phragm\'{e}n-Lindel\"{o}f theorem (see Sec.~\ref{subsec:analyticity}), stating that SL and TL limits of a given FF should correspond. However, on the one hand, as already discussed, relativistic corrections entail important modifications of the asymptotic behavior,
and on the other hand, it seems plausible to consider as center of mass of the SL-TL symmetry not simply $Q^2=0$ but rather a TL value, say $q^2_{\rm CM}$, lying inside the unphysical region. In light of this we should expected
$G_{E,M}^N(Q^2)\simeq |G_{E,M}^N(q^2+2q^2_{\rm CM})|$ (with the argument $Q^2$ we mean SLFF at $|Q^2|=|q^2|$) and using, for instance, $q^2_{\rm CM}=1$ GeV$^2$, the SL-TL discrepancy can be reduced.
%
%  FFs SL vs. TL
%
\noindent
Electric and magnetic FFs of proton and neutron are obtained using the combinations, given in Eq.~(\ref{eq:SV-pn}), of the isospin components, which are the Fourier transforms, see Eq.~(\ref{eq:FTs}), of the two profiles $b(r)$ and $t(r)$, defined in terms of the same chiral angle $F(r)$ through the non-linear differential Eq.~(\ref{eq:b-t}). It follows that, FFs are all non-trivially interconnected. Moreover, as given in Eq.~(\ref{eq:sachs}), $G_E^N$ and $G_M^N$ are also linearly related to the Dirac and Pauli FFs, in such a way that, assuming no singularity at the physical threshold $q^2_{\rm phys}=(2M_N)^2$ for $F_1^N$ and $F_2^N$, the electric and magnetic FFs of each nucleon must coincide at such a $q^2$ value. As explained in Sec.~\ref{subsec:analyticity}, the identity $G_E^{N}(4M_N^2)=G_M^{N}(4M_N^2)$ implies (it is a sufficient condition for) isotropy at the production threshold, i.e., the differential cross section for $e^+e^-\to N\overline{N}$ in the $e^+e^-$ center of mass frame,
\be
\frac{d\sigma_{N\overline{N}}}{d\cos\theta}&=&
\frac{\pi\alpha^2}{2q^2}\sqrt{1-\frac{4M^2_N}{q^2}}
\Bigg\{
\left[1+\cos^2(\theta)\right]\big|G_M^N(q^2)\big|^2
\Bigg.\no\\
&&\Bigg.
+
\frac{4M_N^2}{q^2}\sin^2(\theta)\big|G_E^N(q^2)\big|^2
\Bigg\}\,,
\label{eq:dsigma}\en
loses its dependence on the scattering angle $\theta$ as $q^2\to (q_{\rm phys}^{2})^+$. This also means that, even though parity conservation allows S and D-wave
for the $N\overline{N}$ system, at the production threshold only the S-wave can contribute. So that, by reversing the argument, the violation of the identity\footnote{Being TLFFs complex functions of $q^2$, the equality $G_E^{N}(4M_N^2)=G_M^{N}(4M_N^2)$ is equivalent to two independent identities for the real and the imaginary parts.}
 $G_E^{N}(4M_N^2)=G_M^{N}(4M_N^2)$
 would imply anisotropy, i.e., the presence of a D-wave contribution also at threshold or, equivalently, the presence of singularities in the Born amplitude.%~\cite{d-wave-threshold}.
%
%  GE and GM at threshold
%
\begin{figure}[h]
\footnotesize 
\bm{\columnwidth}
\includegraphics[height=80mm]{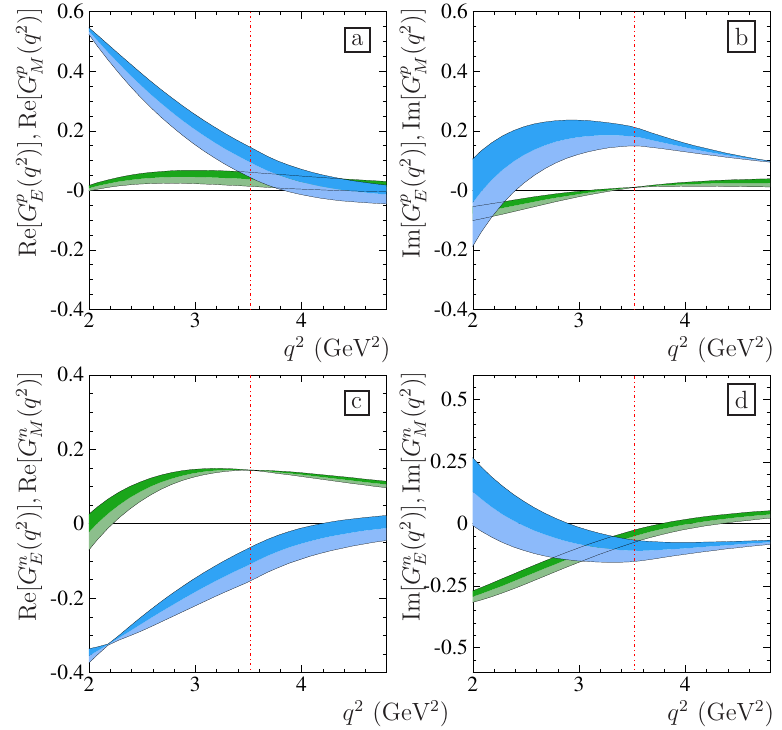}
\em
\vspace{-0mm}
\caption{Panels a and b: real and imaginary part of electric (green band) and magnetic (blue band) FFs of the proton. Panels c and d: real and imaginary part of electric (green band) and magnetic (blue band) FFs of the neutron. The red dash-dotted line indicates the physical threshold $q^2_{\rm phys}=(2M_N)^2$.}
\label{art-fig:GEGMTL}
\end{figure}
Figure~\ref{art-fig:GEGMTL} shows a comparison between real and imaginary parts of electric and magnetic FFs for proton and neutron, in the region of $q^2$ across the physical threshold $q^2_{\rm phys}$ (vertical line).
In order to verify the equality $G_E^{N}(4M_N^2)=G_M^{N}(4M_N^2)$, the two pairs of real parts, as well as the two of pairs of imaginary parts, would coincide at the threshold. Since no one of these identities is verified, there is no coincidence between electric and magnetic FFs at the production threshold (see next section for a detailed discussion).
\begin{figure}[h]
\begin{center}
\includegraphics[width=75mm]{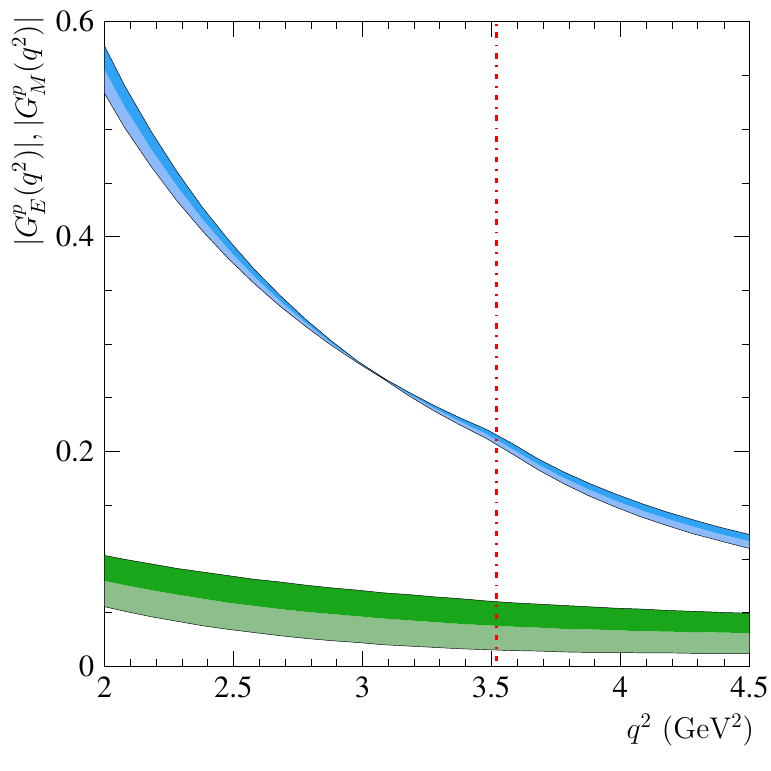}
\\\
\includegraphics[width=75mm]{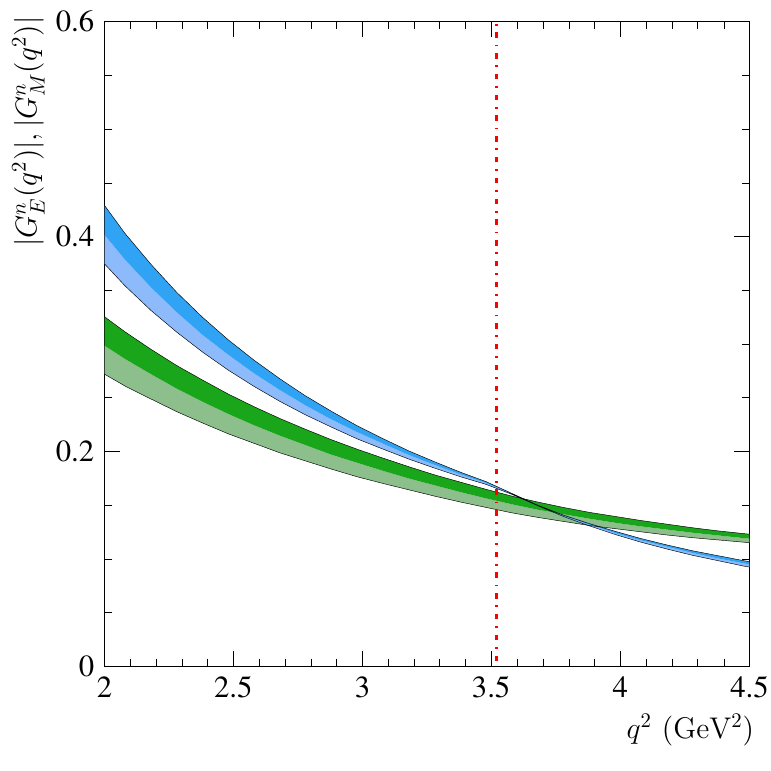}
\vspace{0mm}
\caption{Moduli of the electric (blue band) and magnetic (green band) FFs of the proton (upper panel) and neutron (lower panel).  The red dash-dotted line indicates the physical threshold $q^2=(2M_N)^2$.} 
\label{art-fig:GEGMTL-mod}
\end{center}
\end{figure}
\vspace{0mm}\\
It is interesting to notice that the differences among real and imaginary parts at the threshold are partially compensated when moduli are taken into account, as shown in Fig.~\ref{art-fig:GEGMTL-mod}.
Nevertheless, there is isotropy-violation at the threshold $q^2=\qphy$ as it is shown in Fig.~\ref{art-fig:GSGDTL-mod}, where, in the upper panels, are reported moduli of the S-wave and D-wave, proton and neutron FFs which are defined in terms of Sachs FFs as
\be
G_S^N(q^2)&=&\frac{2 \sqrt{q^2/(4M_N^2)}\,G_M^N(q^2)+G_E^N(q^2)}{3}\,,
\no\\
G_D^N(q^2)&=&\frac{\sqrt{q^2/(4M_N^2)}\,G_M^N(q^2)-G_E^N(q^2)}{3}\,.
\nen
\vspace{-3mm}
\vspace{-0mm}\\
The lower panel of Fig.~\ref{art-fig:GSGDTL-mod} shows the relative contribution, in modulus, of the D-wave with respect to the S-wave FF. It turns out that, in case of the neutron (orange band), the isotropy-violation is stronger, indeed the D-wave is close to the S-wave contribution, in the region around the threshold $q^2_{\rm phys}=4M_N^2$, in particular: 
\be
|G_D^n(4M_N^2)|/|G_S^n(4M_N^2)|\simeq 0.9\,.
\nen
In the proton case, instead, as shown by the violet band on the lower panel of Fig.~\ref{art-fig:GSGDTL-mod}, is the S-wave that gives the main contribution, at the threshold: 
\be
|G_D^p(4M_N^2)|/|G_S^p(4M_N^2)| \simeq 0.4\,.
\nen
Finally to have a comparison with data in the TL region, we consider the so called effective FF, $G_{\rm eff}^N(q^2)$, 
corresponding to the useful working hypothesis of a unique TLFF, that is: $|G_E^N(q^2)|=|G_M^N(q^2)|\equiv G_{\rm eff}^N(q^2)$. Its expression in terms of the Sachs FFs follows by writing the $e^+e^-\to N\overline{N}$ total cross section, obtained from Eq.~(\ref{eq:dsigma}), as
\be
\sigma_{N\overline{N}}\!&=&\!
\sigma_{\rm PL} \cdot \left[G^{N}_{\rm eff}(q^2)\right]^2
\label{eq:sigma}
\\
\!&=&\!
\frac{4\pi\alpha^2}{3q^2}\sqrt{\!1\!-\!\frac{4M^2_N}{q^2}}
\Bigg[\!
\big|G_M^N(q^2)\big|^2\!\!+\!
\frac{2M_N^2}{q^2}\big|G_E^N(q^2)\big|^2
\!\Bigg]\,,
\nen
where $\sigma_{\rm PL}$ represents the cross section in case of point-like fermions in the final state, which is obtained by putting $G_E^N=G_M^N\equiv 1$ in the last expression of Eq.~(\ref{eq:sigma}). 
It follows that the effective FF is 
\be
G_{\rm eff}^N(q^2)\!&=&\!
\sqrt{\frac{\sigma_{N\overline{N}}}{\sigma_{\rm PL}}}
\no\\
\!&=&\!
\sqrt{\frac{q^2\big|G_M^N(q^2)\big|^2+2M_N^2\big|G_E^N(q^2)\big|^2}{q^2+2M_N^2}}\,.
\label{eq:geff}
\en
Figure~\ref{art-fig:Geff} shows the results for the proton (upper panel) and the neutron (lower panel) effective FFs together with all the available data. 
In case of the proton the predictions, in particular the relativistic-uncorrected one, describe quite well the data in the high momentum transfer region, from $q^2\simeq 7$ GeV$^2$ on, while they
fail in reproducing the experimental $G_{\rm eff}^{p}$ at lower $q^2$, close to the physical threshold. Concerning the neutron effective FF, lower panel of Fig.~\ref{art-fig:Geff}, the predicted behavior does not agree with the available data that, however, cover only the near-threshold region.  
Finally, Fig.~\ref{art-fig:RpTL} shows the modulus of the ratio electric to magnetic proton FF in comparison with the data. 
The isotropy-violation is manifest, having at the threshold a non-unitary value. The agreement with data, that favor a constant behavior at high $q^2$, is quite poor, because, both results, corrected and uncorrected, have an increasing behavior, almost linear in $q^2$. This is a consequence of the different high-$q^2$ behaviors predicted for $G_E^p(q^2)$ and $G_M^p(q^2)$, both, in case of uncorrected results, where it is found $G_E^p(q^2)\propto \left(q^2\right)^{-2}$ and $G_M^p(q^2)\propto \left(q^2\right)^{-3}$, see Eqs.~(\ref{eq:asy-sca}) and~(\ref{eq:asy-vec}), and in case of the relativistic predictions, given in Eqs.~(\ref{tlcor}), where $G_E^p(q^2)\propto$ [constant] and $G_M^p(q^2)\propto \left(q^2\right)^{-1}$.
%
%  Gpeff and Gneff
%
%
%
%
%
%   R_p TL
%
\vspace{-0mm}\\
It is just such a failure in predicting the perturbative QCD power-law, see Sec.~\ref{subsec:rel-corr}, that precludes the possibility of drawing any conclusion about the asymptotic regions.
\subsection{Isotropy at the physical threshold}
\label{subsec:isotropy}
\begin{figure}[h!]
\bm{\columnwidth}
\includegraphics[height=80mm]{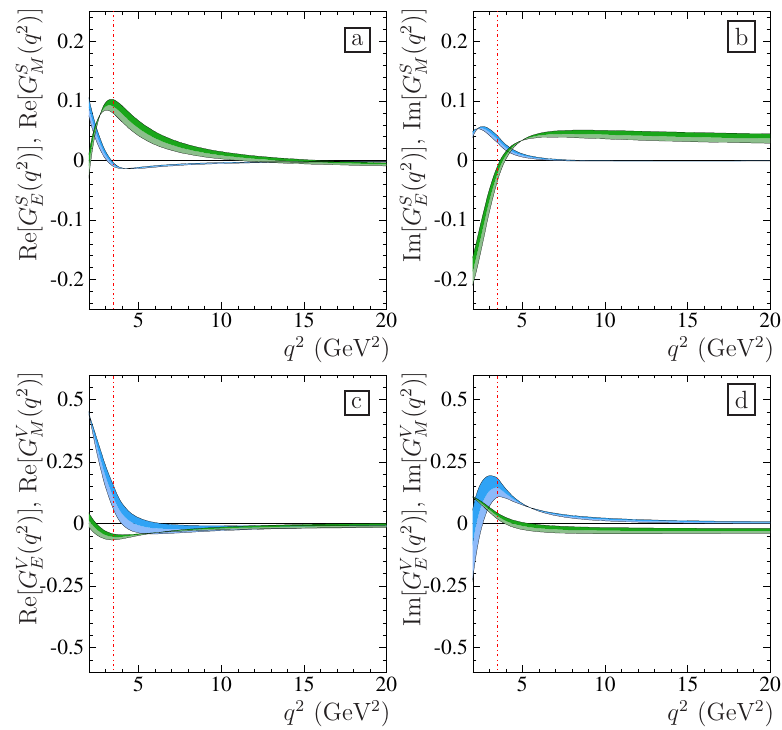}
\em
\vspace{-0mm}
\caption{Panels a and b: real and imaginary part of the isoscalar electric (blue band) and magnetic (green band) FFs. Panels c and d: real and imaginary part of the isovector magnetic (blue band) and magnetic (green band) FFs. The red dash-dotted line indicates the physical threshold $\qphy=(2M_N)^2$.}
\label{art-fig:GSGVTL}
\end{figure}

\begin{figure}[h]
\includegraphics[width=\columnwidth]{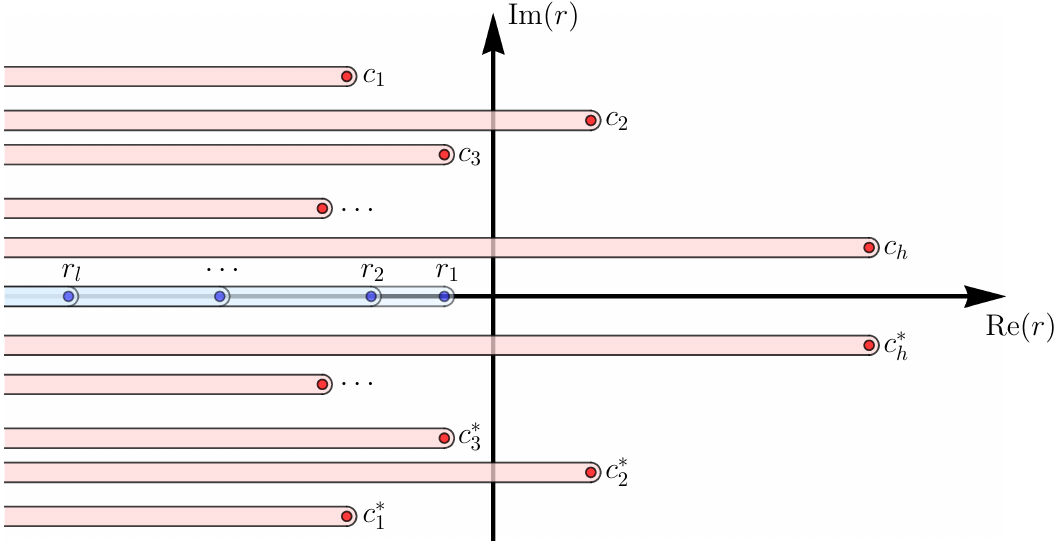}
\vspace{-3mm}
\caption{The analyticity domain of the chiral angle $F(r)$. The blue and red disks indicate, respectively, real and non-real branch points and the shaded bands the corresponding branch cuts, that have been chosen with constant imaginary parts, \ie, parallel to the real axis. Each branch point for $F(r)$ corresponds to a simple pole for the profile function $b(r)$, as a consequence of the definition given in Eq.~(\ref{eq:b-t}).}
\label{art-fig:orizontal-cuts}
\end{figure}

Following the treatment given in Sec.~\ref{subsec:analyticity}, isotropy at the production threshold manifests itself through the identity
\be
G_E^{N}(4M_N^2)=G_M^{N}(4M_N^2)\,,
\label{eq:isotropy-em}
\en
for proton, $N=p$, and neutron, $N=n$. Moreover, being the Sachs FFs (independent) linear combinations of the isospin components, \ie,
$G^{p,n}_{E,M}(q^2)=G_{E,M}^S(q^2)\pm G_{E,M}^V(q^2)$, the identity of Eq.~(\ref{eq:isotropy-em}) is equivalent to
\be
G_E^{S,V}(4M^2_N)=G_M^{S,V}(4M^2_N)\,.
\label{eq:isotropy-sv}
\en
As already discussed in Sec.~\ref{subsec:em-current}, the combination of such
isospin components, that represent our primary outcomes, to obtain proton and neutron Sachs FFs, has to be performed with some care due to the different orders in $1/N_c$ expansion in which they are computed. In particular, from the definitions of Eq.~(\ref{eq:FTs}) and having that both, the mass $M$ and the moment of inertia $\Lambda$ are $\mathcal{O}[N_c]$, we get
\be
\begin{array}{rcl c rcl}
G_E^S\!&=&\!\mathcal{O}\left[N_c^{0}\right]\,, &&
G_M^S\!&=&\!\mathcal{O}\left[N_c^{0}\right]\,,\\
&& && &&\\
G_E^V\!&=&\!\mathcal{O}\left[N_c^{-1}\right]\,,&&
G_M^V\!&=&\!\mathcal{O}\left[N_c\right]\,.\\
\end{array}
\nen
It follows that the more reliable test bed, as necessary condition for the isotropy hypothesis, is the isoscalar identity of Eq.~(\ref{eq:isotropy-sv}). In other words, the violation of such an identity would imply anisotropy.
Figure~\ref{art-fig:GSGVTL} shows real and imaginary parts of the four isospin components of the electric and magnetic FFs in the TL region, across the physical threshold. In every instance, and hence also for the isoscalar FFs, figs.~\ref{art-fig:GSGVTL}a and~\ref{art-fig:GSGVTL}b, the identity is violated, \ie, the curves do not cross each other at the threshold, which is indicated by the vertical red line.
%
%  GS and GV at threshold
%
\\
Let us consider in more detail the constraints imposed by 
 the isoscalar equation. The TL expression of $G_M^S(q^2)$ is obtained by following the procedure, described in~\ref{app:c}, that has been used to compute the expression of $G_E^S(q^2)$ given in Eq.~(\ref{eq:GES-TL}). In particular, considering the same symbols,
  %of Eq.~(\ref{eq:GES-TL}), 
  it reads
\be
G_M^S(q^2)
\!&=&\!
-\frac{\pi M}{\Lambda q^3}
\!\!\sum_{j=1}^m \re\Big\{\!
\tilde{R}_{z_j} H\left[(-q-3M_{\pi})z_j\right](z_j\!+\!1)\Big.
\no\\
&&\Big.+
\tilde{R}_{z_j} H\left[(q-3M_{\pi})z_j\right](z_j-1)
\Big\}\no\\
&&+
\frac{2i\pi^2 M}{\Lambda q^3}\sum_{j=1}^h 
\theta(q-3M_\pi)\theta(-x_j)
\no\\
&&
\times\left[
\tilde R_{c_j} e^{(q-3M_\pi)c_j}\theta(y_j)(c_j+1)\right.\no\\
&&\left.
-\tilde R_{c_j}^*e^{(q-3M_\pi)c_j^*}\theta(-y_j)(c_j^*-1)
\right]\,.
\nen
At the physical threshold, $q=2M_N$, the isoscalar magnetic FF is
\be
G_M^S(4M_N^2)
\!&=&\!
-\frac{\pi }{8M_N^2 \Lambda}
\no\\
&&\times\!\!
\sum_{j=1}^m \re\left\{\!
\tilde{R}_{z_j} H\left[(-2M_N\!\!-3M_{\pi})z_j\right]\!
(1\!+\!z_j)\right.
\no\\
&&\left.
-\tilde{R}_{z_j} H\left[(2M_N-3M_{\pi})z_j\right](1-z_j)
\right\}\no\\
&&+
\frac{i\pi^2}{4M_N^2\Lambda }\sum_{j=1}^h 
\theta(-x_j)
\no\\
&&\times
\left[
\tilde R_{c_j} e^{(2M_N-3M_\pi)c_j}\theta(y_j)(1+c_j)\right.\no\\
&&\left.
+\tilde R_{c_j}^*e^{(2M_N-3M_\pi)c_j^*}\theta(-y_j)(1-c_j^*)
\right]\,,
\nen
while the electric one, from Eq.~(\ref{eq:GES-TL}), 
\be 
G_E^S(4M_N^2)&\!=\!&
-\frac{\pi}{2M_N}\sum_{j=1}^{h} \re\left\{ \tilde{R}_{z_j} H\left[(-2M_N-3M_{\pi})z_j\right]\right.\no\\
&&\left.-
\tilde{R}_{z_j} H\left[(2M_N-3M_{\pi})z_j\right]\right\}
%%\label{eq:ges-easy-tl}
%
\no\\ &&
+\frac{i\pi^2}{M_N}\sum_{j=1}^h \theta(-x_j)\left[\tilde R_{c_j}
e^{(2M_N-3M_\pi)c_j}\theta(y_j)\right.\no\\
&&\left.+\tilde R_{c_j}^*e^{(2M_N-3M_\pi)c_j^*}\theta(-y_j)
\right]\,.
%\label{eq:GES-TL}
\nen
It follows that the isotropy condition of Eq.~(\ref{eq:isotropy-sv}) becomes
\be
&&
\sum_{j=1}^m \re\left\{
\tilde{R}_{z_j} H\left[(-2M_N-3M_{\pi})z_j\right](1\!+\!z_j\!+\!4M_N \Lambda)\right.
\no\\
&&\left.-\tilde{R}_{z_j} H\left[(2M_N-3M_{\pi})z_j\right](1-z_j+4M_N \Lambda)
\right\}=\no\\
&&
2i\pi\!\sum_{j=1}^h \!
\theta(-x_j)\left[\!
\tilde R_{c_j} e^{(2M_N-3M_\pi)c_j}\theta(y_j)(1\!+\!c_j\!+\!4M_N \Lambda)\right.\no\\
&&\left.+\tilde R_{c_j}^*e^{(2M_N-3M_\pi)c_j^*}\theta(-y_j)(1\!-\!c_j^*\!+\!4M_N \Lambda)
\right].
\label{eq:isotropy-S}
\en
It can be interpreted as an implicit relation among poles (they appear in the argument of the $H(z)$ functions, defined in~\ref{app:b}) and the corresponding residues of the function $b_{\rm fit}(r)$, that parametrizes the profile function $b(r)$, see Eq.~(\ref{fitt}).
\\
It is a quite hard task to obtain the identity of Eq.~(\ref{eq:isotropy-S}) from the beginning, \ie, as a condition 
which is automatically fulfilled by any parametrization. 
In fact, the possibility of using the definition of Eq.~(\ref{eq:b-t}) to relate directly the positions of the $b(r)$ poles to the properties of the chiral angle $F(r)$, is prevented by the fact that such a relation holds only for real and positive values of $r$, while the poles $z_j$, $j=1,\ldots,m$, lie in the $r$ complex plane outside the positive real axis. In other words, by solving numerically the Euler-Lagrange equation of the Skyrme model, no information about the complex structure of the chiral angle $F(r)$ can be accessed for $r\not\in(0,\infty)$. 
\\
Following the definition given in Eq.~(\ref{eq:b-t}), 
simple poles of $b(r)$ can be related to branch points of the chiral angle $F(r)$. By considering Eqs.~(\ref{fitt}) and~(\ref{eq:M-L}), and assuming the coincidence between fit function and $b(r)$, we have
\be
b(r) =b_{\rm fit}(r)\!&=&\!
 -\frac{F'(r)}{2\pi^2}\,\frac{\sin^2\left[F(r)\right]}{r^2}\no\\
\!&=&\!e^{-3M_\pi r}\,\sum_{k=1}^m
\frac{R_k }{r-z_k}\,.
\nen
Such a differential equation for $F(r)$ can be integrated and, by using the condition $\sum_{k=1}^m R_k=0$, it is
\be
\int_{F(0)=\pi}^{F(r)} \sin^2\left(\tilde F\right) d\tilde F\!&=&\! -2\pi^2 \sum_{k=1}^m R_k \int_0^r\frac{{r'}^2 e^{-3M_\pi r'}}{r'-z_k}dr'
\no\\
2F(r)-\sin\left[2F(r)\right]
\ug
-8\pi^2\sum_{k=1}^m R_kz_k\Bigg\{\frac{1-e^{-3M_\pi r}}{3M_\pi}
\Bigg.\no\\
&&\Bigg.+
z_k e^{-3M_\pi z_k}\Big[{\rm Ei}\big(3M_\pi(z_k-r)\big)\Big.
\no\\
&&\Big.-{\rm Ei}\left(3M_\pi z_k\right)\Big]\Bigg\}\,,
\label{eq:Ei}\en
where Ei$(z)$ is the multi-valued exponential integral function\footnote{The exponential integral function is defined as~\protect\cite{stegun}
$$
{\rm Ei}(z)=-\int_{-z}^\infty\frac{e^{-t}}{t}dt\,,
$$
the integration is in principal value for real and positive $z$.}, that has branch points in $z=0$ and $z=\infty$. As a consequence, the function in the right-hand-side of Eq.~(\ref{eq:Ei}), besides the one at infinity, has $m$ branch points in each $r=z_k$,
with $k=1,2,\ldots,m$. A similar complex structure is expected for the chiral angle $F(r)$, even though no explicit solution can be obtained due to the implicit nature of the left-hand-side expression. Figure~\ref{art-fig:orizontal-cuts} shows the analyticity domain of $F(r)$ in the case where the branch cut of Ei$(z)$
is placed over the positive real axis. The cuts are obtained by adding to the negative real axis (negative because $r$ appears in the argument of Ei$(z)$ with a minus sign) the points of the set $\{z_k\}_{k=1}^m$, that in the figure are organized in pairs of complex conjugates $\{c_j,c_j^*\}_{j=1}^h$ and real values $\{r_j\}_{j=1}^l$, with $m=l+2h$, as in~\ref{app:c}.
\\
It is interesting to notice that, not only the behavior at the physical threshold, but the entire structure of TLFFs is intimately connected with the analytic extension of the chiral phase $F(r)$ outside the positive real axis, which represents its natural domain. Moreover such an extension drastically changes the character of this function, because, by acquiring a non-vanishing imaginary part, it looses its ``phase'' nature.
%
%	New - April 2021
%
\subsection{The logarithmic nature of branch cuts in the $q^2$ complex plane}
\label{subsec:cuts}
Despite the power of the procedure to reproduce spontaneously the expected analyticity domain and in particular, the presence of branch cuts along the positive real axis of the $q^2$ complex plane, the character of these discontinuities does not fulfill the theoretical requirements.

Indeed, as already pointed out, only logarithmic branch cuts can be generated, while the opening of the $n$-pion intermediate state would manifest themselves as square-root discontinuities, originating at the corresponding production thresholds, $q^2=\lt n M_\pi\rt^2$, while logarithmic branch cuts are expected only in the unphysical Riemann sheets.

On the other hands, however, the logarithmic branch cuts have infinite order, i.e., they generate an infinite tower of unphysical Riemann sheets extending upward and downward. It follows that any of them does have an effect on the first and physical Riemann sheet.

Moreover, being the logarithmic ones the only kind of branch cuts that can be reproduced by our model, they can be interpreted as effective cuts, which account for all the discontinuities due to the opening of all the intermediate channels as described by the optical theorem.
%
%  Gpeff and Gneff
%
\begin{figure}[h!]
\begin{center}
\includegraphics[width=75mm]{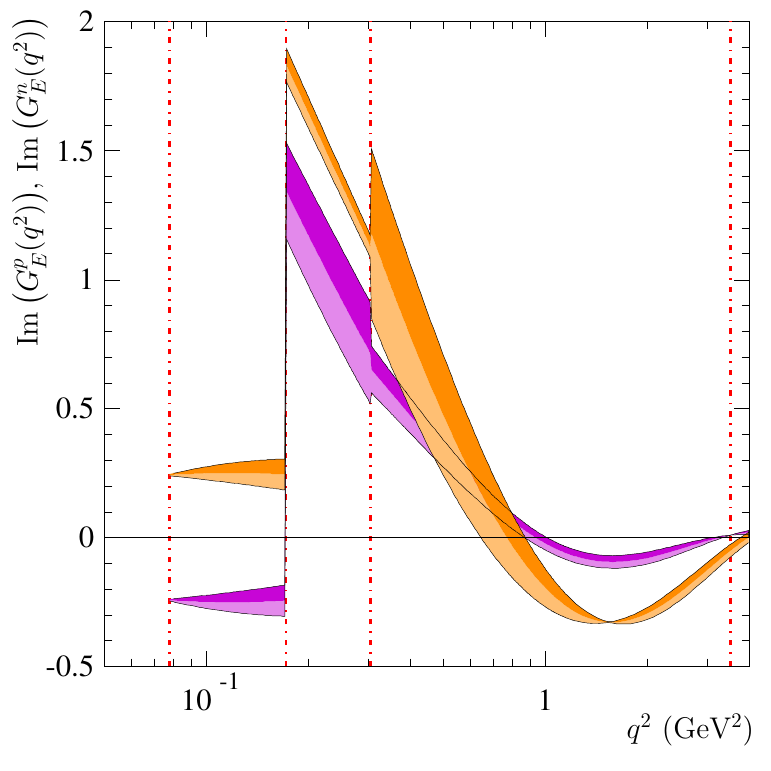}
\\
\includegraphics[width=75mm]{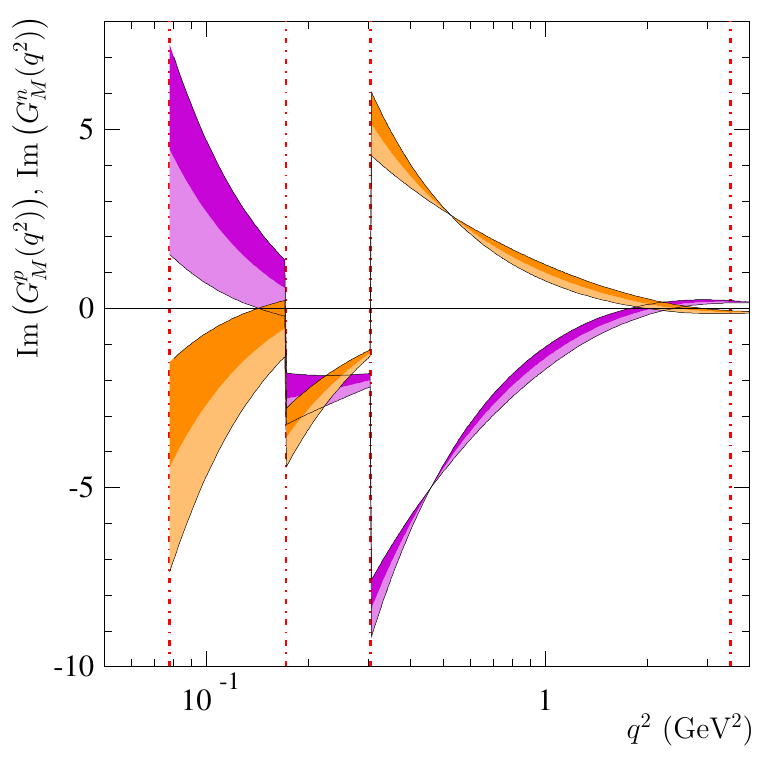}
\vspace{0mm}
\caption{The imaginary parts of the electric, upper panel, and magnetic, lower panel, proton (violet band) and neutron (orange band) FFs. The red dash-dotted lines indicates, from the left to the right: the theoretical thresholds $q^2_{\rm th}=(2M_\pi)^2$,  $q'^2_{\rm th}=(3M_\pi)^2$, $q''^2_{\rm th}=(4M_\pi)^2$ and the physical one $\qphy=(2M_N)^2$.
}
\label{art-fig:cuts}
\end{center}
\end{figure}\\
Figure~\ref{art-fig:cuts} shows the imaginary parts of the electric and magnetic FFs of the proton and the neutron, over the whole unphysical region, in particular at $0\le q^2\le 4$ GeV$^2$. There are three theoretical thresholds corresponding to the opening of two, three and four pion intermediate states. In particular those at $q^2_{\rm th}=(2M_\pi)^2$ and $q''^2_{\rm th}=(4M_\pi)^2$ are related to the isovector amplitude, see Eq.~\eqref{eq:ta-tb}, while $q'^2_{\rm th}=(3M_\pi)^2$ to the isoscalar one. These amplitudes account for the, respectively for the isospin-one and isospin-zero contributions. 

Such contributions, especially the vector meson resonances lying in the unphysical region, that in other models are included in the FFs by hand, described by Breit-Wigner formulae, see, e.g., Ref.~\cite{Pacetti,Belushkin:2006qa,Adamuscin:2016rer,Iachello:2004aq} and references therein, can not be reproduced by our model. Indeed, as a basic version of the Skyrme model, does not entail vector meson fields. Nevertheless, their mean effect is actually accounted for by a kind of duality phenomenon~\cite{Greco}, as proven by the fair agreement with data of the computed FFs in both SL and TL regions. In particular, the magnitude of these contributions is related to the discontinuity of the imaginary parts at the theoretical thresholds, see Fig.~\ref{art-fig:cuts}, which, by their turn, depend on the chiral phase $F(r)$, namely on its poles. This can be clearly seen by looking the expression of the imaginary part of $G_E^S$ given in Eq.~\eqref{eq:im-ges}. 	 
\section{Conclusions}
\label{sec:conclu}
A procedure to compute nucleon TLFFs, starting from integral representations of their SL counterparts, has been defined. Such a procedure consists in modeling the numerical solutions obtained for the nucleon electromagnetic currents in the framework of a generic model of nucleons, explicit calculations have been done in the case of the Skyrme model, with  functions, whose Fourier transforms, not only, are well defined, but they also embody the theoretical features required for the FFs by first principles, i.e., analyticity and unitarity.
\\
The general form, for the functions of the radius $r$, that describe the numerical solutions, is conceived to have automatically 
the expected behaviors in the origin, $r=0$, and in the limit 
$r\to\infty$. 
\\
The results for the nucleon FFs are analytic functions of $Q^2$ or equivalently $q^2$, which are real in the whole SL region and in the small portion of the TL region below the theoretical threshold $q_{\rm th}^2=(2M_\pi)^2$, while are complex elsewhere. Moreover, they also have the branch cut discontinuity $\left((2M_\pi)^2,\infty\right)$, in the $q^2$ complex plane, as expected by assuming analyticity and unitarity.
\\
Once the analytic expressions for all nucleon FFs in the whole $q^2$ complex plane are known, any quantity can be predicted without any further assumption or restriction.
Indeed, the free parameters of the fitting functions can be fixed at any desired degree of precision, since the numerical solutions can be known with an arbitrarily high accuracy.
%
%   F(r) analytic structure
%
\vspace{0mm}\\
It is important to stress that TLFFs are purely relativistic quantities, they can be defined only in the framework of the Quantum Electro-Dynamics (QED) and describe the vertex $\gamma N\overline{N}$, where a virtual photon produces a nucleon-antinucleon pair. In light of that, the only definition of an analytic continuation in TL $q^2$ represents by itself the first relativistic extension of the starting FF expressions given in Eq.~(\ref{eq:FTs}). In spite of that, since such original FF expressions have been obtained for static nucleons, i.e., nucleons in their rest frame, a procedure has to be defined to extend FFs even at relativistic momenta. In the SL region we have adopted the approach described in Ref.~\cite{Ji} and the resulting, relativistically corrected FFs are shown in Eq.~(\ref{slcor}). For TLFFs a modified methodology~\cite{Bardini} has to be used to account for the behavior in the unphysical region, where FFs remain unchanged, and at the production threshold $q^2=\qphy$, see Eq.~(\ref{tlcor}).
Even though such a procedure does not reproduce the asymptotic behaviors expected from perturbative QCD, as it is also discussed in Ref.~\cite{Ji}, the relativistic corrections, in the case of the Skyrme model, improve the agreement with the dipole FF.
\\
The predictions, obtained in such a particular case, i.e., by considering the Skyrme model, have been compared with all the available data in SL and TL region. The fair agreement that is obtained, in most of the cases, in not negligible $q^2$ intervals, appears as a quite encouraging achievement, since these results are based on a microscopic model which contains only pion fields and has no free parameters.
\\
Particular attention has been paid to the TL physical-threshold behavior, to verify if the non-trivial relationship, that exists between the predictions for the electric and magnetic FFs, reproduces the identity $G_E^N(4M_N^2)=G_M^N(4M_N^2)$, expected in case of isotropy and non-singular Dirac and Pauli FFs. We observed that the complex equality is not fulfilled, i.e., the two independent equations for the real and imaginary parts are not verified. By considering the moduli, their differences are partially compensated, nevertheless the isotropy violation at the production threshold
remains an important effect.
\\
In the TL region, especially nearby the threshold, the obtained values of the effective proton and neutron FFs are too small with respect to the data, while, at high $q^2$, especially the relativistic-uncorrected ones, appear in better agreement with data. However, as already stated in Sec.~\ref{sec:skyrme}, the Skyrme model represents an effective approximation of QCD at low energy so that, pushing its predictability at high-$q^2$ goes beyond the aim of model. Moreover, even the failure of relativistic corrections is expected, because it is well known that such corrections do not reproduce the perturbative QCD power law of FFs, which describes quite well the data.
\\
Since the present model contains only pions, the most natural improvement would be the inclusion of vector mesons $\rho$ and $\omega$ as gauge bosons of a hidden symmetry~\cite{Meissner}. This will entail additional degrees of freedom, i.e., further profile functions in terms of which parametrize the nucleon electromagnetic currents and hence the FFs.
\\
Finally, this procedure appears quite suitable to be applied to any other effective model of low-energy QCD, which allows to compute nucleon electromagnetic currents and then SLFFs as their Fourier transforms.
%
%
%
%    Appendices
%
\setcounter{section}{0}
  \setcounter{subsection}{0}
  \setcounter{subsubsection}{0}
  \setcounter{table}{0}
  \setcounter{equation}{0}
  \renewcommand\thetable{\Alph{subsection}\arabic{table}}
  \renewcommand\theequation{\Alph{subsection}.\arabic{equation}}
%
%
%
%\section{}
%
\section*{Appendices}
\subsection{Multipoint Pad\'e approximants}
\label{app:a}
The \P\ rational approximation technique~\cite{pade0} provides a powerful mathematical tool to define analytic expressions for the densities $b(r)$ and $t(r)$, that are known only numerically in terms of the chiral angle $F(r)$ according to Eq.~(\ref{eq:b-t}).
\\
The Pad\'e approximation is usually exploited to describe a function, analytic in a neighborhood of the origin, by means of a ratio of polynomials of arbitrary degrees. The polynomials are completely determined by requiring that the Taylor series in the origin of the difference between the function and the ratio of polynomials has to start from the highest power possible, once the degrees of the polynomials have been fixed. More in detail, given a function $f(z)$, analytic in the origin, and having the Taylor series
\be
f(z)=\sum_{k=0}^\infty f_k\, z^k\,,\hh |z|< \rho\,,
\nen 
where $\rho$ is a finite convergence radius, we define the rational function
\be
\frac{U_p(z)}{D_q(z)}=\frac{\sum_{i=0}^p u_i z^i}{\sum_{j=0}^q d_j z^j}\,,
\nen
with $(p,q)\in\N^2$ and $u_p d_q\not=0$, so that
\be
\frac{U_p(z)}{D_q(z)}=\sum_{k=0}^{p+q} f_k\,z^k+O\left(z^{p+q+1}\right)\,.
\label{eq:pa0}
\en
The rational function $U_p(z)/D_q(z)$ defined through the relation of Eq.~(\ref{eq:pa0}) is called Pad\'e approximant (PA) to $f(z)$ of
orders $(p,q)$. It has $(p+q+1)$ free parameters because, without loss of generality, it has been set $D_q(0)=d_0=1$, i.e., the value of PA in the origin is $u_0$. 
\\
The only numerical knowledge of profile functions $b(r)$ and $t(r)$, and hence the impossibility to obtain their derivatives, prevents the possibility of using the standard PA procedure, so that, to obtain analytic parameterizations, it has been exploited a more general technique, called multipoint Pad\'e approximation~\cite{pade0}. It allows to determine the ratio of polynomials that interpolates values and derivatives of a given function at a finite number of points $\{z_k\}_{k=1}^N$, with $N\in\N$. It essentially consists in finding a PA fulfilling the condition of Eq.~(\ref{eq:pa0})  simultaneously at all the points of the set $\{z_k\}_{k=1}^N$. 
\\
Given a function $f(z)$, analytic in an open and connected set $D$, so that $\{z_k\}_{k=1}^N\subset D$, a multipoint Pad\'e approximant of $f(z)$, with respect to the set of points of $\{z_k\}_{k=1}^N$, is the ratio of polynomials, $U_n(z)/D_m(z)$ of degrees $(n,m)\in\N^2$, whose $(n+m+1)$ coefficients are determined by requiring that, at each point $z_k$, both Taylor series of $U_n(z)/D_m(z)$ and $f(z)$ must coincide up to the order $\mu_k-1\ge 0$, with the condition
\be
\sum_{k=1}^N \mu_k=n+m+1\,.
\nen
 It follows that, $\forall\,k\in\{1,2,\ldots,N\}$,
\be
\!\!
\!\!
\!\!
\!\!\frac{U_n(z)}{D_m(z)}=\sum_{j=0}^{\mu_k-1} f_{k,j}\,(z-z_k)^j+O\lq(z-z_k)^{\mu_k}\rq\,,
\label{eq:pa1}
\en    
where $f_{k,j}$ represents $j$-th coefficient of the $f(z)$ Taylor series, centered at $z_k$, i.e.,
\be
f(z)=\sum_{j=0}^{\infty} f_{k,j}\,(z-z_k)^j\,,\hh
\left\{\begin{array}{l}
	\forall\, z: |z-z_k|<\rho_k\vspace{2mm}\\
	\forall\, k\in\{1,2,\ldots,N\}\\
\end{array}\right.\,,
\nen
where $z_k\in D$ and, being $D$ an open set, the convergence radius $\rho_k$ is not vanishing, i.e., $\rho_k>0$.
\\
The simple case $n=1$, which consists in interpolating the function $f(z)$ and its derivatives at a single point, reproduces the standard PA procedure.
\\
We will consider the special case with $\mu_k=1$, $\forall\,k\in\{1,2,\ldots,N\}$, the so-called Cauchy-Jacobi problem~\cite{pade0}, in which no derivatives are needed. The solution to the Cauchy-Jacobi problem, in terms of determinants, reads~\cite{pade0}
\be\!\!\!\!\!\!\!\!\!\!\!
\frac{U_n(z)}{D_m(z)}=\frac{
\det\lt\begin{array}{cccc}
	g_{2m} & g_{2m-1} & \ldots & g_{m} \\
	g_{2m-1} & g_{2m-2} & \ldots & g_{m-1} \\
	\vdots & \vdots &            & \vdots \\
	g_{m+1} & g_{m} & \ldots & g_{1} \\
	t_{m}(z) & t_{m-1}(z) & \ldots & t_{0}(z) \\
\end{array}\rt
}{\det\lt\begin{array}{cccc}
	g_{2m} & g_{2m-1} & \ldots & g_{m} \\
	g_{2m-1} & g_{2m-2} & \ldots & g_{m-1} \\
	\vdots & \vdots &            & \vdots \\
	g_{m+1} & g_{m} & \ldots & g_{1} \\
	z^{m} & z^{m-1} & \ldots & 1 \\
\end{array}\rt}\,,
\label{eq:mppa}
\en
where the constant $g_k$ and the polynomial $t_k(z)$, of degree $N-1$, depend on the coefficients $f_{j0}=f(z_j)$, i.e., the values of the function $f(z)$ at each $z_j$, with $j\in\{1,2,\ldots,N\}$, and their expressions are 
\be
\left\{\begin{array}{rcl}
g_{k}\!\ug\!\ds\sum_{i=1}^N z_i^{k-1} f_{i0}
\mathop{\prod_{j=1}^N}_{j\not=i}\frac{1}{z_i\!-\!z_j}\,,\no\\
t_{k}(z)\!\ug\!\ds\sum_{i=1}^N z_i^{k} f_{i0}
\mathop{\prod_{j=1}^N}_{j\not=i}\frac{z\!-\!z_j}{z_i\!-\!z_j}\,,\\
\end{array}
\right.
\hspace{4mm} k\in\{1,2,\ldots ,2m\}\,.
\nen
%
%
%
%
%%\newpage
\subsubsection{Convergence of multipoint Pad\'e approximants}
%\label{app:a}
%
Following Ref.~\cite{pade0}, we define the sequence $\left\{z_k^{(N)}\right\}_{k=1}^{N+1}$, so that the interpolation points are $z_k\to z_{k}^{(N-1)}$, with $k\in\{1,2,\ldots,N\}$. We define the point set $E$ (in our case $E$ is the positive real axis, $E=(0,\infty)$), so that its complement $K_E$ in \C, is connected, includes the point at infinity and does not contain the points of the sequence $\left\{z_k^{(N)}\right\}_{k=1}^{N+1}$, as well as its limit value. Consider the sequence of functions
\be
\left\{G_N(z)=\frac{1}{n}\sum_{k=1}^{N+1}\ln\left|z-z_k^{(N)}\right|
\right\}_{N=1}^\infty\,,
\nen
and let $G(z)$ be the function to which the sequence converges, i.e.,
\be
\lim_{N\to\infty}G_N(z)\mathop{=}^{}G(z)\,.
\nen
The Saff's theorem~\cite{pade0} states that: if the convergence is uniform in each bounded and closed subset of $K_E$, the function $G(z)$ defines the regions of convergence of the multipoint PA as it follows: for a given $\sigma>0$, we denote with $E_\sigma$ the interior of the curve $\Gamma_\sigma=\{z:G(z)=\ln(\sigma)\}$, so that it represents the boundary $\Gamma_\sigma=\partial E_\sigma$. For any function $f(z)$, meromorphic in $E_\sigma$, having a total pole multiplicity $m\in\N$, it holds
\be
\lim_{N\to\infty}\n{f-\frac{U_n}{D_m}}=0\,,\hspace{3mm}
\forall\, z\in E_\epsilon, \mbox{with: }\e<\sigma\,,
\nen   
where the ratio of polynomials $U_M(z)/D_N(z)$ is the multipoint PA defined in Eq.~(\ref{eq:mppa}), with matching points $\left\{z_k^{(n+m)}\right\}_{k=1}^{n+m+1}$.
\\
In our case the function has a well known asymptotic behavior, i.e.,
\be
f(z)=O(z^{-h})\,,\hh z\to\infty\,,%
\nen 
with $h\in\N$, this implies that the degrees of polynomials are connected by the relation
\be
m=n+h\,.
\nen
So that, by augmenting $M$, the degrees of both polynomials and hence the number of poles increase linearly. The convergence condition becomes
\be
\lim_{n\to\infty}\n{f-\frac{U_n}{D_{n+h}}}=0\,,
\label{eq:limitPA}
\en 
where the limit is computed on $n$ after the substitution $N = 2n + h+1$ with constant $h$. 
\subsubsection{Stability of multipoint Pad\'e approximants}
%\label{app:a.b}
%
%
\P\ approximants have been used to approximate the three functions
\be
T_0(r)&=&e^{3M_\pi r}b(r)\,,\no\\
T_1(r)&=&e^{2M_\pi r}t_1(r)\,,\label{eq:3T}\\
T_2(r)&=&e^{4M_\pi r}t_2(r)\,,
\nen
where 
\be 
\begin{array}{rcl}
t_1(r)&=&\ds\frac{F_{\pi}^2}{4}\frac{\sin^2\left[F(r)\right]}{r^2}\\
&&\\
t_2(r)&=&\ds\frac{1}{g^2}\frac{\sin^2\left[F(r)\right]}{r^2}\left(\left[F'(r)\right]^2+\frac{\sin^2\left[F(r)\right]}{r^2}\right)\,,\\
\end{array}
\nen
are the two components of the density $t(r)$, i.e., $t(r)=t_1(r)+t_2(r)$. The profiles $b(r)$ and $t(r)$, whose definitions in terms of the chiral angle $F(r)$ are given in Eq.~\eqref{eq:b-t}, are known only numerically. Since, as already mentioned in Sec.~\ref{app:a}, these functions have well known power-law asymptotic behaviors, in order to determine the corresponding PAs it needs, first of all, a criterion to establish the sequence of interpolation points and then the order of the polynomial at the numerator, $n$ in Eq.~\eqref{eq:limitPA}. The order of the polynomial at the denominator is consequently determined by the known power-law asymptotic behavior. For the three functions of Eq.~\eqref{eq:3T} we have
\be
h_0=5\,,\hh	
h_1=4\,,\hh
h_2=6\,.
\nen
The goodness of the approximation is measured by the norm of the difference the normalized function and the PA, i.e.,
\be
\Delta_{n_j}\={\equiv}\n{\frac{T_j}{\n{T_j}} -\frac{U_{n_j}/D_{n_j+h_j}}{\n{U_{n_j}/D_{n_j+h_j}}}}
\no\\
\={=}\lt\int_0^{r_{\rm max}}\left| \frac{T_j(r)}{\n{T_j}} -\frac{U_{n_j}(r)/D_{n_j+h_j}(r)}{\n{U_{n_j}/D_{n_j+h_j}}}\right|^2dr\rt^{1/2},
\nen
with $j=0,1,2$ and where $r_{\rm max}=8$ fm is the upper limit that has been used in the numerical procedure to solve the differential equation that gives the chiral angle $F(r)$, and hence it represents the maximum value of $r$ up to which the numerical solution can be considered reliable. The definition of the norm, $\n{\cdot}$, that corresponds to that of the vector space of square Lebesgue integrable functions in the interval $(0,r_{\rm max})$, i.e., $L^2(0,r_{\rm max})$, is given in the expression of $\Delta_{n_j}$ of the above equation. 
\\
By studying the evolution of $\Delta_{n_j}$ as a function of $n_j\in\N$, $j=0,1,2$, and the texture of zeros and poles of the PA of the functions given in Eq.~\eqref{eq:3T}, the best values for the three parameter $n_0$, $n_1$ and $n_2$ have been determined as
\be
\bar n_0=6\,,\hh	 
\bar n_1=7\,,\hh
\bar n_2=5\,.
\nen
The corresponding norms $\Delta_{n_j}$, normalized to the interval width in order to have adimensional quantities, are
\be
\frac{\Delta_{\bar n_0}}{r_{\rm max}} \ug 2.51\cdot 10^{-13}\,,\no\\
\frac{\Delta_{\bar n_1}}{r_{\rm max}} \ug 1.25\cdot 10^{-9}\,,\no\\
\frac{\Delta_{\bar n_2}}{r_{\rm max}} \ug 3.24\cdot 10^{-10}\,.
\nen 
Their smallness and stability for polynomial degrees $n_j\ge \bar n_j$, $j=0,1,2$, demonstrate the goodness of the PA.\\
The key features of the procedure, which follows that outlined in Ref.~\cite{Masjuan:2007ay}, are listed below, where the index $j=0,1,2$ is used to label the three cases corresponding to the three functions of Eq.~\eqref{eq:3T}.
%
%More in detail~\cite{Masjuan:2007ay}
%
\begin{itemize}
\item The coefficients of the PA $U_{n_j}(r)/D_{n_j+h_j}(r)$ are adimensional quantities, i.e., it is understood that the coefficient of the power $r^n$, $n\in\N$, is divided by $r_0^n$, with $r_0=1$ fm.
\item The PA $U_{n_j}(r)/D_{n_j+h_j}(r)$, $j=0,1,2$, has $n_j$ zeros and $n_j+h_j$ poles that can be either single real numbers or pairs of complex conjugate.
\item %Since the number of free parameters of the $j$-th PA is $N_j=2n_j+h_j$, %because of the constraint of Eq.~\eqref{eq:asy-b}. 
The set of points $\left\{r^{(j)}_k\right\}_{k=1}^{N_j}\subset (0,r_{\rm max})$ needed to solve the Cauchy-Jacobi problem and find the $N_j=2n_j+h_j+1$ coefficients of the $j$-th PA, as in Eq.~\eqref{eq:mppa} with $n=n_j$ and $m=n_j+h_j$, has been optimized by minimizing the norm $\Delta_{n_j}$. %
In order to introduce a systematical error we used two different sets of interpolation points, as mentioned in Sec.~\ref{subsec:results}.
%The optimal choice corresponds to a uniform distribution from the common minimum $r^{(j)}_1=0.0001$~fm up to the common maximum $r^{(j)}_{N_j}= 1.5$~fm. 
%
\item The $j$-th PA can effectively mimic the $T_j(r)$ function of Eq.~\eqref{eq:3T} by generating analytic defects, i.e., the poles. Such poles can be classified as transient and physical. The first ones form a set of unstable artificial poles, their positions undergo large variations when the polynomial degree increases. The physical ones, instead, are those effective poles that multipoint PAs develop in the $r$ complex plane, away from the positive real axis $(0,\infty)$, to reproduce the behavior of the true function in the physical domain, which is indeed the positive $r$ real axis. The divergence of PAs in the neighborhood of poles does not represent a lack of the procedure because the true function is not defined there. The large values of PAs close to the physical poles, which lie far away from the positive real axis, result in only small corrections in the physical domain.
\item The phenomenon of the insurgence of transient poles from a certain value of the polynomial degree $\tilde n_j$ for the $j$-th PA can be used to select the best value of the degree itself. Indeed, such a value determines to the number of physical poles $\bar n_j+h_j$, i.e., the minimum number of poles that the $j$-th PA needs in order to reproduce the $T_j(r)$ function in the physical region.
\item The transient nature of a pole is proven by the fact that it  always appears together with a corresponding zero, so that they cancel out by leaving the PA unchanged.  
\end{itemize}
\subsection{The integral representation of $E_1(z)$}
\label{app:b}
The function $H(\alpha\beta)$ is defined through the integral representation
\be
H(\alpha\beta)\equiv\int_0^\infty\frac{e^{-\alpha r}}{r+\beta}dr\,,
\nen
that, with $\alpha,\beta\in\C$, converges if $\re(\alpha)>0$ and $\beta\not\in(0,\infty)$. Such an integral representation and hence the function $H(\alpha\beta)$ depend on $\alpha$ and $\beta$ only through their product, indeed, by putting $w=\alpha(r+\beta)$ and $z=\alpha\beta$, we have
\be
H(z)=e^{z}\int_{z}^\infty\frac{e^{-w}}{w}dw
\equiv e^{z}E_1(z)\,,
\nen
where $E_1(z)$ is the exponential integral (ExpIntegral) function~\cite{stegun}. A series representation of $E_1(z)$ can be obtained 
by integrating the well known expansion in the origin of its first 
derivative, i.e.,
\be
\frac{d E_1}{d z}=-\frac{e^{-z}}{z}
=-\frac{1}{z}\sum_{k=0}^\infty \frac{(-z)^k}{k!}
=-\frac{1}{z}-\sum_{k=1}^\infty\frac{(-z)^{k-1}}{k!}
\,.
\nen
Indeed, since the series converges uniformly, it can be integrated term by term as
\be
E_1(z)=C-\ln(z)-\sum_{k=1}^\infty \frac{(-z)^k}{kk!}
\,,
\nen
where $C$ is the integration constant whose value is obtained 
by considering the limit $z\to 0$, as
\be
C\ug \lim_{z\to 0}\left[E_1(z)+\ln(z)+\sum_{k=1}^\infty \frac{(-z)^k}{kk!}\right]\no\\
\ug
\lim_{z\to 0}\left[\int_{z}^\infty\frac{e^{- w}}{w}dw+\ln(z)\right]=-\g\,,
\nen
\g\ is the Euler-Mascheroni constant~\cite{stegun}.
In light of these results, the function $H(z)$ has the representation 
\be
H(z)
\ug
e^{z} \left[-\g-\ln(z)-\sum_{k=1}^\infty\frac{(-z)^k}{kk!}\right]
\no\\
\ug
e^{z} \left[-\g-\ln(z)+\sum_{k=1}^\infty\frac{(-1)^{k+1}(z)^k}{kk!}\right]\,.
\nen
It possesses the same properties of $E_1(z)$, i.e., it is analytic in the $z$ complex plane with the cut $(-\infty,0)$, for
 $|\arg(z)|<\pi$. It is real for $z\in(0,\infty)$, which is the interception of its analyticity domain and the real axis, so that it fulfills the Schwarz reflection principle
\be
H(z^*)=H^*(z)\,,\hhh\forall\,z\not\in(-\infty,0)\,.
\nen
\subsection{The branch cut in the $q^2$ complex plane}
\label{app:c}
To study how the logarithmic cut of $E_1(z)$, in the variable $z$, evolves in the variable $Q$, see Eq.~(\ref{gg}), we consider the following cases
\be 
z_\pm =(\pm iQ -3M_\pi)z_j\,, \hspace{10mm} \mbox{with: } z_j=x_j+i\,y_j \nonumber\, ,
\en 
where $x_{j}$ and $y_j$ are the real and the imaginary part of the pole $z_j$. The variable $Q$ ``feels'' the cut when $z_\pm$ crosses the negative real axis, \ie, when $\im(z_\pm)$ changes sign and $\re(z_\pm)<0$. The value $Q_0$ at which the imaginary part of $z_\pm$ vanishes and the corresponding real part, are given by
\be 
\begin{array}{l}
Q_0=\ds\pm \frac{3M_{\pi}y_j}{x_j} \,,\\
 \re(z_\pm^0)=\ds z_\pm^0=-\frac{3M_{\pi}(x_j^2+y_j^2)}{x_j} =-\frac{3M_{\pi}|z_j|^2}{x_j} \,,\\
\end{array}
\label{eq:q0} 
\en
where $z_\pm^0\in\R$ stands for the value of $z_\pm$ corresponding to $Q=Q_0$.
\\
From Eq.~(\ref{eq:q0}) follows that:
$\re(z_\pm^0)<0$ when $x_j>0$ hence, having $Q>0$ by definition, there must be $y_j > 0$ or $y_j<0$ depending on $\pm iQ$. 
Moreover, if $y_j > 0$ ($y_j < 0$) the crossing is from above (below) the cut and so it requires the imaginary part to be increased by $-2\pi$ ($+2\pi$), having $E_1(z)$ a $-\log(z)$ term, see Eq.~(\ref{eq:G2}).
\\
Finally, all these considerations can be summarized in the compact expression 
\be
E_{SL}\left(z_\pm\right)= E_1\left(z_\pm\right)\pm 2\pi i \,\theta \!\left(\!Q\!\mp\! 3M_{\pi}\frac{y_j}{x_j}\!\right)
\theta (x_j)\, \theta (\pm y_j)\,,\hspace{-5mm}\no\\
\label{eq:esl}
\en
where the Heaviside $\theta$ functions select the above conditions and the symbol $E_{SL}$ stands for an ExpIntegral corrected in case of SL momenta, \ie, $Q\in(0,\infty)$.
\\
A similar study can be done also in the TL region.
However, as already discussed, in such a region FF values, at a given $q^2$ above the threshold ($(3M_\pi)^2$ and $(2M_\pi)^2$ for isoscalar and isovector FFs respectively), are obtained as the limits
\be
G(q^2)=\lim_{\epsilon\to 0^+}G(q^2+i\,\epsilon)\,,\hspace{10mm}q^2\ge (3-I)^2M_\pi^2\,,
\no\en
where the symbol $G$ stands for one the four nucleon Sachs FFs and $I=0,1$ is the isospin. It follows that a generic value of $q>(3-I)M_\pi$, is understood as $q+i\eta$, with $\eta\to0^+$.
To obtain TLFFs from the expression of Eq.~(\ref{gg}), we have to make the substitution $Q\to i(q+i\eta)$, and hence the arguments of the ExpIntegral functions become
\be
(\pm iQ-3M_\pi)z_j\!&\to &\! (\mp q-3M_\pi\mp i\eta)z_j
\no\\
\ug
(\mp q-3M_\pi) x_j\!\pm\! \eta y_j\no\\
&&-i\left[(3M_\pi\!\pm\! q)y_j\!\pm\!\eta x_j\right]\,.
\no\en
The imaginary part, being $q$ and $\eta$ positive, vanishes only in the case of ``lower sign'', at
\be
q=q_0\equiv 3M_\pi-\eta\frac{ x_j}{y_j}\,.
\label{eq:q00}
\en
Since in case of $G_E^S$, $q>3M_\pi$, the ratio $x_j/y_j$ must be negative, the pole $z_j$ lies either in the second or in the fourth quarter of the $z$ complex plane. The corresponding real part is
\be
\re[(q_0-3M_\pi+i\eta)z_j]\ug (q_0-3M_\pi)x_j-\eta y_j\no\\
\ug -\eta\,\frac{|z_j|^2}{y_j}\,.
\nen
A correction has to be considered only if such a real part is negative, \ie,
$y_j>0$ and, since $x_j$ and $y_j$ have opposite sign, the only possibility for a pole to generate a correction in the TL region is that it must lie in the second quarter. When $q\to q_0^+$, of Eq.~(\ref{eq:q00}), the imaginary part 
\be
\im[(q-3M_\pi+i\eta)z_j]\ug -(3M_\pi- q)y_j+\eta x_j\no\\
\!&\simeq&\! (q-3M_\pi)y_j
\no\en
vanishes as $\lim_{q\to 3M_\pi^+}(q-3M_\pi)y_j=0^+$ because $y_j>0$.
So, following the previous argument, the imaginary part of the ExpIntegral will be increased by $-2\pi$, hence we can define
\be
E_{TL}\left[(q-3M_{\pi})z_j\right]\ug E_{1}\left[(q-3M_{\pi})z_j\right]\label{fftl}\\&&-
2\pi i \theta (q-3M_{\pi})\theta (-x_j)\theta (y_j)\,.
\nen
These corrections are crucial because, as we will see in more detail,  
they generate the desired complex structure for the FFs.
\\
Having real polynomials with only simple zeros, the poles of $b_{\rm fit}(r)$, $\{z_j\}_{j=1}^m$, can come either as single real negative values $\{r_j\}_{j=1}^l$, or in pairs of complex conjugates $\{c_j,c^*_j\}_{j=1}^h$, and hence
$\{z_j\}_{j=1}^m=\{r_j\}_{j=1}^l\cup \{c_j,c^*_j\}_{j=1}^h$, with: $l+2h=m$.
Moreover, from the definitions given in Eqs.~(\ref{eq:residue0}) and~(\ref{eq:residue}), the residues have the same properties of the corresponding poles, \ie,
\be
\begin{array}{rclcl}
\tilde{R}_{r_j}&\in& \R\,, &\hh&j=1,2\ldots,l\,,\\
%&&&&\\
 \tilde{R}_{c_j} &=& \tilde{R}^*_{c^*_j} && j=1,2,\ldots,h\,.\\
 \end{array}
 \label{eq:tildeR}
\en
In light of this, using the function $H(\alpha\beta)$ of Eq.~(\ref{eq:G2}), in particular its property: $H(z^*)=H^*(z)$ (Schwarz reflection principle) and including the branch cut corrections of Eq.~(\ref{eq:esl}),
the expression of $G_E^S(Q)$ in the SL region, given in Eq.~(\ref{gg}), can be simplified as
\be
G_E^S(Q)\ug
\frac{2\pi}{Q}\sum_{j=1}^{l}\tilde{R}_{r_j}\im\left\{
 H\left[(iQ-3M_{\pi})r_j\right] \right\}\no
 \no\\&&
 +
\frac{2\pi}{Q}\sum_{j=1}^{h}\im\left\{\tilde{R}_{c_j}
 H\left[(iQ-3M_{\pi})c_j\right] \right.\no
 \\
&&\left.- \tilde{R}_{c_j}
 H\left[(iQ-3M_{\pi})^*c_j\right] \right\}\no
 \\ &&
 +
\frac{4\pi^2}{Q}\sum_{j=1}^{h}\theta(x_j)\no\\
&&\times
\re\left[
\theta\!\left(\!Q-3M_\pi\frac{y_j}{x_j}\!\right)\theta(y_j)\tilde R_{c_j}e^{(iQ-3M_\pi)c_j}\right. \no\\
&&+
\left.\theta\!\left(\!Q+3M_\pi\frac{y_j}{x_j}\!\right)\theta(-y_j)\tilde R_{c_j}
e^{(-iQ-3M_\pi)c_j}
\right]\,.
\nen
The TL expression of Eq.~(\ref{gg}), accounting for the corrections of Eq.~(\ref{fftl}), becomes
\be 
G_E^S(iq)\ug 
-\frac{2\pi}{q}\sum_{j=1}^{h} \re\left\{  \tilde{R}_{c_j} H\left[(-q-3M_{\pi})c_j\right]\right.\label{eq:GES-TL}\\
&&\left.- 
\tilde{R}_{c_j}H\left[(q-3M_{\pi})c_j\right]\right\}
\no\\ &&
 -\frac{\pi}{q}\sum_{j=1}^{l} \tilde{R}_{r_j} \Big\{H\left[(-q-3M_{\pi})r_j\right]
\Big.\no\\ 
&&\Big. -H\left[(q-3M_{\pi})r_j\right]\Big\}
\no\\ &&
+\frac{2i\pi^2}{q}\sum_{j=1}^h \theta(q-3M_\pi)\theta(-x_j)
\no\\
&&\!\!\!\!\times\!\!
\left[\tilde R_{c_j}
e^{(q-3M_\pi)c_j}\theta(y_j)
\!+\!
\tilde R_{c_j}^*e^{(q-3M_\pi)c_j^*}\theta(-y_j)
\right]\,.\hspace{-4mm}
\nen
While the SL $G_E^S(Q)$ is real, $G_E^S(iq)$, containing complete $H$ functions (not only their real or imaginary parts), could have a non-zero imaginary part.
\r{%
\subsection{Analyticity checks on the imaginary parts}
\label{app:d}
Analyticity represents one of the load-bearing axes of the procedure to such an extent that it has been conceived in such a way that, by definition, all FF parameterizations are analytic functions of the squared four-momentum transfer $q^2$. Such a property of the parameterizations has been implemented as an inescapable  feature by requiring the validity of the integral representation of Eq.~\eqref{eq:DR}, \ie, the so-called, dispersion relations for the imaginary part. In particular, in the case of the FF $G_E^S$, that we treated extensively, it is self-evident that its SL values given in Eq.~\eqref{eq:G2} can be obtained as the dispersion-relation integral of the TL imaginary part of Eq.~\eqref{eq:ges-easy-tl}.
\\
Nevertheless, consistency checks can be performed in order to have further confirmations that any algebraic manipulation of the parameterizations, that have to carry out to obtain the FFs, does not spoil analyticity. Particularly interesting are those consistency checks which involve the imaginary parts, because they play a pivotal role in the analytic continuation procedure based on dispersion relations.
\\
The first two relationships, which allow to verify analyticity in connection with the structure of the Skyrme model, can be derived by the representations of FFs as Fourier transforms of the baryon and moment-of-inertia radial densities $b(r)$ and $t(r)$ given in Eqs.~\eqref{eq:FTs}, are
\be
G_M^S(Q^2) \ug  -\frac{2M_N}{\Lambda} \frac {d  G_E^S(Q^2)}{dQ^2}\, ,\no\\
G_E^V(Q^2) \ug \frac{1}{M_N \Lambda}\left(\frac{3}{2} + Q^2  \frac{d}{dQ^2}\right) G_M^V(Q^2)\,.
\nen
These identities, that, even though have been obtained for SL momenta, i.e., at $q^2=-Q^2<0$, once the representation is computed, can be extended at all values of $q^2$, must hold also for the imaginary parts. In particular, in the time-like region, $q^2>\qth$, the imaginary parts of the isoscalar and isovector electric and magnetic form factors should verify the following relations
\be
\begin{array}{rcl}
	{\rm Im}\big(G_M^S(q^2)\big) \ug\ds
	 \frac{2M_N}{\Lambda} \frac {d  {\rm Im}\big(G_E^S(q^2)\big)}{dq^2}\, , \\
&&	\\
{\rm Im}\big(G_E^V(q^2)\big) \ug\ds 
\frac{1}{M_N \Lambda}\left(\frac{3}{2} + q^2  \frac{d}{dq^2}\right) {\rm Im}\big(G_M^V(Q^2)\big)\,.\\
\end{array}
\label{eq:first-check}
\en
Figure~\ref{fig:first-check} shows as black empty squares the left-hand-sides and as red disks the right-hand-sides of the first identity of Eq.~\eqref{eq:first-check} in the upper panel, of the second identity in the lower panel, respectively. Besides, the tiny discrepancies for lower-$q^2$ points, due to the limits of the numerical computation of the derivatives, the almost perfect squares-disks superposition does prove the identities of Eq.~\eqref{eq:first-check}, and hence the complete implementation of analyticity in the Skyrme model.
\\
The second check does represent an even more severe test of analyticity. Indeed, it consists in computing the electric-charge and magnetization spatial densities directly from the imaginary parts of the electric and magnetic FFs, i.e., from quantities that are defined in the TL region. This is really interesting because it is only by assuming analyticity that spatial densities can be computed starting from quantities defined in the TL region where their interpretation as Fourier transforms of such spatial densities is no more valid. 
%
%  First check
%
\begin{figure}[h]
\begin{center}
\includegraphics[width=75mm]{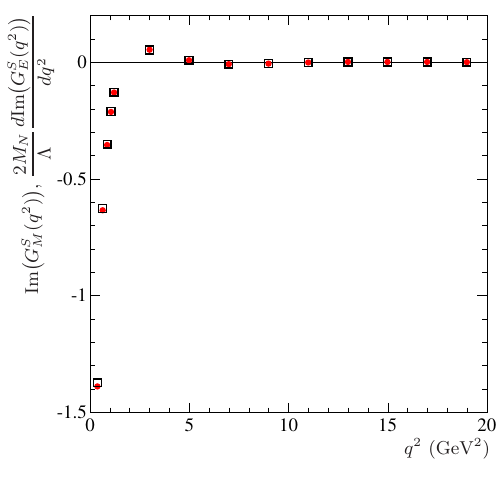}
\\
\includegraphics[width=75mm]{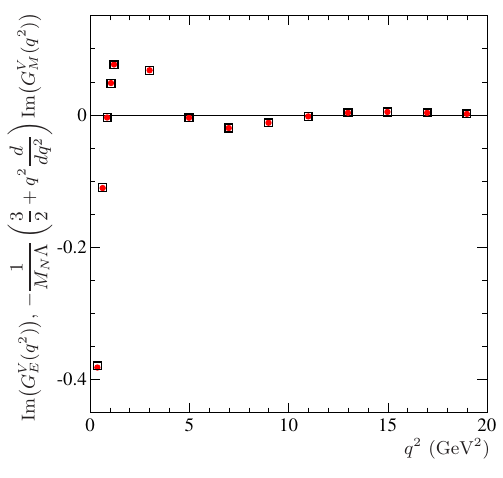}
%\vspace{0mm}
\caption{Upper panel: the empty squares represent the imaginary part of the magnetic isoscalar FF, the red disks the expression, proportional to the imaginary part of the electric isoscalar FF at the second member of the first identity of Eq.~\eqref{eq:first-check}. Lower panel: the empty squares represent the imaginary part of the electric isovector FF, the red disks the expression at the second member of the second identity of Eq.~\eqref{eq:first-check}.}
\label{fig:first-check}
\end{center}
\end{figure}\\
The expressions for the spatial electric charge and magnetization densities follows from their definitions
\be\begin{array}{rcl}
	\rho_E(r)\!\!&=&\!\!\displaystyle
	\frac{1}{2\pi^2}\int_0^\infty G_E(-Q^2)j_0(Qr)Q^2 dQ\,,\\
&&\\
\rho_M(r)\!\!&=&\!\!\displaystyle
	\frac{1}{2\pi^2}\int_0^\infty G_M(-Q^2) j_1(Qr)\frac{Q}{r}Q^2 dQ
\,,\end{array}
	\label{eq:rho}
\en
in terms of Fourier transforms of the electric and magnetic FFs, that, in view of a variable substitution, have been rigorously defined as functions of $q^2=-Q^2$. 
\begin{figure}[h]
	\begin{center}
	\includegraphics[width=75mm]{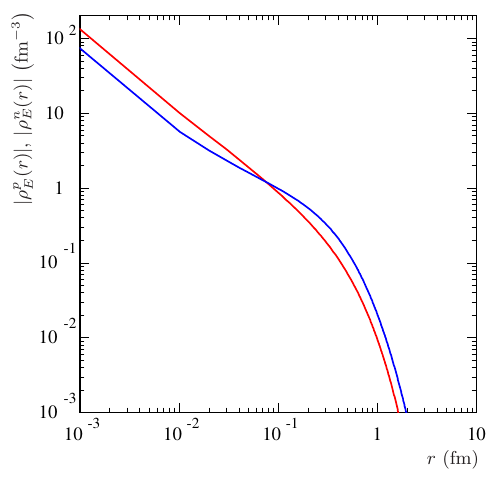}
\\
\includegraphics[width=75mm]{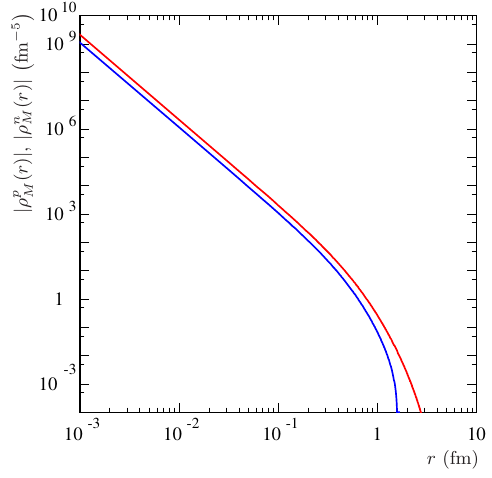}
\caption{\label{fig:answ4}Upper panel: moduli of the electric charge spatial densities of the proton, in red, and neutron in blue. Lower panel: moduli of the magnetization spatial densities of the proton, in red, and neutron in blue.}
	\end{center}
	\end{figure}\\
By considering the explicit form of the Bessel functions we have
\be
	\rho_E(r)\ug
	\frac{1}{2\pi^2}\int_0^\infty G_E(-Q^2)j_0(Qr)Q^2 dQ
	%\no\\\ug
	%\frac{1}{2 \pi^2 r}\int_0^\infty G_E(-Q^2)\frac{\sin(Qr)}{Qr}Q^2 dQ
%	\no\\\ug	\frac{1}{2 \pi^2 r}\int_0^\infty G_E(-Q^2)\sin(Qr)Q dQ	
% \no\\\ug
=\frac{1}{2\pi^2 r} S_E(r)\,,\no\\
\rho_M(r)\ug
	\frac{1}{2\pi^2}\int_0^\infty G_M(-Q^2) j_1(Qr)\frac{Q}{r}Q^2 dQ
%	\no\\\ug \frac{1}{2 \pi^2 r^3 }\int_0^\infty G_M(-Q^2)\left(\frac{\sin(Qr)}{(Qr)^2}-\frac{\cos(Qr)}{Qr}\right)\frac{Q^3}{r}dQ\nonumber\\
%	\no\\\ug\frac{1}{2 \pi^2 r^3 }\int_0^\infty\hspace{-4mm} G_M(-Q^2)\left(\sin(Qr)\!-\!rQ\cos(Qr)\right) QdQ
	\no\\\ug
\frac{1}{2\pi^2 r^3} \big(S_M(r)+r C_M(r)\big)\,,
	\nonumber
\en
where we have defined the two master integrals
\be
S_{E,M}(r)\ug \int_0^\infty G_{E,M}(-Q^2)\sin(Qr)Q dQ\,,\no\\ 
C_{E,M}(r)\ug
-\int_0^\infty G_{E,M}(-Q^2)\cos(Qr)Q^2 dQ\hspace{4mm}
\label{eq:S-C}
\\\ug
-\frac{dS_{E,M}}{dr}\,.
\nen
By exploiting analyticity realized through the dispersion relations, such quantities can be also expressed in terms of integrals of the TL imaginary parts.   
\\
In particular, we use the dispersion relations for the imaginary part of Eq.~\eqref{eq:DR}, that give %relate the SL values of FFs to integrals of their imaginary parts over the upper edge of the branch cut the $(4m_\pi^2,\infty)$ and are  
\be
G_{E,M}(t)=\frac{1}{\pi}\int_{\qth}^\infty\frac{{\rm Im}\left(G_{E,M}(s)\right)}{s-t}ds\,,
\nen
that, with $t=-Q^2<0$, becomes
\be
G_{E,M}(-Q^2)=\frac{1}{\pi}\int_{4m_\pi^2}^\infty\frac{{\rm Im}\left(G_{E,M}(s)\right)}{s+Q^2}ds\,.
\nen
%\cite{Holzwarth1}.
By using these expressions of the SL FFs in the first integral of Eq.~\eqref{eq:S-C}, performing the integration in $dQ$ and making the substitution $s=\mu^2$, we obtain
\be
S_{E,M}(r)
\ug
\frac{1}{\pi}\int_0^\infty dQ \int_{4m_\pi^2}^\infty ds \frac{{\rm Im}\left(G_{E,M}(s)\right)}{s+Q^2}\sin(Qr)Q 
%\no\\\ug\frac{1}{2\pi}\int_{-\infty}^\infty dQ \int_{4m_\pi^2}^\infty ds \frac{{\rm Im}\left(G_{E,M}(s)\right)}{s+Q^2}\sin(Qr)Q 
%
%\no\\\ug\frac{1}{4i\pi}\int_{4m_\pi^2}^\infty ds\,\int_{-\infty}^\infty dQ\,{\rm Im}\left(G_{E,M}(s)\right)\
%\no\\&&\times\frac{e^{iQr}-e^{-iQr}}{(Q-i\sqrt{s})(Q+i\sqrt{s})}Q
%
%\no\\\ug\frac{1}{4i\pi}\int_{4m_\pi^2}^\infty {\rm Im}\left(G_{E,M}(s)\right)2i\pi\no\\&&\times\left(\frac{e^{-\sqrt{s}r}}{2i\sqrt{s}}i\sqrt{s}+\frac{e^{-\sqrt{s}r}}{-2i\sqrt{s}}\left(-i\sqrt{s}\right)\right)ds
%
\no\\\ug
\frac{1}{2}\int_{4m_\pi^2}^\infty 
{\rm Im}\left(G_{E,M}(s)\right)e^{-\sqrt{s}r} ds\no\\
%\nen
%that, with the substitution $s=\mu^2$, becomes
%\be
%S_{E,M}(r)
\ug \int_{2m_\pi}^\infty
{\rm Im}\left(G_{E,M}\left(\mu^2\right)\right)e^{-\mu r}\mu \,d\mu\,.
\nen
Finally, we exploit the second expression of Eq.~\eqref{eq:S-C}, to obtain the master integrals $C_{E,M}(r)$ as the opposite of the first derivative of $S_{E,M}(r)$, i.e.,
\be
C_{E,M}(r)\ug -\frac{d S_{E,M}(r)}{dr}
=%\no\\\ug
\int_{2m_\pi}^{\infty}  \hspace{-4mm}{\rm Im}\left(G_{E,M}\left(\mu^2\right)\right)e^{-\mu r}\mu^2 d\mu\,.
\nen
In the light of these results, the spatial densities of Eq.~\eqref{eq:rho} become
\be
	\rho_E(r)
%\ug\ds	\frac{1}{2\pi^2 r} S_E(r)\\
\ug\ds
\frac{1}{2\pi^2 r}
	\int_{2m_\pi}^\infty
{\rm Im}\left(G_{E}\left(\mu^2\right)\right)e^{-\mu r}\mu \,d\mu\,,
\no\\
&&\label{eq:check}
\\
	\rho_M(r)
%	\ug\ds\frac{1}{2\pi^2 r^3} \left( S_M(r)+rC_M(r)\right)\\
 	\ug\ds\frac{1}{2\pi^2 r^3}
	\int_{2m_\pi}^\infty
{\rm Im}\left(G_{M}\left(\mu^2\right)\right)
\left(1+r\mu\right)  e^{-\mu r}\mu\,d\mu\,.
\nen
Since the FF parameterizations that have been used do respect analyticity and hence the dispersion relation for the imaginary part, indeed, it is self-evident that, for instance, the SL values of $G_E^S$ given in Eq.~\eqref{gg} can be obtained as the dispersion-relation integral of the TL imaginary part of Eq.~\eqref{eq:im-ges}, and, moreover, the FFs are obtained as Fourier transforms of spatial densities, the relations of Eq.~\eqref{eq:check} are automatically fulfilled.
\\
Nevertheless, for completeness, in Fig.~\ref{fig:answ4} we report the moduli of the spatial charge and magnetization densities, upper and lower panel, for proton and neutron, red and blue curves, obtained by numerical and approximate integration because it is truncated at $\mu_{\rm max}=4.5815\,{\rm GeV}\simeq 23.2563$ fm$^{-1}$.
\\
However, in this case, the master proof of the expressions of Eq.~\eqref{eq:check} is their analytic derivation, because the numerical check suffers from the previously highlighted approximations.}
\subsection{The asymptotic behavior} 
\label{app:e}
The integral representations (Fourier transforms) of FFs given in  Eqs.~(\ref{eq:FTs}) can be classified into two species depending on the order of the spherical Bessel function, \ie,
\be
g_0(Q)\={=}\int_0^\infty f(r) j_0(Q r)dr\,,\label{eq:f0}\\
g_1(Q)\={=}\int_0^\infty f(r) j_1(Q r)dr\,,\label{eq:f1}
\en
with% $j_{0,1}(x)$ are the spherical Bessel functions 
\be
j_0(x)=\frac{\sin(x)}{x}\,,\hspace{2mm}
j_1(Q)=\frac{\sin(x)}{x^2}-\frac{\cos(x)}{x}=-\frac{dj_0(x)}{dx}\,,
\nen
and $f(r)$ represents the profile function, which is regular in the origin and vanishes exponentially as $r\to\infty$, in particular
\be
\begin{array}{rcl c rcl}
f(r)\ds\mathop{\propto}_{r\to 0} r^l\,,&\hh &
f(r)\ds\mathop{\propto}_{r\to \infty} \ds\frac{e^{-\mu r}}{r^h}\,,\\\end{array}
\label{eq:constraint}
\en
with $l,h \in\N$ and $\mu >0$. The behavior in $r=0$ is crucial because it determines the asymptotic trend, as $Q\to\infty$, of the functions $g_{1,0}(Q)$. The profile $f(r)$, which is known only numerically, is parametrized as
\be
f(r)=\frac{A_m(r)}{B_{n}(r)}e^{-\mu r}\,,
\nen
where $A_m(r)$ and $B_{n}(r)$ are the real polynomials
\be
A_m(r)=\sum_{k=l}^m a_k r^k\,,\hhh
B_{n}(r)=\sum_{k=0}^{n} b_k r^k\,,
\nen
with $n-m=h$, and $a_l\not=0$, $a_m\not=0$, $b_0=1$, $b_{n}\not=0$, in order to follow the behaviors given in Eq.~(\ref{eq:constraint}).
\\
Assuming the zeros $\{z_k\}_{k=1}^n$ of $B_{n}(r)$ to be
all simple, with $\re(z_k)<0$ and also $A_m(z_k)\not=0$, $\forall\,k\in\{1,2,\ldots, n\}$, the polynomial part of $f(r)$ can be written in terms of the Mittag-Leffler representation
\be
f(r)=e^{-\mu r}\sum_{k=1}^n\frac{R_k}{r-z_k}\,,\hspace{3mm} R_k={\rm Res}\left[
\frac{A_m(r)}{B_{n}(r)},r=z_k\right]\,,
\nen
where $R_k$ is the residue of the simple pole $z_k$. The two polynomials have $n+m-l+1$ real degrees of freedom, while the Mittag-Leffler representation has $2n>n+m-l+1$ free parameters, namely the complex zeros $z_k$ and residues $R_k$, with $k=1,2,\ldots, n$. Indeed, having $B_n(r)$ real coefficients, both, zeros and corresponding residues of the ratio, can be either real or pairs of complex conjugate, hence the real degrees of freedom are only $2n$. Summing up the Mittag-Leffler series we get 
\be
\sum_{k=1}^n\frac{R_k}{r-z_k}=\frac{\sum_{j=0}^{n-1}\alpha_j r^j}{B_n(r)}\,,
\nen
where the coefficients $\alpha_j$ contain zeros and residues. To have, at numerator, a polynomial of $m<n$ degree with a zero of order $l$ in the origin we should impose
\be
&&
\alpha_j=0\no\\
&&\forall\, j\in\{\underbrace{0,1,\ldots,l-1}_{l\,{\rm values}},\underbrace{m+1,m+2,\ldots,n-1}_{n-m-1\,{\rm values}}\}\,,
\nen
these are $l+n-m-1$ constraints. It follows that the number of degrees of freedom reduces to $2n-(l+n-m-1)=n+m-l+1$, which coincides with that of the ratio of polynomials. In particular, 
the zero of order $l$ in $r=0$, first condition of Eq.~(\ref{eq:constraint}), implies \\
$f(0)=\ds\sum_{k=1}^n\frac{R_k}{-z_k}=0\,\Longrightarrow\, C_1\equiv\ds \sum_{k=1}^n\frac{R_k}{z_k}=0\,,$
\\
$f'(0)=\!\!\ds\sum_{k=1}^n R_k\!\left(\!-\frac{1}{z_k^2}\!-\!\frac{\mu}{-z_k}\!\right)\!=\!0\,\Longrightarrow\,C_2\equiv \!\ds\sum_{k=1}^n\frac{R_k}{z_k^2}=0\,,$
\\
$f''(0)=\ds\sum_{k=1}^NR_k\left(-\frac{2}{z_k^3}+\frac{2\mu}{z_k^2}-
\frac{\mu^2}{z_k}\right)=0$\\
\bm{\columnwidth}\hfill$\Longrightarrow\,
C_3\equiv \ds\sum_{k=1}^n\frac{R_k}{z_k^3}=0\,,$\em
\\
so that, at the $j$-th iteration
\be
f^{(j-1)}(0)=0\hspace{2mm}
\Longrightarrow\hspace{2mm}
C_j\equiv\ds \sum_{k=1}^n\frac{R_k}{z_k^{j}}=0\,.
\label{eq:res-cond}
\en
In general, the $C_j$ constants are real, being sum of real and/or pairs of complex conjugate numbers.\\  
The integral representation of Eq.~(\ref{eq:f1}) can be put in a form 
similar to that of Eq.~(\ref{eq:f0}) as 
\be
g_1(Q)\ug\int_0^\infty f(r)j_1(Qr)dr
\no\\\ug
\left.-\frac{j_0(Qr)}{Q}f(r)\right|_0^\infty+\frac{1}{Q}\int_0^\infty j_0(Qr)f'(r) dr
\no\\ 
\ug
\frac{1}{Q}\int_0^\infty j_0(Qr)f'(r) dr\,.
\label{eq:f1-2}
\en
Assuming $l\ge 1$, we define $g(r)=f(r)/r$, with
\be
g(r)\mathop{\propto}_{r\to 0} r^{l-1=l'}\,,\hhh
g(r)\mathop{\propto}_{r\to 0}\frac{e^{-\mu r}}{r^{h+1=h'}}\,.
\nen
and the Mittag-Leffler representation 
\be
g(r)=e^{-\mu r}\sum_{k=1}^{n+1=n'} \frac{D_k}{r-z_k}\,,
\nen
hence the derivative 
\be
f'(r)\ug r g'(r)+g(r)
\no\\\ug
e^{-\mu r}\sum_{k=1}^{n'} D_K\left(-\frac{r}{(r-z_k)^2}
-\frac{\mu r-1}{r-z_k}\right)\,.
\label{eq:fprimo}
\en
\subsubsection{The function $g_0(Q)$}
The analytic expression of $g_0(Q)$ can be obtained 
by integrating the representation of Eq.~(\ref{eq:f0}). In particular we have
\be
g_0(Q)\ug \frac{1}{Q}\int_0^\infty\frac{\sin(Qr)e^{-\mu r}}{r}\sum_{k=1}^N\frac{R_k}{r-z_k}dr\no\\\ug
\frac{1}{Q}\sum_{k=1}^N \frac{R_k}{z_k}\left[
\int_0^\infty \frac{\sin(Qr)e^{-\mu r}}{r-z_k}dr\right.\no\\
&&\left.-
\int_0^\infty \frac{\sin(Qr)e^{-\mu r}}{r}dr\right]\,,
\nen
where the first integral can be computed in terms of the ExpIntegral function, while the second, which does not depend on $z_k$, is equal to the arctangent of $Q/\mu$. Hence we have\be
g_0(Q)\ug\frac{1}{Q}\sum_{k=1}^n\frac{R_k}{z_k}\Bigg[
\frac{
e^{-(\mu-iQ)z_k}E_1[-(\mu-iQ)z_k]}{2i}\Bigg.\no\\
&&\Bigg.-\!\frac{
e^{-(\mu+iQ)z_k}E_1[-(\mu\!+\!iQ)z_k]}{2i}\!-\!
\arctan\left(\!\frac{Q}{\mu}\!\right)\!\!\Bigg]\,.
\nen
The last term vanishes with $l\ge 1$, \ie\ if $f(0)=0$. In general, the profiles that appear in the integrals of Eqs.~(\ref{eq:FTs}) are bounded in the origin and hence, thanks to the factor $r^2$ of the differential $d^3\vec{r}$, the function $f(r)$ has always a zero in $r=0$ of order $l\ge 2$. It follows that the  expression of $g_0(r)$ becomes
\be
g_0(Q)\ug\frac{1}{Q}\sum_{k=1}^n\frac{R_k}{z_k}\Bigg[
\frac{
e^{-(\mu-iQ)z_k}E_1[-(\mu-iQ)z_k]}{2i}\Bigg.\no\\
&&\Bigg.-
\frac{e^{-(\mu+iQ)z_k}E_1[-(\mu+iQ)z_k]}{2i}\Bigg]\,.
\nen
The asymptotic behavior of the ExpIntegral function can be derived from the expansion~\cite{bleistein}
\be
E_1(z)=\frac{e^{-z}}{z}\sum_{j=0}^{N}(-1)^k\frac{j!}{z^j}+\mathcal{O}\left(N!|z|^{-N-1}\right)\,,\hspace{3mm}
z\to\infty\,.
\nen
We consider $g_0(Q)$ in the limit $Q\to\infty$
\be
g_0(Q)\ug
\frac{1}{2iQ}\sum_{k=1}^n\frac{R_k}{z_k}
\sum_{j=0}^{\infty}(-1)^j j!\Bigg[\frac{1}{[-(\mu-iQ)z_k]^{j+1}}
\Bigg.\no\\
&&\Bigg.-\frac{1}{[-(\mu+iQ)z_k]^{j+1}}\Bigg]
\no\\\ug
\frac{1}{2iQ}
\sum_{j=0}^{\infty} j!
\underbrace{\sum_{k=1}^n\frac{R_k}{z_k^{j+2}}}_{C_{j+2}}
\Bigg[\frac{1}{(\mu+iQ)^{j+1}}\Bigg.\no\\
&&\Bigg.-\frac{1}{(\mu-iQ)^{j+1}}\Bigg]\no\\\ug
\frac{1}{Q}
\sum_{j=l-1}^{\infty} j!
C_{j+2}
\frac{\im\left[(\mu-iQ)^{j+1}\right]}{\left(\mu^2+Q^2\right)^{j+1}}\,,
\nen
where the last sum starts from $j=l-1$ because the coefficients $C_j$, defined in Eq.~(\ref{eq:res-cond}), are vanishing for $j\le l$. The
imaginary part can be written in powers of $Q$ so
that
\be
g_0(Q)\ug
\sum_{j=l-1}^{\infty} \frac{j!\,C_{j+2}}{(\mu^2+Q^2)^{j+1}}\no\\
&&\times\!\!\!\!\!
\sum_{s=0}^{{\rm Int}[j/2]}\!\!\!\left(\!\!\begin{array}{c}2s\!+\!1\\ j\!+\!1\\\end{array}\!\!\right)
(-1)^{s+1} Q^{2s}\mu^{j-2s}\,.
\label{eq:g0-sl}
\en
The highest power of $Q$ in the numerator 
coincides with the maximum even number less or equal to $j$.
Hence, for two even-odd consecutive values of $j$, the highest power of $Q$ at numerator remains the same, while that at denominator increases linearly with $j$. This means that the terms at higher orders in $j$
are higher order infinitesimals as $Q\to\infty$.  In particular, the $j$-th term behaves as
\be
\hspace{-2mm}
\mathcal{O}\left[Q^{2{\rm Int}(j/2)-2(j+1)}\right]
=\left\{\begin{array}{lcl}
\ds\mathcal{O}\left[Q^{-4}\right]&\hspace{3mm} &j=1\\
&&\vspace{-3mm}\\
\ds\mathcal{O}\left[Q^{-4}\right]& &j=2\\
&&\vspace{-3mm}\\
\ds\mathcal{O}\left[Q^{-6}\right]& &j=3\\
&&\vspace{-3mm}\\
\ldots && \ldots\\
\end{array}\right.\hspace{-1mm},
\label{eq:power-law}\en
the dominant asymptotic behavior is given by first two terms, with $j=1$ and $j=2$.\\
In the TL region, \ie\ $Q= iq$, with $q>0$,
the expression of Eq.~(\ref{eq:g0-sl}) becomes
\be
g_0(iq)\ug
-\!\!\!\sum_{j=l-1}^{\infty} \frac{j!
C_{j+2}}{\left(\mu^2\!-\!q^2\right)^{j+1}}\!\!\!\!\!
\sum_{s=0}^{{\rm Int}[j/2]}
\left(\!\begin{array}{c}2s\!+\!1\\ j\!+\!1\\\end{array}\!\right)
q^{2s}\mu^{j-2s}\,,\no\\
\label{eq:g0-tl}\en
hence the asymptotic behavior follows the same power law of Eq.~(\ref{eq:power-law}).
\\
More in detail, once the order $l$ (see Eq.~(\ref{eq:constraint})) of the zero, that the profile function posses in $r=0$, is known, 
also the asymptotic behavior in both, SL and TL regions, is obtained as
\be
l \mbox{ odd: } g_0(Q)\!&\ds\mathop{\longrightarrow}_{Q\to\infty}&\!
\frac{(l-1)!\,C_{l+1}}{(\mu^2+Q^2)^{l}}
(-1)^{(l+1)/2} Q^{l-1}
\no\\
&&\sim  
\frac{(l-1)!\,C_{l+1} (-1)^{(l+1)/2}}{Q^{l+1}}\,,
\no\\
l \mbox{ even: } 
g_0(Q)\!&\ds\mathop{\longrightarrow}_{Q\to\infty}&\!
\ds\frac{l!(-1)^{l/2}}{(\mu^2+Q^2)^{l+1}}
\left[Q^l(\mu C_{l+1}-C_{l+2})+\right.\no\\
&&\left.\mu^3 C_{l+1} Q^{l-2} \right]\no\\
&&\sim
\frac{l!(-1)^{l/2}
(\mu C_{l+1}-C_{l+2})}{Q^{l+2}}\,,
%
%\label{eq:asy-0}
\nen
\be
l \mbox{ odd: } 
g_0(iq)\!&\ds \mathop{\longrightarrow}_{q\to\infty}& \!
\ds
-\frac{(l-1)!\,C_{l+1}}{(\mu^2-q^2)^{l}}
q^{l-1}\no\\
&& \sim  \frac{(l-1)!\,C_{l+1} }{q^{l+1}} \,,\no\\
l \mbox{ even: } 
g_0(iq)\!&\ds \mathop{\longrightarrow}_{q\to\infty}& \!
\ds-\frac{l!}{(\mu^2-q^2)^{l+1}}
\left[-q^l(\mu C_{l+1}-C_{l+2})\right.\no\\
&&\left.+\mu^3 C_{l+1} q^{l-2} \right]\no\\
&& \sim- \frac{l! (\mu C_{l+1}-C_{l+2})}{q^{l+2}}\,.
\nen
The electric, isoscalar and isovector, FFs, Eq.~(\ref{eq:FTs}), are obtained through integral representations of type (\ref{eq:f0}) with 
$l=2$ and $l=4$ respectively, and hence
\be
G_E^S(z)\mathop{\sim}_{z\to\infty}z^{-4}\,,\hh
G_E^V(z)\mathop{\sim}_{z\to\infty}z^{-6}\,,
\label{eq:asy-sca}\en
where, the SL and TL limits are considered with $z=Q$ and $z=iq$, respectively.
\subsubsection{The function $g_1(Q)$}
The asymptotic behavior of  $g_1(Q)$ can be achieved by the integral representation of Eq.~(\ref{eq:f1-2}) and the expression of the $f(r)$ derivative given in Eq.~(\ref{eq:fprimo}) as
\be
g_1(Q)\ug
\frac{1}{Q^2}\sum_{k=1}^{n'} D_k
\int_0^\infty \sin(Qr)e^{-\mu r}
\left(-\frac{1}{(r-z_k)^2}\right.\no\\
&&\left.
+\frac{1/z_k-\mu}{r-z_k}-\frac{1}{z_k r}\right)dr
\no\\
\ug\frac{1}{Q^2}\!\!\sum_{k=1}^{N'}\! D_k\Bigg\{\!\!
\frac{ 
(\mu -iQ)e^{-(\mu -iQ)z_k}E_1[-(\mu -iQ)z_k]}{2i}
\Bigg.\no\\
&&-
\frac{(\mu +iQ)e^{-(\mu +iQ)z_k}E_1[-(\mu +iQ)z_k]}{2i}
\no\\
&&+(1/z_k-\mu)\frac{
e^{-(\mu-iQ)z_k}E_1[-(\mu-iQ)z_k]}{2i}\no\\
&&-
(1/z_k-\mu)\frac{e^{-(\mu+iQ)z_k}E_1[-(\mu+iQ)z_k]}{2i}
\no\\
&&-\frac{\arctan(Q/\mu)}{z_k}
\Bigg\}
\,.
\nen
The terms proportional to $\mu$ cancel and that proportional to the arctangent, assuming $l\ge 1$, is vanishing and, using $z=-(\mu-i Q)$, we have
\be
g_1(Q)\ug\sum_{k=1}^{N'} D_k\Bigg\{
-\frac{e^{z z_k}E_1(zz_k)+e^{z^*z_k}E_1(z^*z_k)}{2Q}\no\\
&&+\frac{e^{z z_k}E_1(zz_k)-e^{z^*z_k}E_1(z^*z_k)}{2i Q^2z_k}
\Bigg\}\,.
\nen
By taking advantage from the asymptotic series of the ExpIntegral function, the first term can be written as
\be
&&\ds-\frac{e^{z z_k}E_1(zz_k)+e^{z^*z_k}E_1(z^*z_k)}{2Q}
\no\\
&&\ds\simeq-\frac{1}{2Q}
\sum_{s=0}^\infty (-1)^ss!\left[\frac{1}{(zz_k)^{s+1}}
+\frac{1}{(z^*z_k)^{s+1}}\right]
\no\\
&&\ds\simeq\!\!
\sum_{s=0}^\infty \frac{s!}{z_k^{s+1}|z|^{2s+2}} \!\!\!\!\!\!
\sum_{t=0}^{{\rm Int}[(s+1)/2]}\!\!\left(\!\begin{array}{c}2t\\ s\!+\!1\\\end{array}\!\right)(-1)^t
Q^{2t-1}\mu^{s+1-2t}\,,
\nen 
while for the second term we have
\be
&&\frac{e^{z z_k}E_1(zz_k)-e^{z^*z_k}E_1(z^*z_k)}{2iQ^2 z_k}\no\\
&&\simeq
-
\sum_{s=0}^\infty \frac{s!}{z_k^{s+2}|z|^{2s+2}}\!\!\!\!
\sum_{t=0}^{{\rm Int}[s/2]}\!\!\left(\!\begin{array}{c}2t\!+\!1\\ s\!+\!1\\\end{array}\!\right)
(-1)^tQ^{2t-1}\mu^{s-2t}\,.
\nen
The complete expression is then
\be
g_1(Q)\!&\simeq &\!
\sum_{s=l-1}^\infty \frac{s!C'_{s+1}}{(\mu^2+Q^2)^{s+1}} \no\\
&&\times
\sum_{t=0}^{{\rm Int}[(s+1)/2]}\left(\begin{array}{c}2t\\ s+1\\\end{array}\right)(-1)^t
Q^{2t-1}\mu^{s+1-2t}
\no
\Bigg.\\
&&-\Bigg. 
\sum_{s=l-2}^\infty \frac{s!C'_{s+2}}{(\mu^2+Q^2)^{s+1}}\no\\
&&\times
\sum_{t=0}^{{\rm Int}[s/2]}\left(\begin{array}{c}2t+1\\ s+1\\\end{array}\right)
(-1)^tQ^{2t-1}\mu^{s-2t}
%
%\no\\\={\simeq}
%g_1^{(l-1)}(Q)-g_1^{(l-2)}(Q)
\,,
\nen
where the constants $C'_t$ are defined as those of Eq.~(\ref{eq:res-cond}), but for the residues $D_k$, \ie\ $C'_{t}=\sum_{k=1}^{n'}D_k/z_k^t$ and the lower limits of indexes $s$ account for the behavior of $f(r)$ at $r=0$. The TL asymptotic behavior can be obtained from the previous expression, by setting $Q=iq$, with $q\to \infty$, \ie
\be
g_1(iq)\!&\simeq&\!
-i\sum_{s=l-1}^\infty \frac{s!C'_{s+1}}{(\mu^2-q^2)^{s+1}} \no\\
&&\times 
\sum_{t=0}^{{\rm Int}[(s+1)/2]}\left(\begin{array}{c}2t\\ s+1\\\end{array}\right)
q^{2t-1}\mu^{s+1-2t}
\no\\
&&+i\!\!\!
\sum_{s=l-2}^\infty \frac{s!C'_{s+2}}{(\mu^2\!-\!q^2)^{s+1}}\!\!\!\!
\sum_{t=0}^{{\rm Int}[s/2]}\!\!\left(\!\begin{array}{c}2t\!+\!1\\ s\!+\!1\\\end{array}\!\right)
q^{2t-1}\mu^{s-2t}\,.
\nen
The leading terms are
\be
\mbox{$l$ even: }
g_1(Q)&\ds\mathop{\sim}_{Q\to\infty}&\ds
\frac{l(l-1)! C'_{l} (-1)^{l/2}}{Q^{l+1}}
\no\\
&&\label{eq:asy-01}
\\
\mbox{$l$ odd: }
g_1(Q)&\ds\mathop{\sim}_{Q\to\infty}&
\ds\frac{(-1)^{(l+1)/2}(l-1)!(l+1-\delta_{1,l})}{Q^{l+2}} \no\\
&&\times\Big(
C'_{l+1}-\mu\,C'_l
\Big)
\nen
\be
\mbox{$l$ even: }
g_1(iq)&\ds\mathop{\sim}_{q\to\infty}&
\ds-i\frac{l(l-1)! C'_{l} }{q^{l+1}}
\no\\
\label{eq:asy-02}\\
\mbox{$l$ odd: }
g_1(iq)&\ds\mathop{\sim}_{q\to\infty}&
\ds-i\frac{(l-1)!(l\!+\!1\!-\!\delta_{1,l})\Big(
C'_{l+1}\!-\!\mu\,C'_l
\Big)}{q^{l+2}} 
\nen
The magnetic, isoscalar and isovector, FFs, Eq.~(\ref{eq:FTs}), are obtained through integral representations of type (\ref{eq:f1}) with an additional factor $Q^{-1}$, or $(iq)^{-1}$, and $l=3$ in both cases, it follows that
\be
G_M^S(z)\mathop{\sim}_{z\to\infty}z^{-6}\,,\hh
G_M^V(z)\mathop{\sim}_{z\to\infty}z^{-6}\,,
\label{eq:asy-vec}\en
where, as in the $g_0$ case, SL and TL limits are considered by setting $z=Q$ and $z=iq$, respectively.
%
%
%
%%%%%%%%%%%%%%%%%%%%%%%%%%%%%%%%%%%%%%%%%%%%%%%%%%%%%%%%%%
%
%  Bibliography
%

%
%

\begin{thebibliography}{}
%
%1
\bibitem{sakurai}
J.~J.~Sakurai, {\it Currents and Mesons}, The University of Chicago Press (1969).
%
%2
\bibitem{lattice}
S.~Aoki, Y.~Aoki, C.~Bernard, T.~Blum, G.~Colangelo, M.~Della Morte, S.~DÃ¼rr and A.~X.~El Khadra {\it et al.},
  %``Review of lattice results concerning low-energy particle physics,''
  Eur.\ Phys.\ J.\ C {\bf 74} (2014) 9,  2890
  [arXiv:1310.8555 [hep-lat]].
  %%CITATION = ARXIV:1310.8555;%%
  %105 citations counted in INSPIRE as of 12 Nov 2014
%
%3
\bibitem{chipt}
T.~Bauer, J.~C.~Bernauer and S.~Scherer,
  %``Electromagnetic form factors of the nucleon in effective field theory,''
  Phys.\ Rev.\ C {\bf 86} (2012) 065206
  [arXiv:1209.3872 [nucl-th]] and references therein.
  %%CITATION = ARXIV:1209.3872;%%
  %5 citations counted in INSPIRE as of 12 Nov 2014
%
%4
\bibitem{pedro} 
  P.~Alberto, E.~Ruiz Arriola, M.~Fiolhais, F.~Grummer, J.~N.~Urbano and K.~Goeke,
  %``Nucleon Form-factors in the Projected Linear Chiral Soliton Model,''
  Phys.\ Lett.\ B {\bf 208}, 75 (1988);
  %%CITATION = PHLTA,B208,75;%%
  %19 citations counted in INSPIRE as of 10 Jul 2015
%\cite{RuizArriola:1989ca}
%\bibitem{RuizArriola:1989ca} 
  E.~Ruiz Arriola, P.~Alberto, J.~N.~Urbano and K.~Goeke,
  %``Hedgehog Structures In General Quark - Meson Lagrangians,''
  Z.\ Phys.\ A {\bf 333}, 203 (1989);
  %%CITATION = ZEPYA,A333,203;%%
  %4 citations counted in INSPIRE as of 10 Jul 2015
%\cite{Alberto:1990ru}
%\bibitem{Alberto:1990ru} 
  P.~Alberto, E.~Ruiz Arriola, M.~Fiolhais, K.~Goeke, F.~Grummer and J.~N.~Urbano,
  %``Form-factors in the projected linear chiral sigma model,''
  Z.\ Phys.\ A {\bf 336}, 449 (1990);
  %%CITATION = ZEPYA,A336,449;%%
  %11 citations counted in INSPIRE as of 10 Jul 2015
%\cite{Alberto:1990zs}
%\bibitem{Alberto:1990zs} 
  P.~Alberto, E.~Ruiz Arriola, J.~N.~Urbano and K.~Goeke,
  %``Form-factors in the projected chiral soliton model with vector mesons,''
  Phys.\ Lett.\ B {\bf 247}, 210 (1990);
  %%CITATION = PHLTA,B247,210;%%
  %3 citations counted in INSPIRE as of 10 Jul 2015
%\cite{RuizArriola:1995zr}
%%\bibitem{RuizArriola:1995zr} 
  E.~Ruiz Arriola, P.~Alberto, J.~N.~Urbano and K.~Goke,
  %``The Projected chiral soliton model with vector mesons,''
  Nucl.\ Phys.\ A {\bf 591}, 561 (1995).
  %%CITATION = NUPHA,A591,561;%%
  %1 citations counted in INSPIRE as of 10 Jul 2015
%
%5
\bibitem{largeNc}
P.~Masjuan, E.~Ruiz Arriola and W.~Broniowski,
  %``Meson dominance of hadron form factors and large-Nc phenomenology,''
  Phys.\ Rev.\ D {\bf 87} (2013) 014005
  [arXiv:1210.0760 [hep-ph]] and references therein.
  %%CITATION = ARXIV:1210.0760;%%
  %20 citations counted in INSPIRE as of 12 Nov 2014
%
%6
\bibitem{holographic}
D.~K.~Hong, M.~Rho, H.~U.~Yee and P.~Yi,
  %``Nucleon form-factors and hidden symmetry in holographic QCD,''
  Phys.\ Rev.\ D {\bf 77} (2008) 014030
  [arXiv:0710.4615 [hep-ph]] and references therein.
  %%CITATION = ARXIV:0710.4615;%%
  %75 citations counted in INSPIRE as of 12 Nov 2014 
%
%7
\bibitem{Perdrisat} 
C.~F.~Perdrisat, V.~Punjabi and M.~Vanderhaeghen,
  %``Nucleon Electromagnetic Form Factors,''
  Prog.\ Part.\ Nucl.\ Phys.\  {\bf 59} (2007) 694
  [hep-ph/0612014].
  %%CITATION = HEP-PH/0612014;%%
  %226 citations counted in INSPIRE as of 12 Nov 2014%5
%
%8  
\bibitem{egle-simone}
S.~Pacetti, R.~Baldini Ferroli and E.~Tomasi-Gustafsson,
  %``Proton electromagnetic form factors: Basic notions, present achievements and future perspectives,''
  Phys.\ Rept.\  {\bf 550-551} (2014) 1.
  %%CITATION = PRPLC,550-551,1;%%
  %1 citations counted in INSPIRE as of 31 mar 2015
%
%9
\bibitem{Salme} A.~Denig and G.~Salm\`e,
  %``Nucleon Electromagnetic Form Factors in the Timelike Region,''
  Prog.\ Part.\ Nucl.\ Phys.\  {\bf 68} (2013) 113
  [arXiv:1210.4689 [hep-ex]].
  %%CITATION = ARXIV:1210.4689;%%
  %6 citations counted in INSPIRE as of 12 Nov 2014%7
%
%10
\bibitem{besiii}
D.~M.~Asner, T.~Barnes, J.~M.~Bian, I.~I.~Bigi, N.~Brambilla, I.~R.~Boyko, V.~Bytev and K.~T.~Chao {\it et al.},
  %``Physics at BES-III,''
  Int.\ J.\ Mod.\ Phys.\ A {\bf 24} (2009) S1
  [arXiv:0809.1869 [hep-ex]].
  %%CITATION = ARXIV:0809.1869;%%
  %167 citations counted in INSPIRE as of 12 Nov 2014
%
%11
\bibitem{snd}
M.~N.~Achasov, A.~Y.~Barnyakov, K.~I.~Beloborodov, A.~V.~Berdyugin, D.~E.~Berkaev, A.~G.~Bogdanchikov, A.~A.~Botov and D.~A.~Bukin {\it et al.},
  %``First results of spherical neutral detector (SND) experiments at VEPP-2000,''
  Prog.\ Part.\ Nucl.\ Phys.\  {\bf 67} (2012) 594.
  %%CITATION = PPNPD,67,594;%%
  %3 citations counted in INSPIRE as of 12 Nov 2014
%
%12
\bibitem{cmd3}
G.~V.~Fedotovich [CMD-3 Collaboration],
  %``CMD-3 detector for VEPP-2000,''
  Nucl.\ Phys.\ Proc.\ Suppl.\  {\bf 162} (2006) 332.
  %%CITATION = NUPHZ,162,332;%%
  %16 citations counted in INSPIRE as of 12 Nov 2014
%
%13
\bibitem{panda}
E.~Tomasi-Gustafsson {\it et al.}  [PANDA Collaboration],
  %``Electromagnetic proton form factors: perspectives for PANDA,''
  EPJ Web Conf.\  {\bf 66} (2014) 06024.
  %%CITATION = 00776,66,06024;%%
%
%14
\bibitem{Holzwarth1} 
G.~Holzwarth,
  %``Electromagnetic nucleon form-factors and their spectral functions in soliton models,''
  Z.\ Phys.\ A {\bf 356} (1996) 339
  [hep-ph/9606336].
  %%CITATION = HEP-PH/9606336;%%
  %94 citations counted in INSPIRE as of 12 Nov 2014%8
%
%15
\bibitem{Holzwarth2} 
G.~Holzwarth,
  %``Electromagnetic form-factors of the nucleon in chiral soliton models,''
  hep-ph/0511194.
  %%CITATION = HEP-PH/0511194;%%
  %6 citations counted in INSPIRE as of 12 Nov 2014
%
%16
\bibitem{Hammer} 
H.~W.~Hammer,
  %``Nucleon form-factors in the space - like and time - like regions,''
  eConf C {\bf 010430} (2001) W08
  [hep-ph/0105337].
  %%CITATION = HEP-PH/0105337;%%
  %8 citations counted in INSPIRE as of 12 Nov 2014
%
%17
\bibitem{Egle} 
 E.~A.~Kuraev, E.~Tomasi-Gustafsson and A.~Dbeyssi,
  %``A Model for space and time-like proton (neutron) form factors,''
  Phys.\ Lett.\ B {\bf 712} (2012) 240
  [arXiv:1106.1670 [hep-ph]].
  %%CITATION = ARXIV:1106.1670;%%
  %2 citations counted in INSPIRE as of 12 Nov 2014
%  
%18
\bibitem{Pacetti} E.~L.~Lomon and S.~Pacetti,
  Phys.\ Rev.\ D85 (2012) 113004
  [Erratum-ibid.\ D86 (2012) 039901]
  [arXiv:1201.6126 [hep-ph]] and references therein.

%20
\bibitem{Bardini} A. Bardini and A. Drago {\itshape Fattori di Forma
    Elettromagnetici del Nucleone nella Regione Tempo}, Master Degree thesis, Universit\`a Degli Studi di Ferrara A. A. 2000/2001, not published. 
%
%21
\bibitem{Braaten} 
E.~Braaten, S.~M.~Tse and C.~Willcox,
 Phys. Rev. Lett. {\bf 56} (1986) 2008;
 E. Braaten, S. M. Tse and C. Willcox, Phys. Rev. D {\bf 34} (1986) 1482.
%
%
%19
\bibitem{Skyrme} 
T. H. R. Skyrme, Proc. Roy. Soc. {\bf A} 260 (1961) 127.
%
%22
%\bibitem{mathematica} 
%Wolfram Research, Inc., Mathematica, Version 7.0, Champaign, IL (2010).
%
%23
\bibitem{Foldy} 
L.~L.~Foldy, Phys. Rev. {\bf 87} (1952) 688.
%
%24
\bibitem{Sachs} 
  F.~J.~Ernst, R.~G.~Sachs and K.~C.~Wali,
  %``Electromagnetic form factors of the nucleon,''
  Phys.\ Rev.\  {\bf 119} (1960) 1105;
  %%CITATION = PHRVA,119,1105;%%
  %116 citations counted in INSPIRE as of 25 Apr 2014
R.~Sachs,
  %``Asymptotic symmetries in gravitational theory,''
  Phys.\ Rev.\  {\bf 128} (1962) 2851.
  %%CITATION = PHRVA,128,2851;%%
  %151 citations counted in INSPIRE as of 25 Apr 2014
%
%25
\bibitem{polarization-tl} 
 A.~Z.~Dubnickova, S.~Dubnicka and M.~P.~Rekalo,
  %``Investigation of the nucleon electromagnetic structure by polarization effects in e+ e- ---> N anti-N processes,''
  Nuovo Cim.\ A {\bf 109} (1996) 241;
  %%CITATION = NUCIA,A109,241;%%
  %49 citations counted in INSPIRE as of 12 Nov 2014
E.~Tomasi-Gustafsson, F.~Lacroix, C.~Duterte and G.~I.~Gakh,
  %``Nucleon electromagnetic form-factors and polarization observables in space-like and time-like regions,''
  Eur.\ Phys.\ J.\ A {\bf 24} (2005) 419
  [nucl-th/0503001].
  %%CITATION = NUCL-TH/0503001;%%
  %47 citations counted in INSPIRE as of 12 Nov 2014
%
%26
\bibitem{Drell} 
S. D. Drell, D. J. Levy and T.-M. Yan, Phys. Rev. {\bf 187}
  (1969) 2159.
%
%27
%\bibitem{Drell2} 
%S. D. Drell and F. Zachariasen, {\itshape
%    Electromagnetic Structure of the Nucleons}, Oxford Univeristy
%  Press, London (1961).
%
%28
\bibitem{Brodsky} 
V. A. Matveev, R. M. Muradian, and
  A. N. Tavkhelidze, Lett. Nuovo Cim. {\bf 7} (1973) 719;
S. J. Brodsky and G. R. Farrar, Phys. Rev. Lett. {\bf 31} (1973) 1153;
S. J. Brodsky and G. P. Lepage, Phys. Rev. D {\bf 22} (1980) 2157.
%
%29
\bibitem{Titmarch} 
E. C. Titchmarsh, {\itshape Theory of functions}, Oxford University Press, London (1939).
%
%30
\bibitem{arfken} 
G.~B.~Arfken, H.~J.~Weber and F.~E. Harris, 
{\it Mathematical Methods for Physicists: A Comprehensive Guide}, Academic Press, Oxford (2012).
%
%31
\bibitem{noi-meissner}
R.~Baldini, S.~Pacetti, A.~Zallo and A.~Zichichi,
  %``Unexpected features of e+ e- ---> p anti-p and e+ e- ---> Lambda anti-Lambda cross sections near threshold,''
  Eur.\ Phys.\ J.\ A {\bf 39} (2009) 315
  [arXiv:0711.1725 [hep-ph]];
  %%CITATION = ARXIV:0711.1725;%%
  %29 citations counted in INSPIRE as of 29 juil. 2015
  %
  R.~Baldini Ferroli, S.~Pacetti, A.~Zallo and A.~Zichichi,
  %``Experimental evidence for Pointlike Baryons at ${q^2}= 4M_\mathcal{B}^2$,''
  Subnucl.\ Ser.\  {\bf 47} (2011) 155;
  %%CITATION = SUSEE,47,155;%%
  %
  R.~Baldini Ferroli, S.~Pacetti and A.~Zallo,
  %``Time-like baryon form factors near threshold: Status and perspectives,''
  Nucl.\ Phys.\ Proc.\ Suppl.\  {\bf 219-220} (2011) 32;
  %%CITATION = NUPHZ,219-220,32;%%
  %2 citations counted in INSPIRE as of 29 juil. 2015
  %
  R.~Baldini Ferroli and S.~Pacetti,
  %``Baryon form factors at threshold,''
  Nucl.\ Phys.\ Proc.\ Suppl.\  {\bf 225-227} (2012) 211;
  %%CITATION = NUPHZ,225-227,211;%%
  %1 citations counted in INSPIRE as of 29 Jul 2015
  %
  R.~Baldini Ferroli, S.~Pacetti and A.~Zallo,
  %``No Sommerfeld resummation factor in $e^+e^- -> p \bar{p}$?,''
  Eur.\ Phys.\ J.\ A {\bf 48} (2012) 33
  [arXiv:1008.0542 [hep-ph]];
  %%CITATION = ARXIV:1008.0542;%%
  %8 citations counted in INSPIRE as of 29 Jul 2015
  %
J.~Haidenbauer, X.-W.~Kang and U.-G.~Meissner,
  %``The electromagnetic form factors of the proton in the timelike region,''
  Nucl.\ Phys.\ A {\bf 929} (2014) 102
  [arXiv:1405.1628 [nucl-th]];
  %%CITATION = ARXIV:1405.1628;%%
  %4 citations counted in INSPIRE as of 29 Jul 2015
%
    A.~Bianconi and E.~Tomasi-Gustafsson,
  %``Periodic interference structures in the timelike proton form factor,''
  Phys.\ Rev.\ Lett.\  {\bf 114} (2015) 23,  232301
  [arXiv:1503.02140 [nucl-th]];
  %%CITATION = ARXIV:1503.02140;%%
  %1 citations counted in INSPIRE as of 29 juil. 2015
  I.~T.~Lorenz, H.-W.~Hammer and U.-G.~Meissner,
  %``New structures in the proton-antiproton system,''
  arXiv:1506.02282 [hep-ph].
  %%CITATION = ARXIV:1506.02282;%%
  %2 citations counted in INSPIRE as of 29 juil. 2015
%
%32
\bibitem{BaBar} 
B. Aubert {\it et al.} (BaBar Collaboration), Phys. Rev. D {\bf 73} (2006) 012005. 
%
%33
\bibitem{PS170} 
G. Bardin {\it et al.} (PS170 Collaboration),
  Nucl. Phys. B {\bf 411} (1994) 3.
%
%34
\bibitem{Witten1} 
E. Witten, Nucl. Phys. B {\bf 160} (1979) 57.
%
%35
\bibitem{Hofft} 
G.~'t Hooft, Nucl. Phys. B {\bf 72} (1974) 461.
%
%36
\bibitem{Adkins1} 
G. S. Adkins, C. R. Nappi and E. Witten, Nucl. Phys. B {\bf 228}
  (1983) 552.
%
%37
\bibitem{Adkins2} 
G. S. Adkins and C. R. Nappi, Nucl. Phys. B {\bf 233} (1984) 109.
%
%38
\bibitem{sigma-model} 
M.~Gell-Mann and M.~Levy,
  %``The axial vector current in beta decay,''
  Nuovo Cim.\  {\bf 16} (1960) 705.
  %%CITATION = NUCIA,16,705;%%
  %1496 citations counted in INSPIRE as of 02 May 2014
%
%42
\bibitem{pdg}
K. A. Olive {\it et al.} (Particle Data Group), Chin. Phys. C {\bf 38} (2014) 090001.
%39
\bibitem{Zumino} 
J. Wess and B. Zumino, Phys. Lett. B {\bf 37} (1971) 95.
%
%40
\bibitem{Witten3} 
E. Witten, Nucl. Phys. B {\bf 223} (1983) 422.
%
%41
\bibitem{Finkel} 
D. Finkelstein, J. Rubinstein, J. Math. Phys. {\bf 9} (1968) 1762.
%
%
%43
\bibitem{Ji} 
X. Ji, Phys. Lett. B {\bf 254} (1990) 456.
%
%44
\bibitem{ahlfors} 
L.~Ahlfors, {\it Complex analysis}, McGraw Hill, (1979).
%
%45
\bibitem{stegun} 
M.~Abramowitz and I.~A.~Stegun, {\it Handbook of Mathematical Functions: with Formulas, Graphs, and Mathematical Tables}, Courier Dover Publications (2012).
%
%46
\bibitem{bleistein} 
N.~Bleistein and R.~A.~Handelsman,
{\it Asymptotic Expansions of Integrals}, Dover Books on Mathematics Series, Dover Publications (1975).
%
%47
\bibitem{babar-ratio}
  J.~P.~Lees {\it et al.}  [BaBar Collaboration],
  %``Study of $e^+e^- \to p \bar{p}$ via initial-state radiation at BABAR,''
  Phys.\ Rev.\ D {\bf 87} (2013) 9,  092005
  [arXiv:1302.0055 [hep-ex]].
  %%CITATION = ARXIV:1302.0055;%%
  %11 citations counted in INSPIRE as of 10 Nov 2014
%
%48
\bibitem{lear-ratio}
  G.~Bardin, G.~Burgun, R.~Calabrese, G.~Capon, R.~Carlin, P.~Dalpiaz, P.~F.~Dalpiaz and J.~Derre {\it et al.},
  %``Determination of the electric and magnetic form-factors of the proton in the timelike region,''
  Nucl.\ Phys.\ B {\bf 411} (1994) 3.
  %%CITATION = NUPHA,B411,3;%%
  %127 citations counted in INSPIRE as of 10 Nov 2014
%
%%\cite{Belushkin:2006qa}
\bibitem{Belushkin:2006qa}
M.~A.~Belushkin, H.~W.~Hammer and U.~G.~Meissner,
%``Dispersion analysis of the nucleon form-factors including meson continua,''
Phys. Rev. C \textbf{75} (2007), 035202
%doi:10.1103/PhysRevC.75.035202
[arXiv:hep-ph/0608337 [hep-ph]].
%211 citations counted in INSPIRE as of 07 Apr 2021
%
%\cite{Adamuscin:2016rer}
\bibitem{Adamuscin:2016rer}
C.~Adamuscin, E.~Bartos, S.~Dubnicka and A.~Z.~Dubnickova,
%``Numerical values of $f^F$, $f^D$, $f^S$ coupling constants in $SU(3)$ invariant interaction Lagrangian of vector-meson nonet with $1/2^+$ octet baryons,''
Phys. Rev. C \textbf{93} (2016) no.5, 055208
%doi:10.1103/PhysRevC.93.055208
[arXiv:1601.06190 [hep-ph]].
%9 citations counted in INSPIRE as of 07 Apr 2021
%
%\cite{Iachello:2004aq}
\bibitem{Iachello:2004aq}
F.~Iachello and Q.~Wan,
%``Structure of the nucleon from electromagnetic timelike form factors,''
Phys. Rev. C \textbf{69} (2004), 055204.
%doi:10.1103/PhysRevC.69.055204
%86 citations counted in INSPIRE as of 07 Apr 2021
%
%49
%
%\cite{Greco}
\bibitem{Greco}
M.~Greco, G.~Penso and Y.~Srivastava,
%``{QCD} and Duality in $e^+ e^-$ Annihilation,''
Phys. Rev. D \textbf{21} (1980), 2520.
%oi:10.1103/PhysRevD.21.2520
%26 citations counted in INSPIRE as of 07 Apr 2021
%
\bibitem{Meissner} 
U.-G. Meissner, N. Kaiser and W. Weise,
  Nucl. Phys. A {\bf 466} (1987) 685;
 U.-G. Meissner, Phys. Rep. {\bf 161} (1988) 213.
%
 \bibitem{pade0} G.~A.~Baker and P.~Graves-Morris, {\it Pad\'e Approximants}, Encyclopedia of Mathematics and its Applications, Cambridge Univ. Press 1996;
	G.~A.~Baker, {\it Essentials of Pad\'e Approximants}, Academic Press 1975.
%	
%\cite{Masjuan:2007ay}
\bibitem{Masjuan:2007ay}
  P.~Masjuan and S.~Peris,
  %``A Rational approach to resonance saturation in large-N(c) QCD,''
  JHEP {\bf 0705} (2007) 040
  doi:10.1088/1126-6708/2007/05/040
  [arXiv:0704.1247 [hep-ph]].
  %%CITATION = doi:10.1088/1126-6708/2007/05/040;%%
  %98 citations counted in INSPIRE as of 21 May 2019
%
\end{thebibliography}
\end{document}